\documentclass[3p,a4paper]{elsarticle}
\usepackage[utf8]{inputenc}
\usepackage{subfigure}

\title{EPi-cKANs: Elasto-Plasticity Informed   Kolmogorov-Arnold Networks Using Chebyshev Polynomials}
\author[myUaddress]{Farinaz Mostajeran}
\author[myUaddress]{Salah A Faroughi \corref{mycorrespondingauthor}}

\cortext[mycorrespondingauthor]{Corresponding author: salah.faroughi@utah.edu}

\address[myUaddress]{Department of Chemical Engineering, University of Utah, 
 Salt Lake City, Utah  84112, USA
}

\date{\today}

\usepackage{mwe}

\usepackage{setspace} 
\usepackage{tabularx}
\usepackage{lineno,hyperref}
\usepackage{amsmath}
\usepackage{bm}
\usepackage{amssymb}
\usepackage[]{amsmath}
\usepackage{upgreek}

\let\today\relax
\makeatletter
\def\ps@pprintTitle{%
    \let\@oddhead\@empty
    \let\@evenhead\@empty
    \def\@oddfoot{\footnotesize\itshape
         {Submitted preprint- October 2024} \hfill\today}%
    \let\@evenfoot\@oddfoot
    }
\makeatother

\usepackage{pgfplots}
\pgfplotsset{compat=1.5}
\usepgfplotslibrary{units}
\usetikzlibrary{spy}
\usetikzlibrary{pgfplots.groupplots}

\usepackage{epstopdf}
\usepackage{units}
\usepackage{lscape}
\usepackage{booktabs}
\usepackage{gensymb}
\usepackage{multirow}

\usepackage{graphicx}
\usepackage{caption}
\usepackage{subcaption}

\usepackage{booktabs}
\usepackage{multirow}
\usepackage{caption}
\usepackage{subcaption}
\usepackage{float} 

\usepackage{soul,color}
\soulregister\cite7
\soulregister\ref7
\soulregister\pageref7

\usepackage[flushleft]{threeparttable}
\usepackage{tikz}
\usepackage{placeins}
\usepackage{color}
\usepackage{hyperref}

\usepackage{stackengine}

\usepackage{array}
\usepackage{tabularx}
\usepackage{pdfpages}
\usepackage{natbib}
\usepackage{textcomp, gensymb}
\usepackage{booktabs} 

\begin{document}

\begin{abstract}

Multilayer perceptron (MLP) networks are predominantly used to develop data-driven constitutive models for granular materials. They offer a compelling alternative to traditional physics-based constitutive models in predicting nonlinear responses of these materials, e.g., elasto-plasticity, under various loading conditions. To attain the necessary accuracy, MLPs often need to be sufficiently deep or wide, owing to the curse of dimensionality inherent in these problems. To overcome this limitation, we present an elasto-plasticity informed Chebyshev-based Kolmogorov-Arnold network (EPi-cKAN) in this study. This architecture leverages the benefits of KANs and augmented Chebyshev polynomials, as well as integrates physical principles within both the network structure and the loss function. The primary objective of EPi-cKAN is to provide an accurate and generalizable function approximation for non-linear stress-strain relationships, using fewer parameters compared to standard MLPs.  To evaluate the efficiency, accuracy, and generalization capabilities of EPi-cKAN in modeling complex elasto-plastic behavior, we initially compare its performance with other cKAN-based models, which include purely data-driven parallel and serial architectures. Furthermore, to differentiate EPi-cKAN's distinct performance, we also compare it against purely data-driven and physics-informed MLP-based methods. Lastly, we test EPi-cKAN's ability to predict blind strain-controlled paths that extend beyond the training data distribution to gauge its generalization and predictive capabilities. Our findings indicate that, even with limited data and fewer parameters compared to other approaches, EPi-cKAN provides superior accuracy in predicting stress components and demonstrates better generalization when used to predict sand elasto-plastic behavior under blind triaxial axisymmetric strain-controlled loading paths.

\end{abstract}

\begin{keyword}
    Physics-informed Neural Networks\sep%
    Kolmogorov-
Arnold Network \sep%
    Chebyshev Polynomials \sep%
    Elasto-plasticity \sep%
     Constitutive Models\sep%
    Scientific Machine Learning  
\end{keyword}

\maketitle

\section{Introduction}\label{sec:Intro}

The study of elasto-plastic constitutive models for granular materials is critical in civil engineering, especially geotechnical engineering, due to their complex and nonlinear behaviors under different loading conditions \cite{Choo2SH, Pietruszczak47H}. 
These models help predict the stress-strain relationship and the transition from elastic to plastic deformation, which is critical for the design of foundations, retaining structures, and other civil infrastructure \cite{ur2021three, halabian2022unified, wijesinghe2024slope}. 
Sand, as a granular material, exhibits complex mechanical behaviors, including anisotropy, dilatancy, and strain-softening, making developing accurate models challenging \cite{xue2021three,
yang2022analysis, datta2024angle, wang2024strength}. 
Understanding the elasto-plastic behavior of sand enables better predictions of load-bearing capacity and failure thresholds, thereby minimizing risks in construction and improving the safety and longevity of engineering projects.
In particular, the realistic simulation of both elastic and plastic responses under varying loading conditions is crucial for ensuring the reliability and safety of structures exposed to earthquakes, loading, and environmental changes \cite{Kim, Yao}.
The relevance of these models is highlighted in numerous studies, including the development of constitutive models for sand behavior under cyclic loading, the incorporation of critical state soil mechanics into elasto-plastic frameworks, and the modeling of anisotropic and strain-hardening characteristics in sand \cite{WangR, Wang}. Despite advances in the area of elasto-plastic models for sand, several critical obstacles continue to challenge researchers. 
One of the main difficulties is the accurate representation of the highly non-linear behavior of sand under varying stress conditions \cite{WangY}. 
Traditional models often struggle to capture the transition between the elastic and plastic states, particularly when considering factors such as strain localization and liquefaction \cite{Oliveira, Gao}. In fact, parameter identification for elasto-plastic models is often complicated by the inherent heterogeneity of sand and the difficulty of obtaining reliable experimental data \cite{Liu40H, Najjar44H, Zhang60H}. Furthermore, the main challenge lies in the computational demands of these models, as simulations of large-scale problems require significant computing resources \cite{Matou4SH}. Overcoming these challenges is vital to advance the application of elasto-plastic models in academic and practical engineering contexts.

To address computational demands, several mesh-based methods (e.g., the finite element method, FEM, \cite{de, Lisjak}) and mesh-free approaches (e.g., smoothed particle hydrodynamics, SPH, \cite{Liu, Bui,hoang2024development}) have been developed. These methods attempted to accurately capture the elasto-plastic behavior of sand under various loading conditions, while also optimizing computational resources when handling complex geometries and boundary conditions. For instance, in \cite{Bammer}, the authors analyzed a higher-order mixed-FEM for elasto-plasticity with linear kinematic hardening and introduced a Lagrange multiplier to address non-differentiability in the plasticity functional.
In \cite{GPU}, the authors also introduced a novel GPU-based parallel framework for elasto-plastic FEM, significantly improving computational efficiency for large-scale problems through optimized memory access and fine-grain parallelism.\cite{jiao2024numerical} recently used SPH technique to efficiently analysing large deformations in sand-like granular soils. The Gudehus-Bauer hypoplastic model is implemented within the SPH framework to capture the pressure- and density-dependent behavior of the soil, along with an adaptive stress integration scheme to ensure stability and accuracy. However, despite these improvements, classical methods still face significant challenges when dealing with highly non-linear behavior and the inherent heterogeneity of sand and parameter identification for elasto-plastic models. A major limitation of these classical approaches is their reliance on accurate material models and parameter calibration, which often require extensive experimental data. Due to the inherent variability and complexity of sand, this task is difficult, and classical methods may struggle with generalizing across different sand types or environmental conditions. 
To address these limitations, researchers are increasingly turning to data-driven approaches, including neural networks  \cite{patino2023back, datta2023multihead, soltanmohammadi2023comparative} and physics-guided \cite{pawar2024geo}, physics-informed \cite{faroughi2023physics, datta2022physics, Haghighat, Karapiperis32H, raissi2019physics}, and physics-encoded \cite{faroughi2024physics} neural networks, which are capable of handling large datasets and simulations while incorporating physical laws into machine learning models to some extent. These approaches offer significant potential for reducing the dependency on experimental data while improving generalization across different materials and conditions, marking a promising shift away from traditional numerical methods.

The shift toward the use of neural networks (NNs) and physics-informed neural networks (PiNNs) for stress estimation in sand has been particularly effective \cite{Zhang, EPNN}. This transformation is driven by advances in machine learning platforms such as TensorFlow and PyTorch, as well as improvements in computational hardware, which have enabled the modeling of complex systems with greater accuracy and speed \cite{Abadi, Paszke}. Data-driven techniques not only remove several constraints associated with traditional methods, but also improve the generalization over various material properties and conditions. Therefore, they signify a progression in the domain, shifting away from conventional numerical methods that are highly dependent on substantial experimental data \cite{Rezaei, Zhang}. For example, \cite{Rezaei},  applied PiNNs to solve constitutive relations for non-linear, path-dependent materials, bypassing the need for iterative methods like Newton's method, thus reducing computational effort while maintaining thermodynamic consistency. Moreover, this method has shown promising results when integrated with finite element methods, making it an efficient tool for simulating rate-independent processes such as plasticity and damage models. Moreover, \cite{EPNN} introduced a PiNN architecture that incorporates the underlying physics of elasto-plasticity, demonstrating that this integration significantly enhances the model's ability to improve the accuracy of  stress approximation and extrapolate beyond the training data for unseen loading conditions.

Despite the promising advances in data-driven techniques like PiNN-based schemes that employ multilayer perceptron  networks (MLPs), a significant gap remains in the ability to efficiently handle highly non-linear, multi-scale, multi-resolution material behaviors, particularly when dealing with real-world high-dimensional path-dependent processes. 
MLPs may struggle to efficiently approximate high-dimensional functions due to the curse of dimensionality and require deeper or wider architectures to achieve higher accuracy \cite{KAN, PiKAN, cKAN}. Therefore, we need more powerful architectures that address these limitations. In this regard, the most recent trend is the development of explainable neural network models using Kolmogorov-Arnold networks (KANs), which facilitate model calibration and discovery while requiring minimal data \cite{KAN, PiKAN, cKAN}. The benefits of KAN-based neural techniques, including their interpretable framework for function approximation, superior accuracy with fewer parameters than traditional MLPs, and efficiency in managing complex functions and non-linear interactions, have inspired us to develop a suit of new architectures in this paper to model sand elasto-plasticity, bridging previously identified gaps as well as predicting unseen strain-controlled loading paths for sands.

The main objective of this study is to design and implement an elasto-plasticity informed Chebyshev-based Kolmogorov-Arnold network, aka EPi-cKAN. This architecture draws inspiration from \cite{EPNN}'s work on replacing traditional elasto-plastic constitutive models, yet it differentiates itself by incorporating KAN alongside augmented Chebyshev polynomials and maintaining core physical principles. Consequently, EPi-cKAN achieves high accuracy and manages complex functional forms more efficiently with fewer parameters than current methods, thereby advancing the domain of physics-informed neural networks for constitutive modeling. To assess the efficiency, accuracy, and generalization abilities of EPi-cKAN in modeling complex elasto-plastic behavior, we first compare its performance against other cKAN-based models, including both parallel and serial data-driven architectures. Additionally, to differentiate the EPi-cKAN's performance, we contrast it with MLP-based approaches, such as EPNN and parallel and serial MLPs. We further test all models' ability to predict blind stress-strain paths that are not part of the training data, demonstrating the generalization and predictive capabilities of the EPi-cKAN architecture.




\section{Underlying Physics}

This section begins with an introduction to the classical elasto-plastic constitutive model, providing the fundamental framework for understanding material behavior under stress, particularly focusing on plastic and elastic deformations. Following this, the concept of sand elasto-plasticity is explored within the context of small deformation theory, which is essential for capturing the stress-strain relationship in granular materials like sand. Lastly, we discuss the dataset generation process, detailing the creation of both training data and testing data focused on blind paths to assess the generalization and predictive capabilities of the proposed models.

\subsection{Classical Elasto-plastic Constitutive Model}

The mechanical behavior of granular materials, such as sands and soils, is highly non-linear and characterized by irreversible plastic deformations when subjected to loading \cite{Lubliner41H, Borja5H}. A classical elasto-plastic constitutive model is used to describe this behavior, accounting for both elastic (reversible) and plastic (irreversible) deformations \cite{Sim053H}. In this framework, at small deformation, the total strain $\varepsilon_{ij}$ is decomposed into elastic and plastic components,
\begin{equation}
    \varepsilon_{ij} = \varepsilon_{ij}^e + \varepsilon_{ij}^p,
\end{equation}
where $\varepsilon_{ij}$ is the total strain tensor, $\varepsilon_{ij}^e$ is the elastic strain tensor, and $\varepsilon_{ij}^p$ is the plastic strain tensor. The elastic behavior of granular materials is typically described by a linear elastic constitutive relation, which can follow general elasticity, hyperelasticity, or hypoelasticity \cite{Chen, Klein30SH}. For small strains, the stress-strain relationship in the elastic range can be written as,
\begin{equation}\label{Eq.elasto}
    \dot{\sigma}_{ij} = C_{ijkl} \dot{\varepsilon}_{kl}^e,
\end{equation}
where $\sigma_{ij}$ represents the Cauchy stress tensor, $C_{ijkl}$ is the fourth-order elastic stiffness tensor, defined as,
\begin{equation}
    C_{ijkl} = \lambda \delta_{ij} \delta_{kl} + \mu \left( \delta_{ik} \delta_{jl} + \delta_{il} \delta_{jk} \right),
\end{equation}
where $\lambda$ and $\mu$ are the Lamé constants \cite{Slaughter}, and $\delta_{ij}$ is the Kronecker delta. These constants are related to the bulk and shear moduli of the material, which define its resistance to volumetric and distortional deformations, respectively.

The plastic behavior in such materials is activated when the stress reaches a certain level, known as the yield stress. This threshold is defined by a yield function $F(\sigma_{ij}, \kappa)$, which describes the boundary between purely elastic and elasto-plastic behavior. The yield function is typically expressed as,
\begin{equation}
    F(\sigma_{ij}, \kappa) \leq 0,
\end{equation}
where $\kappa$ is a set of internal state variables that represent the material's history, such as hardening variables. When $F(\sigma_{ij}, \kappa) = 0$, the material is at the onset of yielding, and plastic deformations begin to occur. For $F(\sigma_{ij}, \kappa) < 0$, the material behaves elastically. The evolution of the plastic strain is governed by the plastic flow rule, which defines the rate of plastic strain increment based on the stress state. This is typically written in a non-associative form as,
\begin{equation}
    \dot{\varepsilon}_{ij}^p = \dot{\lambda} \frac{\partial P(\sigma_{ij}, \kappa)}{\partial \sigma_{ij}},
\end{equation}
where $\dot{\lambda}$ is a non-negative plastic multiplier that scales the rate of plastic deformation, and $P(\sigma_{ij}, \kappa)$ is the plastic potential function. The plastic potential governs the direction of plastic flow in stress space. For associative plasticity, the plastic potential function $P$ coincides with the yield function $F$, but for granular materials, a non-associative flow rule is often employed to better capture the material behavior. The plastic multiplier $\dot{\lambda}$ is determined using the consistency condition, which ensures that the stress state remains on the yield surface during plastic flow, 
\begin{equation}
    \dot{\lambda} \geq 0, \quad F(\sigma_{ij}, \kappa) \leq 0, \quad \dot{\lambda} F(\sigma_{ij}, \kappa) = 0.
\end{equation}
This set of conditions, known as the Kuhn-Tucker conditions \cite{Kuhn}, enforces that plastic flow only occurs when the material is yielding ($F = 0$), and ensures that no further plastic deformation occurs when the stress state is inside the yield surface ($F < 0$).

The evolution of the yield surface, as the material undergoes plastic deformation, is controlled by hardening (or softening) laws. These laws define how the internal variables $\kappa$ evolve with plastic strain. In many granular materials, hardening behavior is modeled to capture the material strengthening under shear, while softening is used to represent post-peak behavior where the material loses strength due to factors like dilation or particle rearrangement. The hardening law can be written as,
\begin{equation}
    \dot{\kappa} = H(\kappa, \sigma_{ij}, \varepsilon_{ij}),
\end{equation}
where $H(\kappa, \sigma_{ij}, \varepsilon_{ij})$ is a function that governs the rate of change of the internal variables. Common forms of hardening include isotropic hardening, where the yield surface expands uniformly in all directions in stress space, and kinematic hardening, where the yield surface translates in stress space. In the case of granular materials, a combination of both isotropic and kinematic hardening may be used to model their complex behavior.

\subsection{Sand Elasto-plasticity — Small Deformation}\label{WGModel}

To model the elasto-plasticity behavior of sand experiencing small deformation, we use the model proposed by Wan and Guo \cite{Guo1998, Guo1999}, so-called the WG model hereafter. The WG model for sands builds upon the Mohr-Coulomb framework and incorporates critical state theory. It also leverages Rowe's stress-dilatancy relation to account for plastic flow and dilatancy, which are critical in modeling the behavior of sands \cite{Rowe15WG, Rowe16WG}. It has been used to describe key characteristics of sand such as dilatancy, barotropy, and pyknotropy, as well as to investigate strain localization and failure precursors in sands \cite{Gong}. The elasticity behavior in the WG model follows hypo-elasticity, in which 
the elastic stiffness operator is defined as,
\begin{equation}\label{Eq.C}
    C_{ijkl} = G\Big{(}(R-\frac{2}{3})\:\delta_{ij}\delta_{kl} + (\delta_{ik}\delta_{jl} + \delta_{il}\delta_{jk})\Big{)},
\end{equation}
where the tangent shear modulus \(G\) and parameter \(R\) are given by, 
\begin{equation}\label{Eq.GR}
\begin{array}{cc}
    G=G^0 \displaystyle{\frac{(2.17-e)^2}{1+e}}\:\sqrt{p^0 p}, &  R = \displaystyle{\frac{2\:(1+\nu)}{3\:(1-2\nu)}},
\end{array}
\end{equation}
in which \(G^0\) is the reference shear modulus, \(\nu\) is the Poisson’s ratio,
is the mean stress, \(p^0\) is the reference mean stress set at 1 kPa. For the plasticity behavior, assuming the plastic strain and void ratio as internal state variables for sands, the WG model's yield function \(F\) and plastic potential \(P\) are defined as,
\begin{equation}\label{Eq.Q}
    \begin{array}{cc}
        F = q - M\: p \leq 0, & P = q - N\: p,
    \end{array}
\end{equation}
where  $q$ is the deviatoric stress,  and $p$ is the mean effective stress,
\begin{equation}\label{Eq.QP}
    q = \sqrt{\frac{3}{2} s_{ij} s_{ij}}, \quad p = \frac{tr(\boldsymbol{\sigma})}{3},
\end{equation}
where $s_{ij} = \sigma_{ij} - p\: \delta_{ij}$ represents the deviatoric stress tensor. In Eq. \eqref{Eq.Q}, $M$ is the mobilized friction coefficient, and $N$ is the dilatancy coefficient, defined as, 
\begin{equation}\label{Eq.MN}
    \begin{array}{cc}
        M =\displaystyle{\frac{2 \mu}{(1+\mu)-(1-\mu)\:t}}\: M^{tc}, & 
        N = \displaystyle{\frac{\sin \psi - (e/ e^{cs})^{n_f}\: \sin \psi^{cs}}{1-(e/ e^{cs})^{n_f}\: \sin \psi^{cs}\:\sin \psi}},
    \end{array}
\end{equation}
where 
\begin{equation}
        M^{tc} = \displaystyle{\frac{6\: \sin \psi}{3-\sin \psi}}, ~~~ t = \displaystyle{\frac{J_3}{2}(\frac{3}{J_2})^{3/2}},   
\end{equation}

Here, $n_f$ is a material constant, $\psi$ is the mobilized friction angle that can vary depending on the stress state and the void ratio $e$, $\mu$ is a constant showing the ratio of the mobilized friction coefficient, $M$, along the triaxial extension to triaxial compression, $M^{tc}$, directions, $e^{cs}$ and $\psi^{cs}$ denote the void ratio and friction angle in the critical state, respectively. The critical state friction angle, $\psi^{cs}$, is typically constant for a given material. Lastly, in the calculation of the parameter \(t\), we note that \(J_2=q^2/3\) and \(J_3=\det(s_{ij})\) are the second and third invariants of the deviatoric stress.

The WG model also uses a hardening law to describe the evolution of the yield surface. This is crucial to capture the effects of plastic strain accumulation. The hardening modulus $h$ is given as,
\begin{equation}
    h = H_0 \left( 1 - \frac{p}{p^0} \right),
\end{equation}
where $H_0$ is the initial hardening modulus. Additionally, the  critical state void ratio, $e^{cs}$, and friction angle evolve as,
\begin{equation}\label{Eq.e1}
    \begin{array}{cc}
         e^{cs} = e^{cs,0} \exp \left( - d^{cs} \left( \displaystyle{\frac{p}{p^0}} \right)^{n^{cs}} \right),
        &  
         \sin \psi = \left( \displaystyle{\frac{e}{e^{cs}}} \right)^{-n_m}\displaystyle{\frac{\gamma^p}{\alpha_0 + \gamma^p}}\: \sin \psi^{cs}, 
    \end{array}
\end{equation}
where $e^{cs,0}$ is the initial critical state void ratio,  $n^{cs}$ controls the rate of evolution of the void ratio with the mean stress, and $d^{cs}$, $n_m$, and \(\alpha_0\) are material constants. In Eq. \eqref{Eq.e1}, \(\gamma^p\) denotes the equivalent plastic shear strain defined as, 
\begin{equation}\label{Eq.gammap}
    \gamma^p = \displaystyle{\sqrt{\frac{2}{3}\eta^p_{ij}\: \eta^p_{ij}}},
\end{equation}
where \(\eta^p_{ij} = \varepsilon^p_{ij} - \left(\varepsilon^{v,p}/3\right) \delta_{ij}\) represents the deviatoric plastic strain. Here, \(\varepsilon^{v,p} = \varepsilon^p_{ii}\) denotes the plastic volumetric strain. Finally, we need the changes in void ratio over time defined as, 
\begin{equation}\label{Eq.e2}
    \dot{e} = -(1+e)\: \dot{\varepsilon}^v,
\end{equation}
 where $\dot{\varepsilon}^v$ is the volumetric strain rate.

\subsection{Dataset Generation}\label{Sec.DataG}

The WG model, as elaborated in Section \ref{WGModel} through Eqs.~\eqref{Eq.C}-\eqref{Eq.e2}, incorporates ten calibration parameters to map the elasto-plastic behavior of sands.  
These parameters have been meticulously calibrated for a variety of sands, as described in detail in \cite{Pinheiro}. For the purposes of this study, similar to the work of \cite{EPNN}, Ottawa sand is selected as the representative material for generating synthetic datasets due to its well-documented behavior and established calibration parameters. The specific calibrated values for Ottawa sand are drawn from Chapter 3, Table 3-4 of \cite{Pinheiro}, and are concisely summarized in Table \ref{Tab.SettingPinheiro}-row(a). This allows for a robust and realistic simulation of the sand's elasto-plastic behavior under various loading conditions covering both drained and undrained scenarios.

To simulate sand behavior using the WG model, two key initial conditions must be specified: the initial confining pressure, \(p^{\text{in}}\), and the initial void ratio, \(e^{\text{in}}\). These initial conditions are systematically varied across a predefined grid, with the specific ranges for \(p^{\text{in}}\) and \(e^{\text{in}}\) provided in Table \ref{Tab.SettingPinheiro}-row(b). For each combination of these initial values, a series of 20 strain-controlled tests are carried out to simulate the sand's response to the applied loads. 
The simulation process to generate ground truth datasets using the WG formulation involves employing an implicit integration scheme, specifically a backward Euler method. To maintain accuracy during this implicit integration, a return mapping algorithm (as outlined in \cite{Sim053H}) is implemented. This algorithm ensures proper enforcement of the plasticity constraints during the stress update process, which is essential for capturing the elasto-plastic behavior of the sand accurately.
In each case, proportional strain path directions (SPD) are applied in a three-dimensional stress-strain space, defined by the strain principal components \(\{\varepsilon_{11}, \varepsilon_{22}, \varepsilon_{33}\}\). These strain paths are applied over a sequence of up to 200 steps, representing a gradual application of deformation. The direction of the strain path is randomly selected at the beginning of each test and remains fixed for all 200 steps, ensuring consistency in the direction of loading.  The magnitude of the strain increments varies randomly between 0.0 and 0.0016 at each step. Such randomness maintains first-order homogeneity in the incremental stress-strain relationship, ensuring a realistic simulation of the behavior of the material under loading \cite{EPNN}.
Through this procedure, a comprehensive dataset is generated, which captures the sand's complex behavior under different initial conditions and loading paths. This dataset is crucial for training and validating our models designed to approximate elasto-plastic responses.

\begin{table}[H]
\caption{\label{Tab.SettingPinheiro} Datasets generation using WG model for Ottawa sand. }
\centering
\begin{tabular}{ll c c c c c c c c c c}
\toprule
\multicolumn{12}{l}{\textbf{(a) Material parameters of the WG model for Ottawa sand}} \\
\midrule
\textbf{Parameter} & {$G^0$} & {$\nu$} & {$\mu$} & {$\sin \psi^{cs}$} & {$n_m$} & {$\alpha_0$} & {$e^{cs,0}$} & {$d^{cs}$} & {$n^{cs}$} & {$n_f$} & \\
\textbf{Units}     & {(kPa)} & {} & {} & {} & {} & {} & {} & {} & {} & {} & \\
\textbf{Value}     & 900 & 0.30 & 0.80 & 0.53 & 1.3 & 0.008 & 0.74 & 0.005 & 0.40 & 1.5 & \\
\midrule
\midrule
\multicolumn{12}{l}{\textbf{(b) Initial variable range}} \\
\midrule
\multirow{3}{*}{\textbf{Variable}} & $p^{\text{in}}~$(kPa) & \multicolumn{10}{c}{[50.0, 500.0]} \\
                                   & $e^{\text{in}}$ & \multicolumn{10}{c}{[0.5, 0.74]} \\
                                   & SPD & \multicolumn{10}{c}{Randomly selected} \\
\midrule
\midrule
\multicolumn{12}{l}{\textbf{(c) Input and output data statistics}} \\
\midrule
& \textbf{Variable} & {$e$} & {tr($\boldsymbol{\sigma}$)} & {$\varepsilon^{v}$} & {$\varepsilon^{v,p}$} & {$\Delta \varepsilon^{v}$} & {$\Delta e$} & {$\Delta \varepsilon^{v,p}$} & {tr($\Delta \boldsymbol{\sigma}$)}  & & \\
\cmidrule(lr){2-12}
\multirow{4}{*}{\textbf{Inputs}} & \textbf{Min} & 3.2e-1 & 3.0e+0 & -5.8e-2 & -4.9e-2 & -1.2e-3 & & & & & \\
                                 & \textbf{Max} & 7.6e-1 & 2.9e+5 & 1.8e-1 & 1.9e-2 & 2.7e-3 & & & & & \\
                                 & \textbf{Mean} & 5.6e-1 & 3.3e+4 & 4.0e-2 & 6.9e-5 & 5.1e-4 & & & & & \\
                                 & \textbf{Std} & 8.6e-2 & 4.6e+4 & 4.1e-2 & 6.6e-3 & 5.7e-4 & & & & & \\
\cmidrule(lr){2-12}
\multirow{4}{*}{\textbf{Outputs}} & \textbf{Min} & & & & & & -4.6e-3 & -6.0e-4 & -2.5e+2 & & \\
                                  & \textbf{Max} & & & & & & 1.9e-3 & 4.7e-4 & 8.4e+3 & & \\
                                  & \textbf{Mean} & & & & & & -7.7e-4 & -1.2e-5 & 6.6e+2 & & \\
                                  & \textbf{Std} & & & & & & 8.7e-4 & 8.4e-5 & 9.9e+2 & & \\
                                
\bottomrule
\end{tabular}
\end{table}

\subsubsection{Training Data} \label{sec:trdata}

The input dataset (\(\textbf{x}\)) is structured as an \(m \times 13\) tensor, where \(m\) represents the total number of data points. The dataset includes essential features that characterize the elasto-plastic behavior of sand. Specifically,  the input is expressed as \(\textbf{x} = \{e, \boldsymbol{\sigma}, \boldsymbol{\varepsilon}, \boldsymbol{\varepsilon}^{\text{p}}, \Delta \boldsymbol{\varepsilon}\}\), where \(e\) is the void ratio, \(\boldsymbol{\sigma}\) represents the stress tensor, \(\boldsymbol{\varepsilon}\) corresponds to the total strain tensor, \(\boldsymbol{\varepsilon}^{\text{p}}\) refers to the plastic strain tensor, and \(\Delta \boldsymbol{\varepsilon}\) denotes the strain increments applied during each load step.
The corresponding output dataset (\(y\))  is defined as \(y = \{\Delta e^*, \Delta \boldsymbol{\sigma}^*, \Delta \boldsymbol{\varepsilon}^{\text{p*}}\}\), where \(\Delta e^*\) captures the increment in void ratio, \(\Delta \boldsymbol{\sigma}^*\) represents the change in the stress, and \(\Delta \boldsymbol{\varepsilon}^{\text{p*}}\) describes the evolution of plastic strain during the load step.
To ensure consistency and improve the training process of the neural network, all features in both input and target datasets undergo min-max normalization, scaling values to the range \([-1.0, 1.0]\). The normalization process relies on the minimum and maximum values for each feature, which are detailed in Table \ref{Tab.SettingPinheiro}-row(c). This normalization step helps stabilize the learning process by preventing features with larger ranges from disproportionately influencing the model.
These synthetic data sets will be used to train, validate, and evaluate neural network models, providing a rich foundation for accurately predicting the behavior of sand under various loading conditions. For more information on the dataset generation process, readers can refer to \cite{EPNN}.

\subsubsection{Testing Data: Blind Paths}\label{Sec.Blindpath}

To further evaluate the predictive capabilities of our proposed methods, we apply them to simulate general axisymmetric triaxial loading paths. These paths, known as``blind paths," are specifically designed to not include any prior information from the training data, allowing us to effectively assess the model’s ability to generalize and predict behavior in unseen scenarios.
\\
In these paths, the strain increments in the two lateral principal directions are equal, denoted as \(\varepsilon_{11} = \varepsilon_{22} = \varepsilon\), while the axial strain increment, \(\varepsilon_{33}\), is scaled by a proportionality factor \(\xi\), resulting in \(\varepsilon_{33} = \xi \varepsilon\). 
Given the initial conditions for confining pressure, \(p^{\text{in}}\), and the initial void ratio, \(e^{\text{in}}\), the corresponding initial stress and void ratio values are calculated as \(\sigma^0_{ii} = p^{\text{in}}\) and \(e^{0} = e^{\text{in}}\), respectively. Additionally, it is crucial to determine the final values of the strain components, particularly \(\varepsilon_{ii}^{\text{final}}\), since this allows us to calculate the strain increments for each loading step using the formula \(\Delta \boldsymbol{\varepsilon}_{ii}^n = \varepsilon_{ii}^{\text{final}} / n_{\text{step}}\), where \(n_{\text{step}}\) represents the total number of loading steps.
The volumetric strain, \(\varepsilon^v\), is calculated as \(\varepsilon^v = (\xi + 2)\varepsilon\), while the equivalent shear strain, \(\gamma\), is determined by \(\gamma = \frac{2}{3} |(1 - \xi) \varepsilon|\). Throughout the simulation, the strain paths follow a linear progression, with the material being incrementally loaded based on this relationship.
By varying the strain paths in a controlled manner, we assess how effectively the model can predict the material’s response under different loading conditions. The detailed values for \(p^{\text{in}}\), \(e^{\text{in}}\), \(\xi\), and \(\varepsilon_{ii}^{\text{final}}\) are provided in Section \ref{Sec.res}. This approach offers a robust way to validate the model’s performance in scenarios that were not part of the original training dataset.

\section{Methodology}\label{Sec.method}

This section outlines the methodologies employed in this study. It begins by introducing key neural network frameworks: MLPs, KAN, and Chebyshev-based Kolmogorov-Arnold networks (cKAN), laying the foundation for understanding the models used in this study. Following this, we discuss regular architectures, including both the parallel and serial cKAN model. Finally, we present our proposed method, the EPi-cKAN architecture, which integrates cKAN principles with elasto-plastic theory to enhance predictive accuracy and efficiency, particularly in capturing the non-linear behavior of sand under stress.
\subsection{Multilayer Perceptron Newworks}\label{Sec.MLP}

Multilayer Perceptron networks are fundamental components of feed-forward artificial neural networks, consisting of an input layer, one or more hidden layers, and an output layer, which enable the learning of complex, nonlinear relationships in data with the use of fixed activation functions. According to the Universal Approximation Theorem, MLPs are capable of approximating any continuous function on compact subsets of \(\mathbb{R}^n\), given a sufficient number of neurons and appropriate activation functions. 

Mathematically, the output \(y\) of an MLP can be rewritten using the composition operator \(T \circ T \circ \dots \circ T\), where each layer transformation can be viewed as applying a function \(T\) in sequence, hence,
\begin{equation}\label{Eq.MLP}
    y_{_\text{MLP}}(\boldsymbol{x}) = T^{(L)} \circ T^{(L-1)} \circ \dots \circ T^{(1)}(\boldsymbol{x}),
\end{equation}
where \(\boldsymbol{x} = (x_1, x_2, \dots, x_n)\) is the input vector, and the subtitle MLP is  used here to make a distinction between MLP and other type of neural networks. In Eq. \eqref{Eq.MLP}, each \(T^{(l)}\) represents the operation performed by the \(l\)-th layer, including the application of the weights, biases, and activation function. Specifically, for each layer, we have,
\begin{equation}\label{Eq.layerMLP}
    T^{(l)}(\boldsymbol{x}) = \sigma(\boldsymbol{W}^{(l)}\: \boldsymbol{x} + \boldsymbol{b}^{(l)}),
\end{equation}
where \(\boldsymbol{W}^{(l)}\) and \(\boldsymbol{b}^{(l)}\) are the weights and biases of the \(l\)-th layer, respectively, and \(\sigma\) is the activation function. 
We denote the set of trainable parameters of an MLP by \(\boldsymbol{\theta}_{_\text{MLP}} = \{ \boldsymbol{W}^l, \boldsymbol{b}^l \}\), where the number of parameters is around \(O(n_l H^2)\), with \(n_l\) and \(H\) representing the number of hidden layers and neurons per hidden layer, respectively \cite{raissi2019physics,PiKAN}. 
Thus, the entire MLP can be seen as a composition of transformations \(T^{(l)}\), applied layer by layer, from the input to the output. Despite their power, MLPs suffer from interpretability challenges, as well as issues such as overfitting, vanishing gradients, and scalability limitations.

\begin{figure}[!h]
\centering
\includegraphics[width=.85\textwidth]{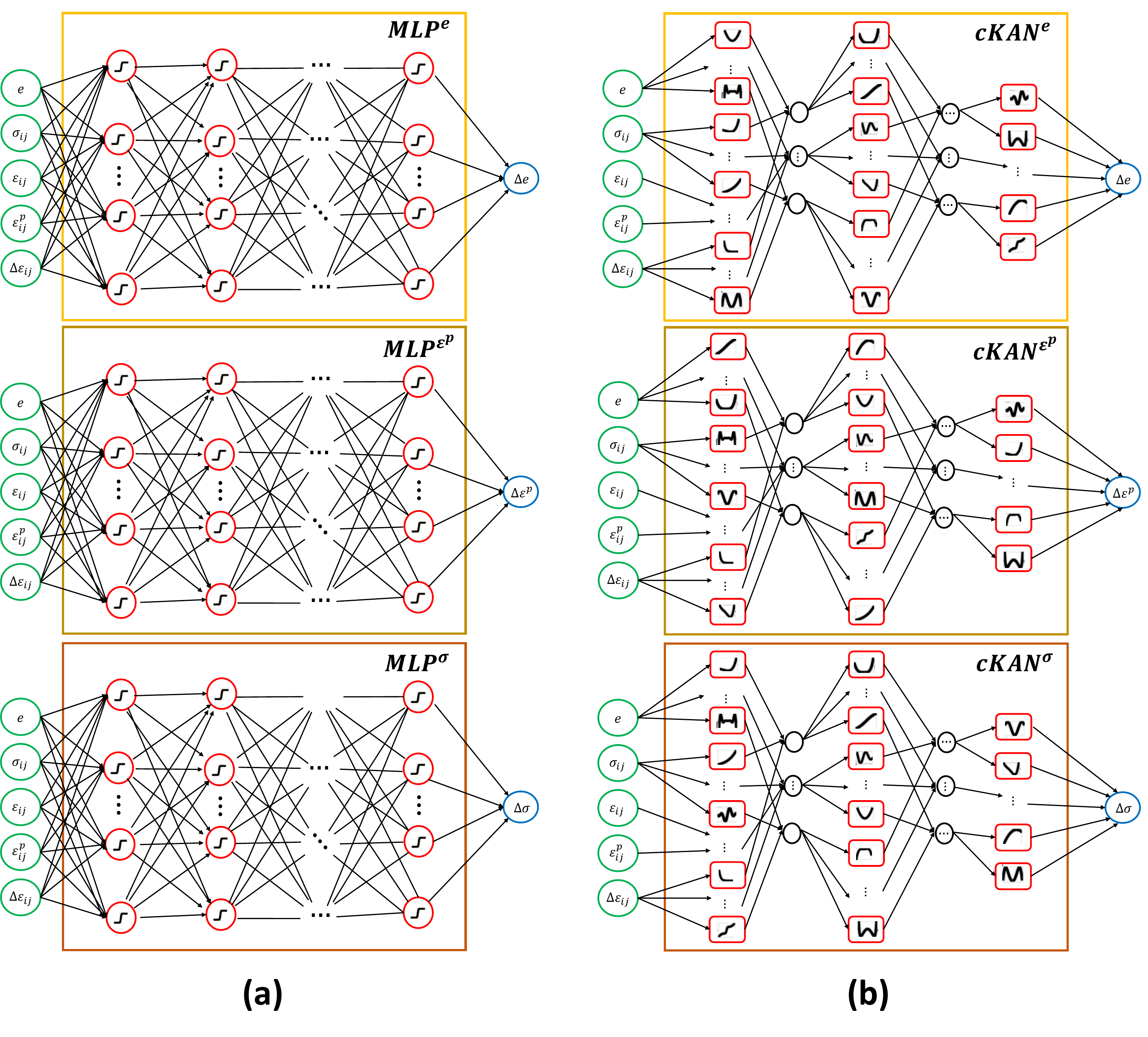}\hspace{.1cm}
\caption{Parallel architectures to model sands non-linear stress-strain behavior. Panel (a): parallel MLP architecture, panel (b): parallel cKAN architecture.} \label{Pic.Parallel}
\end{figure}
\begin{figure}[!h]
\centering
\includegraphics[width=.85\textwidth]{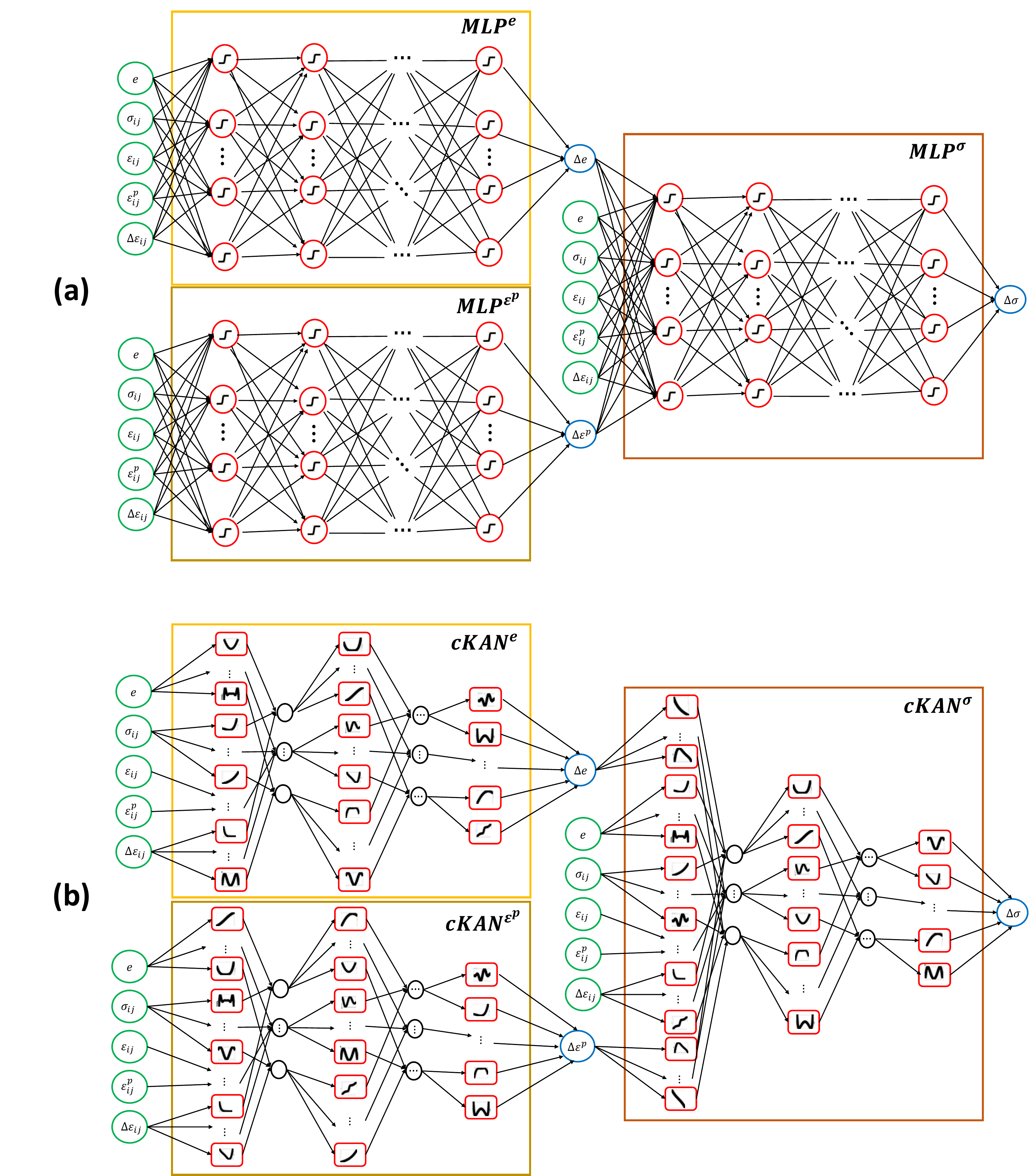}\hspace{.1cm}
\caption{Serial architectures to model sands non-linear stress-strain behavior. Panel (a): serial MLP architecture, panel (b): serial cKAN architecture.} \label{Pic.Serial}
\end{figure}

\subsection{Kolmogorov-Arnold Networks}\label{Sec.kan}

Kolmogorov-Arnold Networks have been proposed recently as  promising alternatives for MLPs, opening opportunities to further improve data-driven models. KANs follow the Kolmogorov-Arnold representation theorem, a foundational result in approximation theory, states that any continuous multivariate function on a bounded domain can be expressed as a superposition of univariate functions and linear operations \cite{KAN}. Mathematically, for a continuous function \( f : [0,1]^{n} \to \mathbb{R}\) on the \(n\)-dimensional unit hypercube, the theorem ensures the existence of continuous univariate functions \( \Phi_j \) and \( \phi_{i,j} \) such that,
\begin{equation}\label{Eq.KAN}
    f(\boldsymbol{x}) = \sum_{j=0}^{2n} \Phi_j \left( \sum_{i=1}^{n} \phi_{i,j}(x_i) \right),
\end{equation}
where \(\boldsymbol{x} = (x_1, x_2, \dots, x_n)\) is the input vector. According to this theorem, KANs is developed as a new type of neural network architecture that approximates multivariate functions by composing univariate functions across layers.
A general KAN is structured as a composition of \(L\) layers. Given an input vector \(\boldsymbol{x} \in \mathbb{R}^{n}\), the output of the KAN can be expressed as,
\begin{equation}\label{Eq.general_KAN}
    y_{_\text{KAN}}(\boldsymbol{x}) = (\boldsymbol{\Phi_{L}} \circ \boldsymbol{\Phi_{L-1}} \circ \cdots \circ \boldsymbol{\Phi}_2 \circ \boldsymbol{\Phi}_1)(\boldsymbol{x}),
\end{equation}
where each \(\boldsymbol{\Phi}_l\) represents a KAN layer with \(n_{l}\)-dimensional inputs and \(n_{l+1}\)-dimensional outputs and can
be defined as a matrix of 1-dimensional functions
\begin{equation}\label{Eq.KANlayer}
    \boldsymbol{\Phi}_l = \{\phi_{l, i,j}\},\:\: l=1, \cdots , L,\:\: i=1,\cdots , n_{l}, \:\: j=1,\cdots , n_{l+1}.
\end{equation}

In the original implementation of KAN \cite{KAN}, \(\phi\) is defined as a weighted combination of a basis function \(b\) and B-splines,
\begin{equation}\label{Eq.phi}
    \phi (x) = w_b b(x) + w_s \text{spline}(x),
\end{equation}
where \(b(x)\) and \(\text{spline}(x)\) are defined as,
\begin{equation}
    \begin{array}{cc}
       b(x) = \displaystyle{\frac{x}{1+\exp(-x)}},  &   
       \text{spline}(x) = \displaystyle{\sum_i c_i\: B_i(x)},
    \end{array}
\end{equation}
where, \(c_i\), \(w_b\), and \(w_s\) are trainable parameters, and we denote them by \(\boldsymbol{\theta}_{_\text{KAN}}= \{c_i, w_b, w_s\}\). 
The splines \(B_i\), are characterized by the polynomial order \(k\), and the number of grid points \(g\).

In this study, we use KANs, but using Chebyshev polynomials  following  \cite{cKAN}, rather than using the standard basis and spline functions. Chebyshev polynomials  are a sequence of orthogonal polynomials playing a crucial role in approximation theory and numerical analysis \cite{Cesarano, Karageorghis}. They can be  defined on the interval \([-1, 1]\)  satisfying the recurrence relation as,
\begin{equation}
    \begin{array}{l}
         T_0(x) = 1,  \:\:\: T_1(x) = x,\\
         T_n(x) = 2\:x\:T_{n-1}(x)-T_{n-2}(x),\:\: n\geq 2.
    \end{array}
\end{equation}
Therefore, for cKAN, we have
\begin{equation}\label{Eq.cheblayer}
    \phi (x) =  \displaystyle{\sum_{i} c_i\: T_i(x)},
\end{equation}
where \(c_i\) is the learnable parameter.

KANs and cKANs share a similar structure in approximating complex functions, but they differ in the number of parameters and their dependency on grid points. In KANs, the total number of trainable parameters is influenced by the input size, number of hidden layers, neurons per layer, grid size, and polynomial order, as expressed by \(|\boldsymbol{\theta}|_{\text{KAN}} \sim O(n_l H^2 (k + g))\), where \(g\) is the grid size and \(k\) is the polynomial order. In contrast, cKANs eliminate the need for grid points, reducing the number of parameters to \(|\boldsymbol{\theta}|_{\text{cKAN}} \sim O(n_l H^2 k)\), which simplifies the model and improves efficiency by focusing solely on the polynomial order \(k\) \cite{PiKAN}.

\subsection{Parallel and Serial  Architectures}\label{Sec.parackan}

In this study, we explore two distinct cKAN architectures for modeling elasto-plastic behavior: the ``parallel" and ``serial" cKAN architectures. In the parallel cKAN architecture, as shown in Fig.~\ref{Pic.Parallel}, independent neural networks are employed to separately predict increment of stress, plastic strain, and void ratio, enabling simultaneous processing of these variables.
In contrast, the serial cKAN architecture, as shown in Fig.~\ref{Pic.Serial}, first predicts the increments in the void ratio and plastic strain, and then these predictions are sequentially fed into another network to predict the stress increments.
Each architecture offers unique benefits: the parallel approach allows for efficient simultaneous handling of different processes, while the serial architecture captures the interdependence of variables by using each prediction to inform the next. The serial MLP and parallel MLP architectures were originally proposed in \cite{Masi} and \cite{EPNN}, respectively.

In the parallel cKAN architecture, see Fig.~\ref{Pic.Parallel}, the network consists of three separate sub-networks: the void ratio sub-network (cKAN\(^{e}\)), the plastic strain sub-network (cKAN\(^{\varepsilon^{p}}\)), and the stress sub-network (cKAN\(^{\sigma}\)).
Each sub-network takes as input relevant variables such as the void ratio \(e\), stress \(\sigma_{ij}\), strain \(\varepsilon_{ij}\), plastic strain \(\varepsilon^{p}_{ij}\), and incremental strain \(\Delta\varepsilon_{ij}\). The void ratio, the plastic strain, and the stress sub-networks predict \(\Delta e\), \(\Delta \varepsilon^{p}\), and \(\Delta \sigma\), respectively.
To facilitate simultaneous training of these sub-networks, we define a total cost function \(CF\) as the sum of the individual cost functions for each sub-network, i.e. \(CF = CF^{e} + CF^{\varepsilon^{p}} + CF^{\sigma}\), where,
\begin{equation}\label{Eq.CFs}
    \begin{array}{ccc}
       CF^{e}= \text{MSE}(\Delta e,  \Delta e^*), &
       CF^{\varepsilon^{p}}= \text{MSE}(\Delta \varepsilon^{p},  \Delta \varepsilon^{p,*}), &
       CF^{\sigma}= \text{MSE}(\Delta \sigma,  \Delta \sigma^*),
    \end{array}
\end{equation}
where MSE denotes the mean squared error, and \(\Delta e^*\),\(\Delta \varepsilon^{p,*}\), and \(\Delta \sigma^*\)  represent the corresponding ground truth data.


In the serial cKAN architecture, see Fig.~\ref{Pic.Serial}, the network is structured in a sequential manner, where the outputs of one sub-network serve as inputs to the next. Initially, the void ratio sub-network (cKAN\(^{e}\)) predicts the change in void ratio \(\Delta e\) based on inputs such as the current void ratio \(e\), stress \(\sigma_{ij}\), strain \(\varepsilon_{ij}\), plastic strain \(\varepsilon^{p}_{ij}\), and incremental strain \(\Delta\varepsilon_{ij}\). 
The plastic strain sub-network (cKAN\(^{\varepsilon^{p}}\)) uses the same set of input variables as the void ratio sub-network to predict the incremental plastic strain \(\Delta \varepsilon^{p}\), independently of the predicted \(\Delta e\).
Finally, the outputs from both the void ratio and plastic strain sub-networks, along with the input variables, are passed into the stress sub-network (cKAN\(^{\sigma}\)) to predict the incremental stress \(\Delta \sigma\). The total cost function \(CF\) for this architecture is the sum of the cost functions of the three sub-networks, \(CF = CF^{e} + CF^{\varepsilon^{p}} + CF^{\sigma}\), similar to what  described for the parallel architecture in Eq. \eqref{Eq.CFs}.

Figures 
\ref{Pic.Parallel} and \ref{Pic.Serial} additionally illustrate the parallel and serial architectures utilizing MLPs, providing comparisons to cKAN's configurations. In the MLP architectures, the same activation function is applied uniformly across all neurons, resulting in a fixed nonlinearity throughout the network. In contrast, the cKAN architectures leverage Chebyshev polynomials as basis functions, providing a more flexible and adaptive approach to representing the solution space. This flexibility enables cKANs to capture complex patterns with greater accuracy and efficiency compared to MLPs. 




\begin{figure}[!h] 
\centering
\includegraphics[width=.85\textwidth]{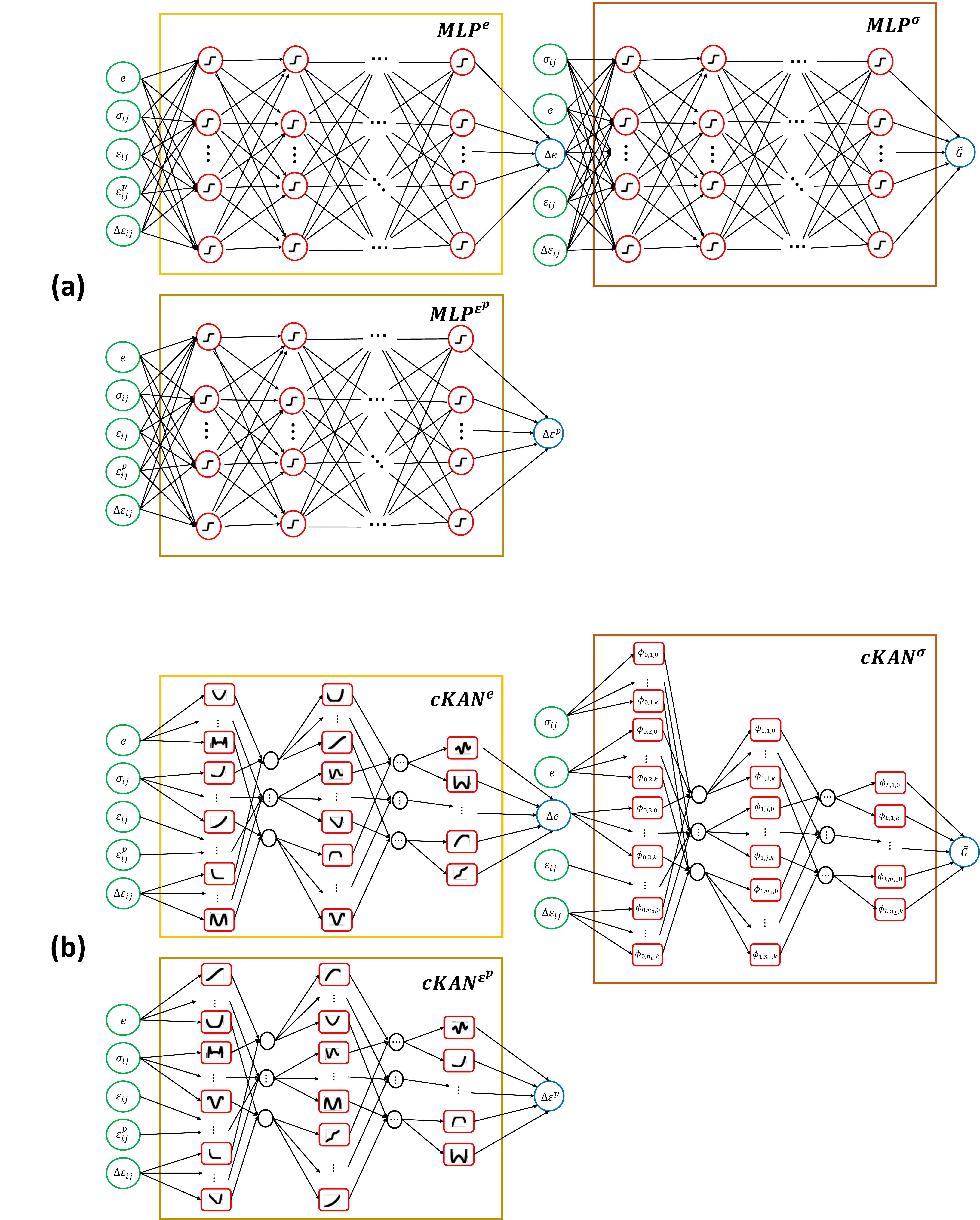}\hspace{.1cm}
\caption{Elasto-plastic based architectures to model sands non-linear stress-strain behavior. Panel (a): EPNN architecture, panel (b): EPi-cKAN architecture.} \label{Pic.EPbased}
\end{figure}

\subsection{Elasto-plastic cKAN Architecture}\label{Sec.EPckan}
In this section, we introduce the Elasto-Plastic informed cKAN (EPi-cKAN) architecture, inspired by the success of both the EPNN framework \cite{EPNN} and the strong potential of cKAN. As discussed in Section \ref{Sec.parackan}, the parallel and serial models rely solely on data-driven methods, using input-output pairs to approximate void ratio, plastic strain, and stress increments. In contrast, the EPi-cKAN architecture improves upon this by integrating the essential principles of elasto-plasticity. We specifically incorporate governing stress-strain relationship, which is vital for accurately depicting material behavior. Following this, EPi-cKAN can attain a more precise and robust representation of complex elasto-plastic responses, thus bridging the gap between purely data-driven models and traditional physics-based formulations.

Figure \ref{Pic.EPbased} shows the EPi-cKAN architecture and contracts it against the MLP version proposed by \citet{EPNN}. As shown, the EPi-cKAN architecture consists of three interconnected sub-networks: the plastic strain sub-network (cKAN\(^{\varepsilon^{p}}\)), the void ratio sub-network (cKAN\(^{e}\)), and the stress sub-network (cKAN\(^{\sigma}\)). 
The cKAN\(^{\varepsilon^{p}}\) sub-network predicts the incremental plastic strain \(\Delta \varepsilon^{p}\) using inputs such as the current void ratio \(e\), stress \(\sigma_{ij}\), strain \(\varepsilon_{ij}\), plastic strain \(\varepsilon^{p}_{ij}\), and strain increment \(\Delta \varepsilon_{ij}\). This sub-network operates independently of the other two sub-networks.
The cKAN\(^{e}\) sub-network, on the other hand, predicts the change in void ratio \(\Delta e\) based on the same set of input variables as the the plastic strain sub-network.
The outputs from the void ratio sub-network, along with all input variables except for plastic strain, are passed to the cKAN\(^{\sigma}\) sub-network. In contrast to the previous two structures, where the output of the cKAN\(^{\sigma}\) sub-network was the stress increment \(\Delta \sigma\), the output in this structure is the tensor \(\tilde{G}\). By using \(\tilde{G}\) in conjunction with the parameter \(\tilde{R}\), we can predict the secant elasticity tensor \(\tilde{C}\), which is critical for calculating the stress increment \(\Delta \sigma_{ij}\) according to the incremental stress-strain relationship outlined in Eq.~\eqref{Eq.elasto}. The symbols with tildes (e.g., \(\tilde{C}\), \(\tilde{G}\), and \(\tilde{R}\)) represent their estimated values obtained by the network. Once these variables are estimated, we use the following equation governing the relationship between stress and strain,
\begin{equation}\label{Eq.elasto2}
    \Delta \sigma_{ij} = \tilde{C}_{ijkl}:(\Delta \varepsilon_{kl} - \Delta \varepsilon^{p}_{kl}),
\end{equation}
to obtain the stress increment \(\Delta \sigma_{ij}\). Here, the operation shown by `:' indicates the summation over the \(k\) and \(l\) indices,
\(\tilde{C}_{ijkl}\) is the estimated secant elasticity tensor, \(\Delta \varepsilon_{kl}\) is the total strain increment, and \(\Delta \varepsilon^{p}_{kl}\) is the plastic strain increment. 
As explained in more detail in Eq.~\eqref{Eq.C}, the elasticity tensor \(\tilde{C}\) is derived from the estimated values of \(\tilde{G}\) and \(\tilde{R}\), where \(\tilde{G}\) is computed from the stress sub-network's output, and \(\tilde{R}\) is an independently estimated scalar introduced within the model to account for additional non-linearity or material properties.







The cost function for the EPi-cKAN architecture is designed to incorporate the physics of elasto-plasticity, following a structure similar to the EPNN framework. Specifically, the total cost function is defined as \(CF = CF^{e} + CF^{\varepsilon^{p}} + CF^{\sigma}\), where \( CF^{e}\) and \(CF^{\varepsilon^{p}}\) follow the same form as in Eq.~\eqref{Eq.CFs}, while \( \text{CF}^\sigma \) is given by,
\begin{equation}\label{Eq.CFsigma}
\text{CF}^\sigma = \text{MSE}(\tilde{C}_{ijkl} : (\Delta \varepsilon_{kl} - \Delta \varepsilon^{p}_{kl}), \Delta \sigma_{ij}^*).
\end{equation}
This total cost function thus includes the plastic strain error \( CF^{\varepsilon^{p}} \), which accounts for the plastic strain increments, the void ratio error \( CF^{e} \), which tracks changes in the void ratio, and the stress-related error \( CF^\sigma \), which captures discrepancies in stress increments while accounting for the physical principles of elasto-plasticity. 

Furthermore, the cKAN architecture used to compute the elasticity matrix relies on the estimation of \(\tilde{G}\) through the following \(\phi\) formulation, rather than using Eq.~\eqref{Eq.cheblayer}:
\begin{equation}\label{Eq.improvedPHI}
\phi(x) = \Gamma(x) + \text{Chebyshev}(x),
\end{equation}
where
\(\Gamma\) represents a weighted ReLU-based function with learnable weights \(\boldsymbol{\omega}\), while \(\text{Chebyshev}\) is a Chebyshev polynomial expansion. This combination of a weighted baseline function and Chebyshev polynomials, thus \(\boldsymbol{\theta}_{_\text{Epi-cKAN}}= \{\boldsymbol{\omega}, c_i\}\), provides the model with enhanced flexibility to capture complex material behaviors. 
 This blend of data-driven and physics-informed elements enables more accurate representation of elasto-plastic responses.

As shown in Fig.~\ref{Pic.EPbased},  both the EPNN, panel (a), and EPi-cKAN, panel (b), architectures are designed to model elasto-plastic behavior. Both approaches leverage physics-informed techniques to capture material responses, yet they differ in their implementation. Specifically, EPi-cKAN employs two sub-networks based on the cKAN architecture, which enhances their capability to model complex material behavior. In addition, one of the EPi-cKAN sub-networks utilizes an augmented basis function as described in Eq. \eqref{Eq.improvedPHI}, combining a ReLU function with Chebyshev polynomials.


\section{Results and Discussion}\label{Sec.res}

This section demonstrates   the performance of our proposed architectures using Ottawa sand with elasto-plastic behavior and well-documented mechanical properties in geotechnical studies. The section is organized into two main subsections to provide a comprehensive evaluation of the models. \\
In Section \ref{Sec.firstphase}, we focus on training, validating, and evaluating the cKAN and MLP architectures using datasets generated in Section \ref{sec:trdata} for Ottawa sand, which captures its behavior under various loading conditions. We then conduct a detailed comparison between the cKAN architectures and traditional MLP models to assess their respective abilities to approximate the material's stress-strain relationships. \\
In Section \ref{Sec.secondphase}, we evaluate the predictive power of trained networks by simulating blind axisymmetric triaxial loading paths, a critical test to validate constitutive models in geomechanics. The datasets used in this phase is generated following instruction given in Section \ref{Sec.Blindpath}. This phase of the study demonstrates how well trained networks can generalize beyond training data, providing insight into the robustness of the EPi-cKAN architecture in predicting complex, unseen loading paths.


\subsection{First Phase:  Training and Validation}\label{Sec.firstphase}
In this section, we start by optimizing the hyperparameters to find the ideal settings for the proposed network architectures. After determining these parameters, we assess the performance of our cKAN architectures against both parallel and serial MLPs, and also against the EPNN model. At this point, we focus solely on training, validating, and evaluating the networks, pinpointing the top-performing configurations that will be utilized in the next phase to approximate blind loading paths. The key hyperparameters that need to be identified include the polynomial order \(k\), the number of layers \(L\), and the number of neurons per layer \(N\). Selecting the appropriate values for these parameters is crucial for enhancing the accuracy and efficiency of the models. 

As outlined in Eqs.~\eqref{Eq.CFs} and \eqref{Eq.CFsigma}, we utilize the mean square error (MSE) as the primary cost function to evaluate the models during training and validation. By minimizing the total cost function, we aim to improve the models' ability to capture the underlying patterns in the data, leading to better generalization in the prediction phase. Moreover, in our implementation, we employ ADAM with a learning rate of 0.001 for all three sub-networks. This learning rate strikes a balance between allowing the model to learn effectively without overshooting minima during the optimization process. A proper value for the learning rate is crucial, as too high a value can lead to unstable training, while too low a value can result in slow convergence  \cite{bengio2017deep, sutskever, lecun2002, smith2017cyclical, popel}.

\begin{figure}[!h]
\centering
\subfigure[$N=20$]
{ \label{Pic.N20diffL}\includegraphics[width=.45\textwidth]{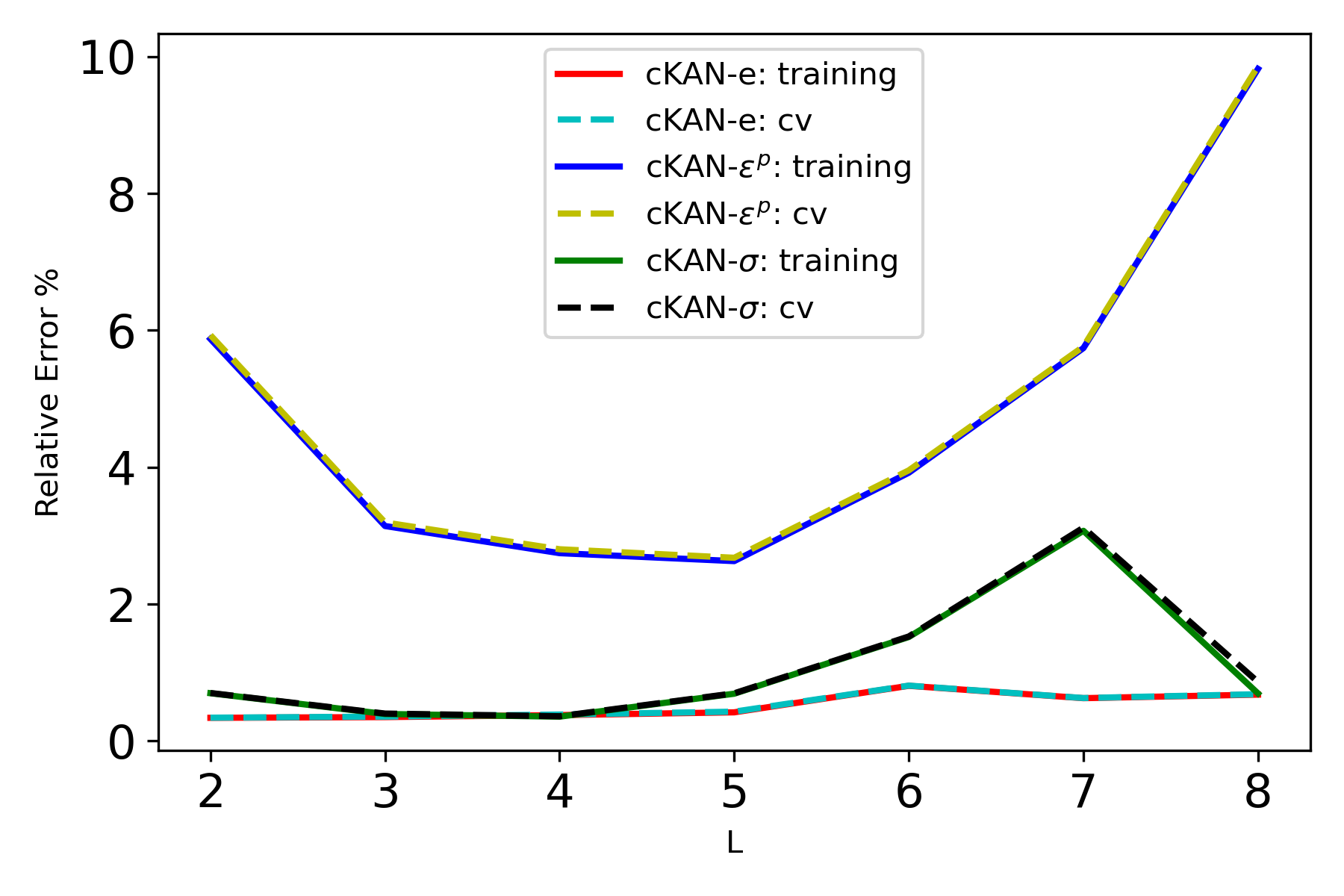}\hspace{.1cm}}
\subfigure[$N=30$]
{ \label{Pic.N30diffL}\includegraphics[width=.45\textwidth]{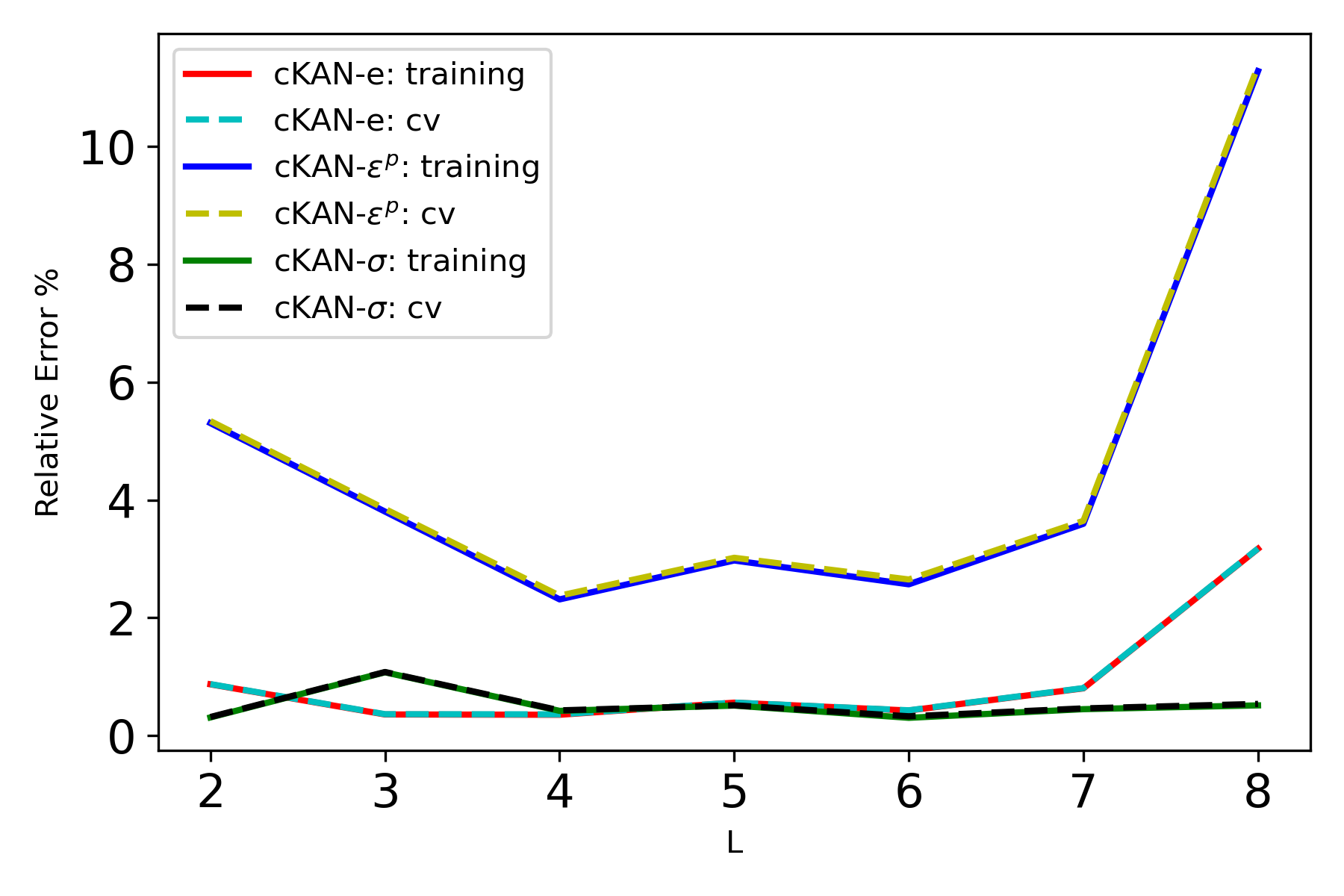}\hspace{.1cm}}
\caption{Relative $\mathcal{L}^2$ errors versus the number of layers for the sub-networks of the parallel cKAN architecture with a polynomial order of  \(k=3\) on the training and cross-validation (cv) sets for different numbers of neurons per layer ($N$) after 2e4 epochs of training.} \label{Pic.fixedNdiffL}
\end{figure}

\begin{figure}[!h]
\centering
\subfigure[$L=3$]
{ \label{Pic.L3diffN}\includegraphics[width=.45\textwidth]{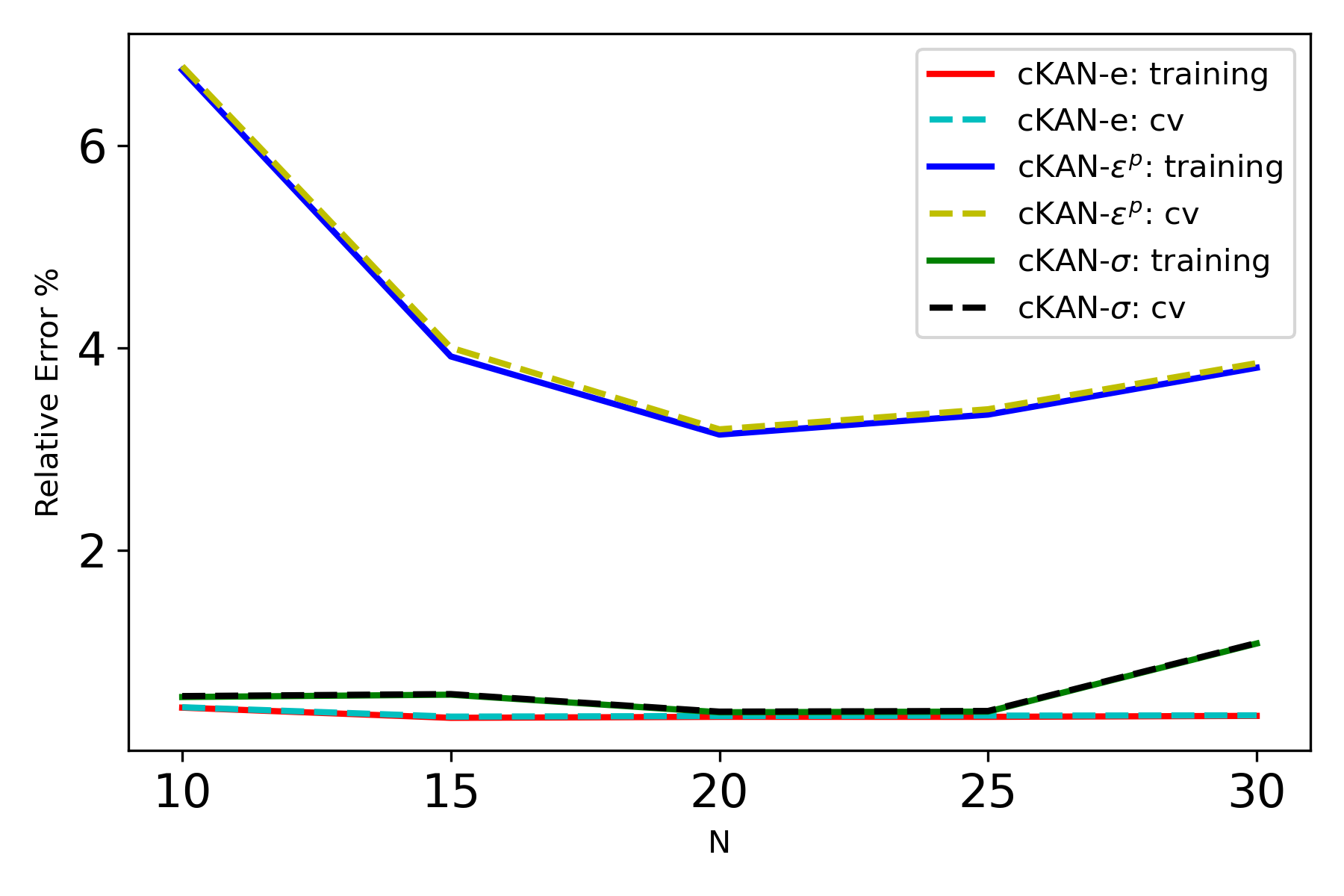}\hspace{.1cm}}
\subfigure[$L=4$]
{ \label{Pic.L4diffN}\includegraphics[width=.45\textwidth]{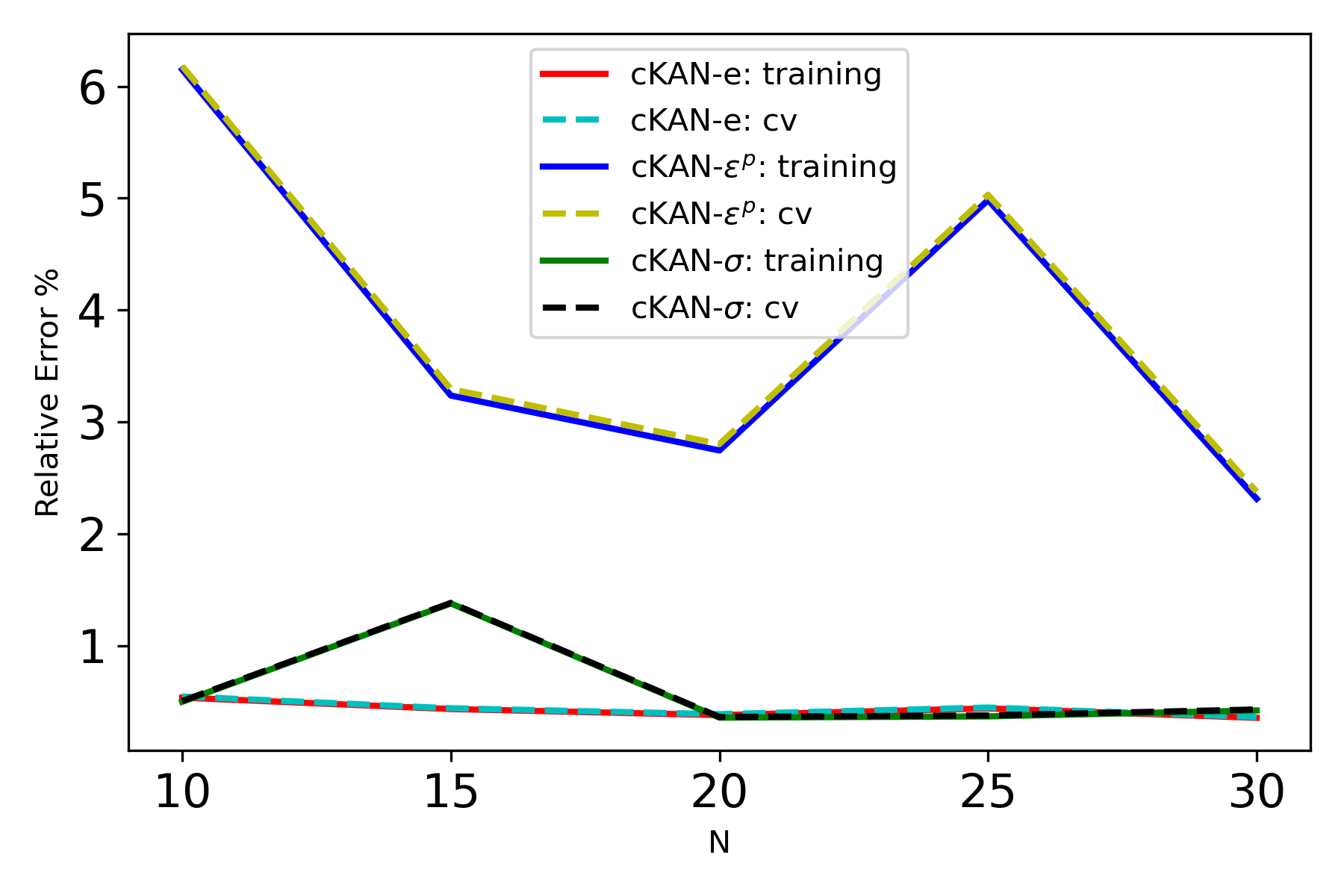}\hspace{.1cm}}
\caption{Relative $\mathcal{L}^2$ errors versus the number of neurons per layer for the sub-networks of the parallel cKAN architecture with a polynomial order of  \(k=3\) on the training and cross-validation (cv) sets for different numbers of layers ($L$) after 2e4 epochs of training.} \label{Pic.fixedLdiffN}
\end{figure}
\begin{figure}[!h]
\centering
\includegraphics[width=.45\textwidth]{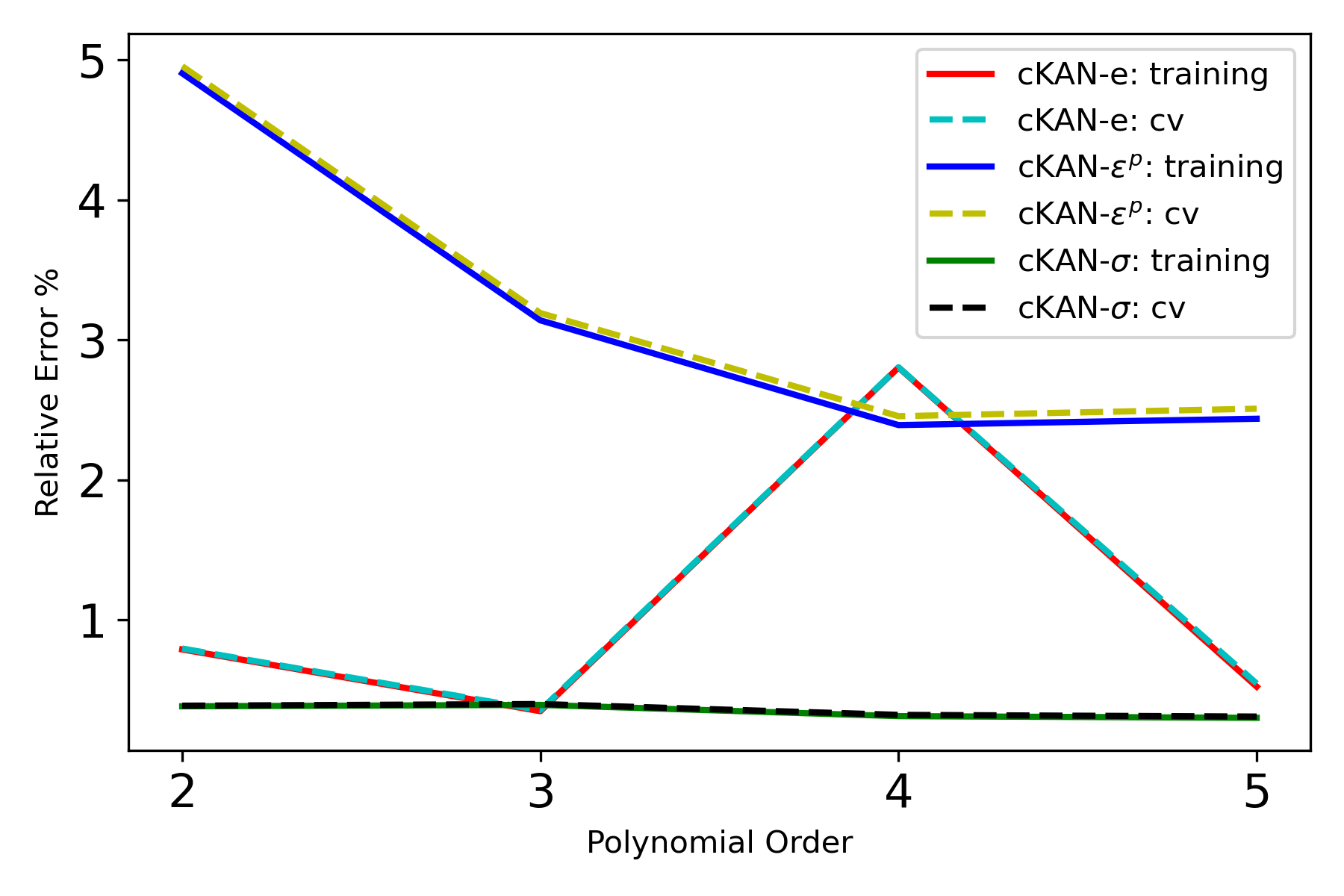}\hspace{.1cm}
\caption{Relative $\mathcal{L}^2$ errors versus polynomial order \(k\) of each sub-networks of the parallel cKAN architecture with $L=3$ and $N=20$ on the training and cross-validation (cv) sets after 2e4 epochs of training.} \label{Pic.DL3N20}
\end{figure}

Details regarding the generation of synthetic data are presented in Section \ref{Sec.DataG}. 
The dataset consists of approximately 312,000 samples (\(m = 312,000\)). To ensure robust model training, validation, and evaluation, we randomly shuffle the data and split it into three subsets. We allocate around 187,000 samples to the training set, which is used to train the models. Additionally, we reserve approximately 62,000 samples for the cross-validation set, which helps in tuning the hyperparameters and preventing overfitting. The remaining samples are designated as the evaluation set, used for assessing the final performance of the trained models on unseen data, but within the training range. This balanced split allows for a comprehensive model assessment at each stage of development.

In this study, the error is quantified using the relative \(\mathcal{L}^2\) norm between the output matrix of each sub-network and the corresponding ground truth data. In EPi-cKAN and EPNN, the output of the stress sub-network is further processed through the problem’s underlying physics. After calculating the stress increment, this result is compared with the corresponding ground truth data.
The hyperparameters tuning is performed for the parallel cKAN architecture, and results are shown in Figs.~\ref{Pic.fixedNdiffL} through \ref{Pic.DL3N20}. Since in the parallel method, each of the three sub-networks is trained independently and their behaviors do not affect each other, we use this approach to tune the hyperparameters and examine the behavior of the sub-networks. Subsequently, the tuned hyperparameters will be transferred to other architectures when comparing the performance of different architectures.

Figure \ref{Pic.fixedNdiffL} presents the final relative \(\mathcal{L}^2\) error as a function of the number of layers for both the training and cross-validation sets, using a polynomial order of \(k=3\) after 20,000 training epochs. Panel (a) in Fig.~\ref{Pic.fixedNdiffL} illustrates the results for \(N=20\), where the relative \(\mathcal{L}^2\) error for the void ratio sub-network consistently remains below 1\% across all layers. In the plastic strain sub-network, the error decreases as the number of layers increases from 2 to 3, reaches a minimum, and then begins to rise again beyond 4 layers. Similarly, the stress sub-network shows an increase in error after 5 layers. 
Panel (b) in Fig.~\ref{Pic.fixedNdiffL}, plotted for \(N=30\), reveals that the relative \(\mathcal{L}^2\) error for both the void ratio and stress sub-networks remains below 2\% for all layers. In contrast, for the plastic strain sub-network, the error decreases as the number of layers increases from 2 to 4, reaching its lowest point at \(L = 4\), but then increases again for \(L > 6\).

In the next experiment, we fix the number of layers and analyze the behavior of the networks by varying numbers of neurons per layer. Figure \ref{Pic.fixedLdiffN} presents the final relative \(\mathcal{L}^2\) error on the training and cross-validation sets after 20,000 epochs, using a polynomial order of \(k=3\), with the error plotted against the number of neurons per layer. 
Panel (a) in Fig.~\ref{Pic.fixedLdiffN}, corresponding to \(L=3\), shows that the void ratio sub-network maintains a relative \(\mathcal{L}^2\) error below 1\% across all number of neurons per layer. For the plastic strain sub-network, the error decreases as \(N\) increases from 10 to 20, reaching its minimum at \(N = 20\), but then starts to increase for \(N > 20\). Similarly, the stress sub-network shows a slight increase in error for \(N > 25\).
Panel (b) in Fig.~\ref{Pic.fixedLdiffN}, for \(L=4\), reveals that the stress sub-network achieves an error below 1\% at \(N = 10\), 20, 25, and 30. In contrast, for the void ratio and plastic strain sub-networks, the error initially starts high, but decreases steadily until \(N = 20\), after which it begins to rise again.

In the final experiment, we fix the number of layers and neurons per layer, while varying the polynomial order. Figure \ref{Pic.DL3N20} shows the final relative \(\mathcal{L}^2\) error on both the training and cross-validation sets after 20,000 epochs, with the network consisting of 3 layers and 20 neurons per layer. As observed, the relative error remains below 1\% for all polynomial orders in the stress sub-network and also in the void ratio sub-network at polynomial order 5. The error generally decreases for most networks as the polynomial order increases up to 3, but in the void ratio sub-network, there is a noticeable increase at order 4 before it decreases again at order 5.

According to Figs. \ref{Pic.fixedNdiffL} through \ref{Pic.DL3N20}, we can observe that the relative \(\mathcal{L}^2\) error on the cross-validation set consistently remains very close to that on the training set across all three sub-networks. For both the void ratio and stress sub-networks, the errors predominantly stay around or below 1\(\%\), regardless of the number of layers, neurons per layer, or polynomial orders.
Considering the relative \(\mathcal{L}^2\) errors on the training and cross-validation sets, we select a 2-layer network with 20 neurons per layer and a polynomial order of 3 for the void ratio sub-network. For the other two sub-networks, we choose a 3-layer network with 20 neurons per layer and a polynomial order of 4. Table \ref{Tab.Settings} summarizes the network configurations along with the total number of parameters. Moreover, in the last column of Table \ref{Tab.Settings}, we report the the computation time of training per iteration in milliseconds.
\begin{table}[!h]
\centering
\caption{\label{Tab.Settings}Network architectures, layer configurations \((L, N, k)\) for the void ratio \((e)\), plastic strain \((\varepsilon^p)\), and stress \((\sigma)\) sub-networks, along with the total number of parameters (\(\vert \boldsymbol{\theta}\vert\)) and the computation time per iteration for each architecture. \(L\) represents the number of layers, \(N\) is the number of neurons per layer, and \(k\) is the polynomial order (for MLP-based architectures, which do not use polynomial orders, a dash is indicated for \(k\)). Times are measured on a system with NVIDIA RTX A6000 GPU with 47.5 GB of dedicated GPU memory.}
\renewcommand{\arraystretch}{1.2}
\setlength{\tabcolsep}{12pt}
\begin{tabular}{l|ccc|c|c}
\toprule
\textbf{Architecture} & \(\boldsymbol{(L, N, k)_e}\) & \(\boldsymbol{(L, N, k)_{\varepsilon^p}}\) & \(\boldsymbol{(L, N, k)_{\sigma}}\) & \(\boldsymbol{\vert \theta \vert}\) & \textbf{Time} (ms)\\
\midrule
EPi-cKAN & (2, 20, 3) & (3, 20, 4) & (3, 20, 4) & 14,560 & 51 \\
Parallel cKAN & (2, 20, 3) & (3, 20, 4) & (3, 20, 4) & 13,920 & 32\\
Serial cKAN & (2, 20, 3) & (3, 20, 4) & (3, 20, 4) & 14,020 & 34 \\
\midrule
EPNN & (2, 45, -) & (5, 35, -) & (5, 35, -) & 13,435 & 18\\
Parallel MLP & (2, 45, -) & (5, 35, -) & (5, 35, -) & 13,935&21\\
Serial MLP & (2, 45, -) & (5, 35, -) & (5, 35, -) & 13,610 &21\\
\bottomrule
\end{tabular}
\end{table}

\begin{table}[!h]
\centering
\caption{\label{Tab.TestCompare1}Relative \(\mathcal{L}^2\) errors on the evaluation set. The columns represent the errors in predicting void ratio \((\Delta e)\), plastic strain \((\Delta \varepsilon^p)\), and stress \((\Delta \sigma)\).}
\renewcommand{\arraystretch}{1.2}
\setlength{\tabcolsep}{12pt}
\begin{tabular}{l|ccc}
\toprule
\textbf{Architecture} & \(\mathcal{L}^2_{\boldsymbol{e}}\) & \(\mathcal{L}^2_{\boldsymbol{\varepsilon^p}}\) & \(\mathcal{L}^2_{\boldsymbol{\sigma}}\)  \\
\midrule
EPi-cKAN & 0.32\% & 2.55\% & \textbf{0.13\%} \\
Parallel cKAN & 0.47\% & 2.57\% & 0.35\%  \\
Serial cKAN & 0.35\% & \textbf{2.41\%} & 0.34\%  \\
\midrule
EPNN & 0.85\% & 3.21\% & 0.24\%  \\
Parallel MLP & 0.34\% & 3.21\% & 0.65\%  \\
Serial MLP & \textbf{0.23\%} & 3.41\% & 0.53\%  \\
\bottomrule
\end{tabular}
\end{table}
\begin{figure}[!h]
\centering
\subfigure[cKAN architectures]
{ \label{Pic.DiffSizecKANs}\includegraphics[width=.45\textwidth]{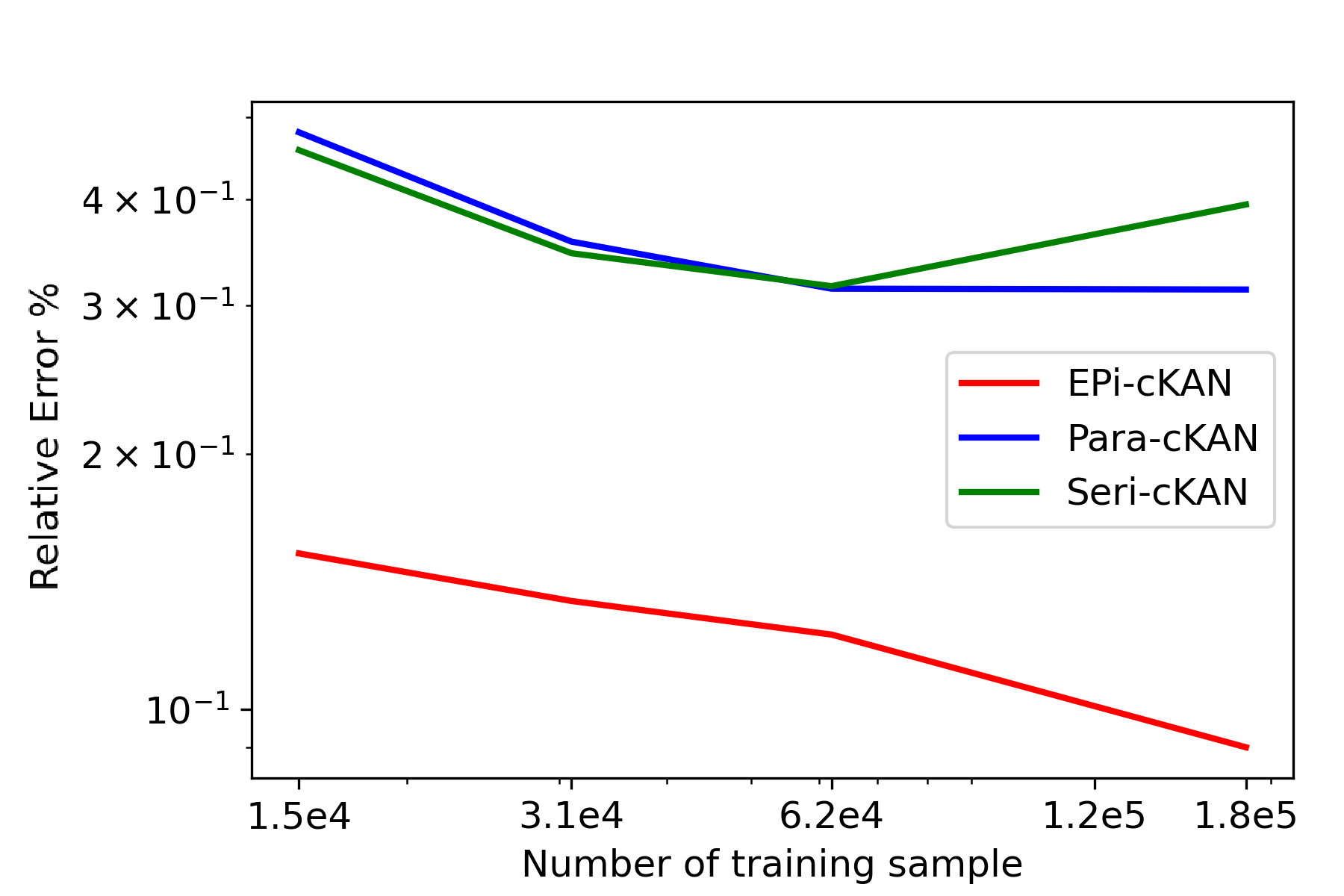}\hspace{.1cm}}
\subfigure[Elasto-plastic based architectures]
{ \label{PicDiffSizeEPs}\includegraphics[width=.45\textwidth]{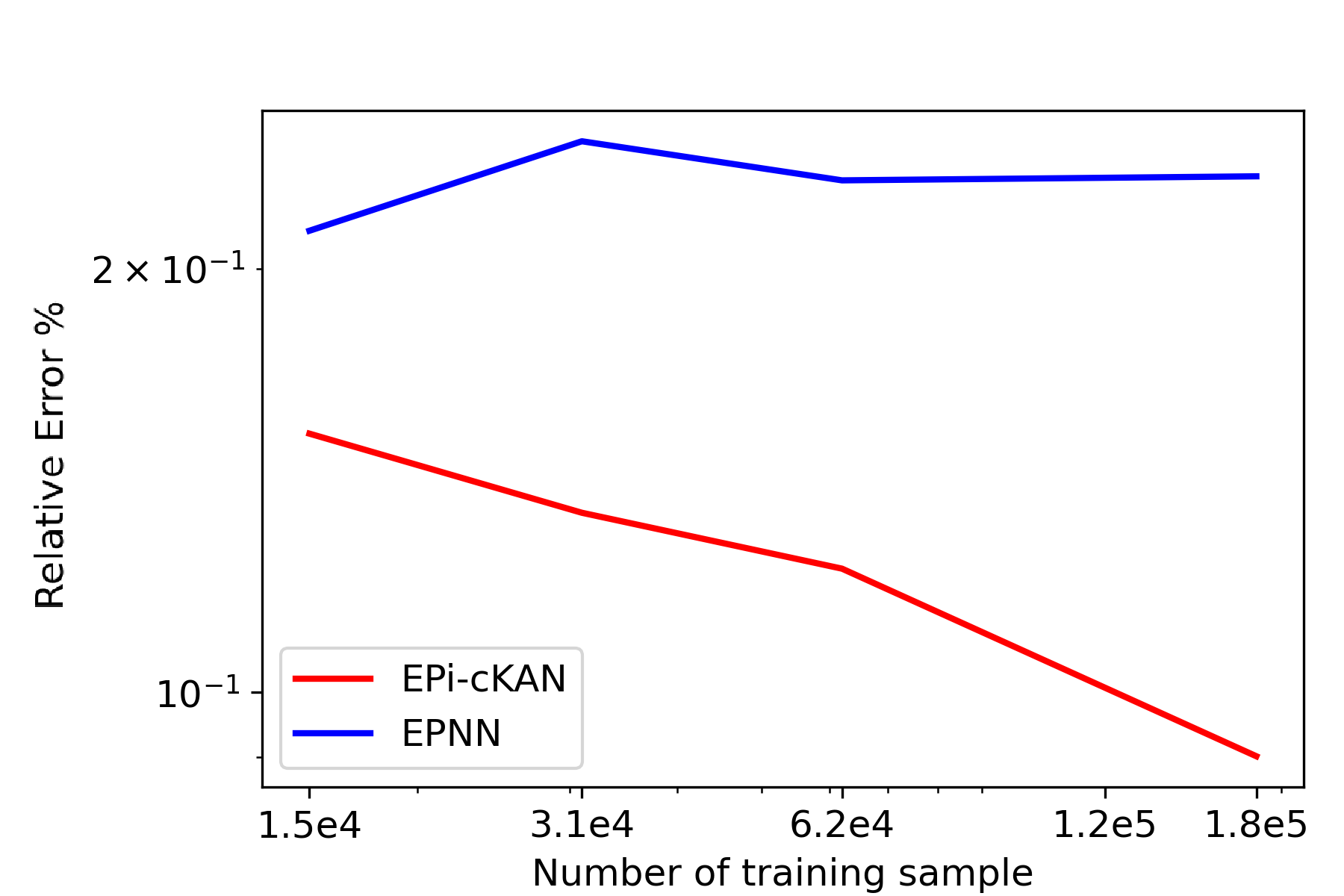}\hspace{.1cm}}
\caption{Relative $\mathcal{L}^2$ errors of the evaluation set for the stress sub-network across different architectures, plotted against the number of training samples. The results are obtained after 20,000 epochs of training, with configurations specified in Table \ref{Tab.Settings}. The axes are presented on a logarithmic scale to highlight variations in error magnitudes.} \label{Pic.DiffSize}
\end{figure}
\begin{figure}[!h]
\centering
\subfigure[cKAN architectures]
{ \label{Pic.LosscKAN}\includegraphics[width=.45\textwidth]{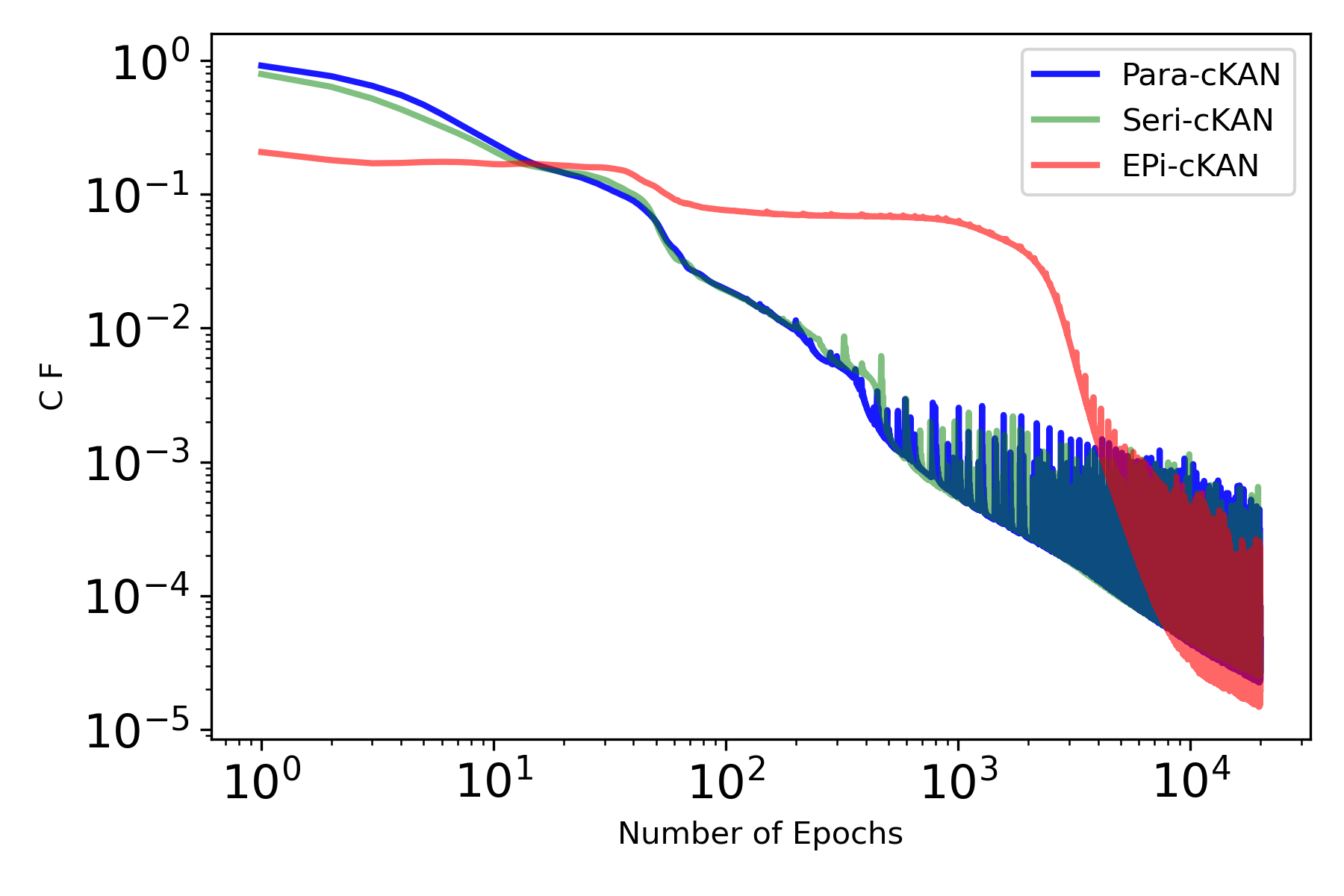}\hspace{.1cm}}
\subfigure[MLP architectures]
{ \label{Pic.LossMLP}\includegraphics[width=.45\textwidth]{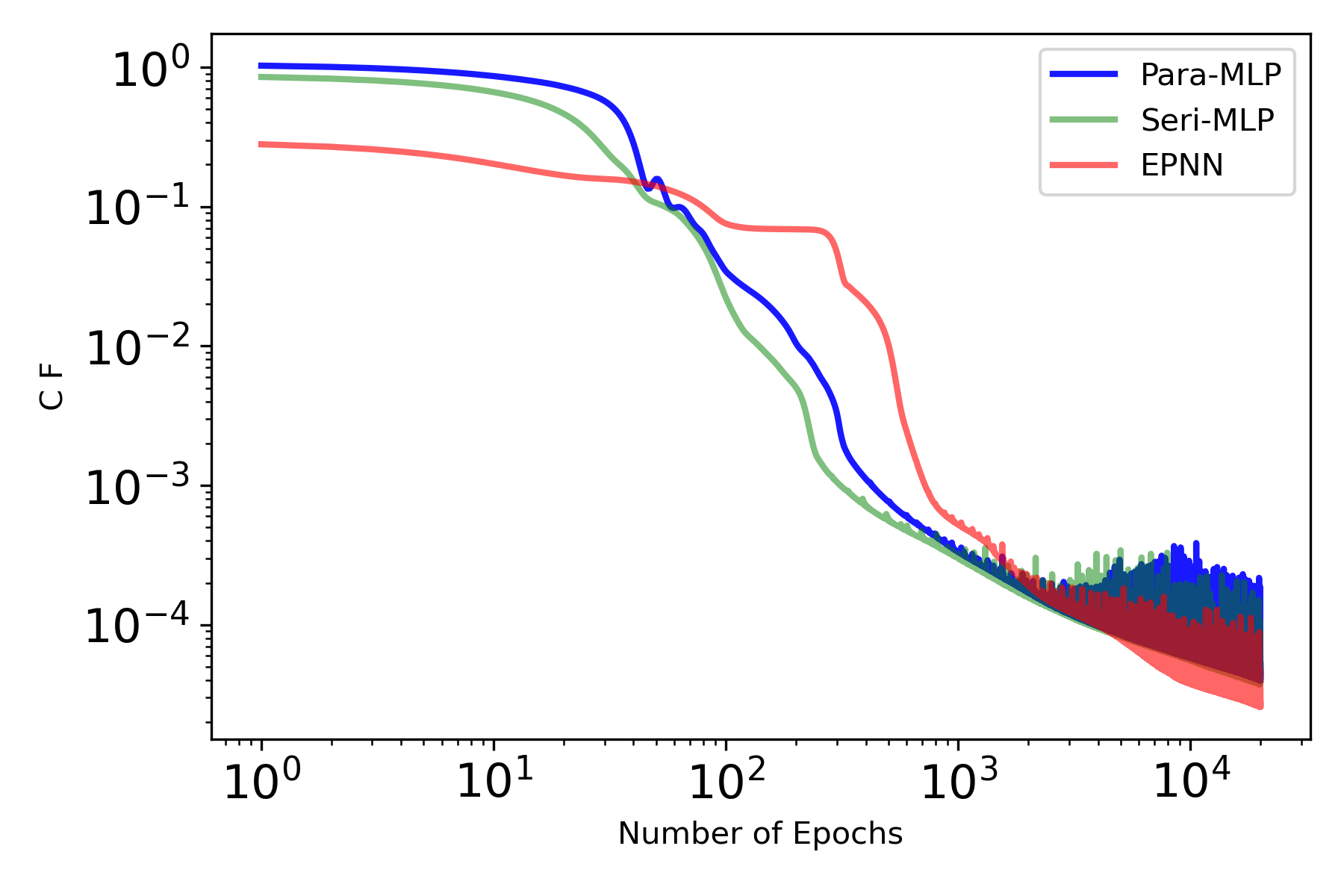}\hspace{.1cm}}
\caption{Cost function versus number of training epochs for 187000 training samples.} \label{Pic.Losses}
\end{figure}

Next, we compare the performance of the cKAN-based and MLP-based architectures. Table \ref{Tab.Settings} outlines the configurations of each architecture along with their total number of parameters. To ensure a fair comparison, the MLP network structures have been designed such that the total number of parameters is approximately equal across all architectures. We use 31,000 training data points—roughly twice the number of parameters.
Table \ref{Tab.TestCompare1} presents the relative \(\mathcal{L}^2\) errors of the sub-networks on the unseen evaluation set after 20,000 training iterations.
According to the results reported in Table \ref{Tab.TestCompare1}, the cKAN-based architectures generally outperform the MLP-based architectures in terms of accuracy, particularly in predicting stress and plastic strain, though they require more computation time. By comparing the results in Table \ref{Tab.TestCompare1} and the training time per iteration in Table \ref{Tab.Settings}, we can infer that cKAN models are preferable for applications demanding high accuracy and robustness, while MLP models may be better suited for scenarios where faster computations are prioritized, even at the cost of a slight reduction in accuracy.

To examine the behavior of stress with different training data sizes, we evaluated different architectures using diverse training sample sizes, with results presented in Fig. \ref{Pic.DiffSize}.
In panel \ref{Pic.DiffSizecKANs} in Fig. \ref{Pic.DiffSize}, we compare different cKAN architectures. The plot clearly demonstrates that EPi-cKAN has a significant advantage in reducing relative error as the number of training samples increases, while Parallel cKAN and Serial cKAN perform comparably, albeit less effectively.
Panel \ref{PicDiffSizeEPs} in Fig. \ref{Pic.DiffSize} compares the elasto-plastic based architectures, EPi-cKAN and EPNN. The results indicate that EPi-cKAN consistently outperforms EPNN, achieving lower relative errors and demonstrating a more stable improvement as the training data increases.
Overall, EPi-cKAN consistently surpasses all the compared models across both sets of figures, demonstrating superior accuracy and generalization with increasing training data. In Fig. \ref{Pic.Losses}, we present the evolution of the cost function over the number of training epochs, using 187000 training data. This plot illustrates how the model's performance improves during the training process, as reflected by the decreasing cost function.

\begin{figure}[!h]
\centering
\subfigure[\(\varepsilon^{v,p}\) vs \(\varepsilon_{33}\)]
{ \label{Pic.Fig24_cKAN_a}\includegraphics[width=.30\textwidth]{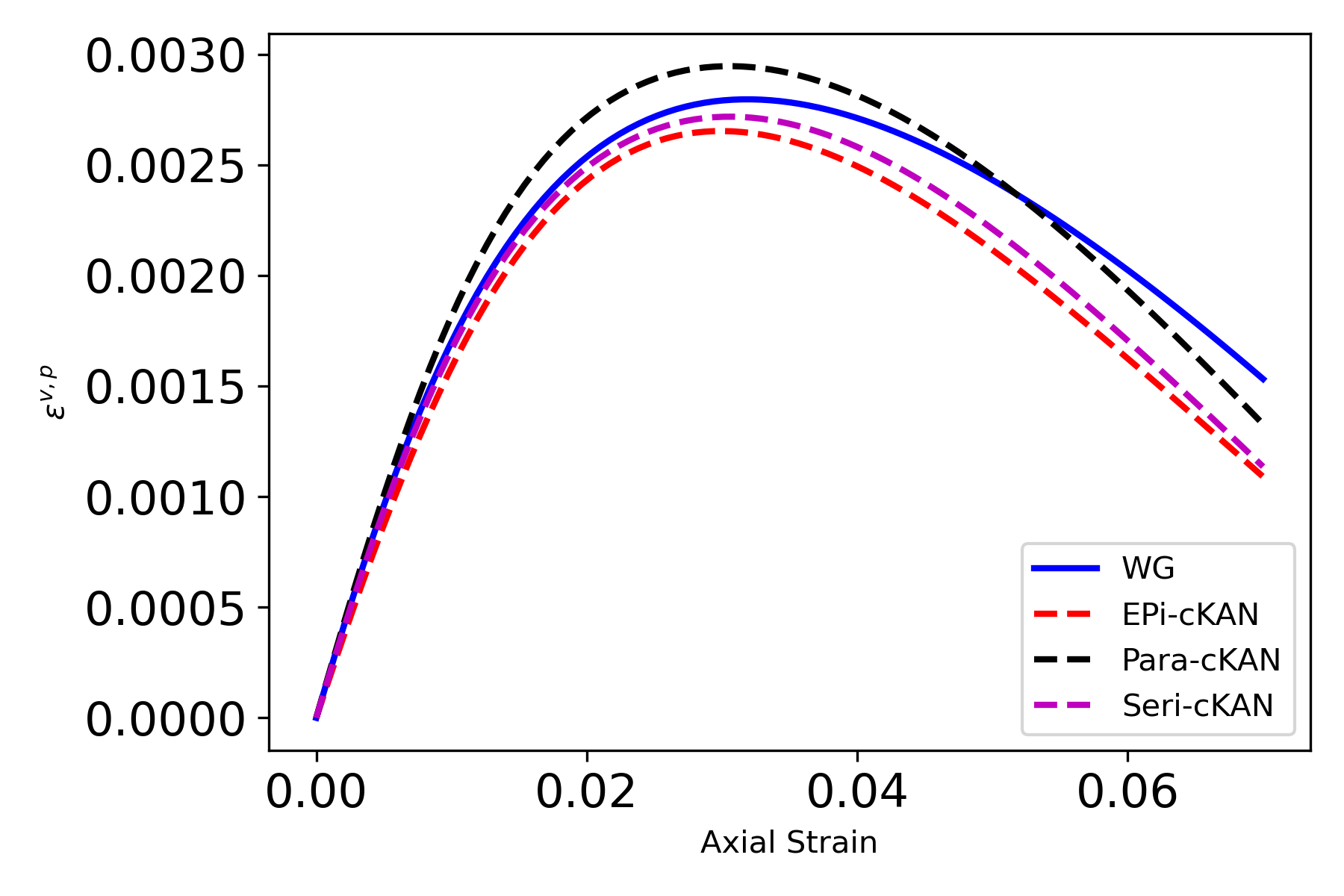}\hspace{.1cm}}
\subfigure[\(\gamma^{p}\) vs \(\varepsilon_{33}\)]
{ \label{Pic.Fig24_cKAN_b}\includegraphics[width=.30\textwidth]{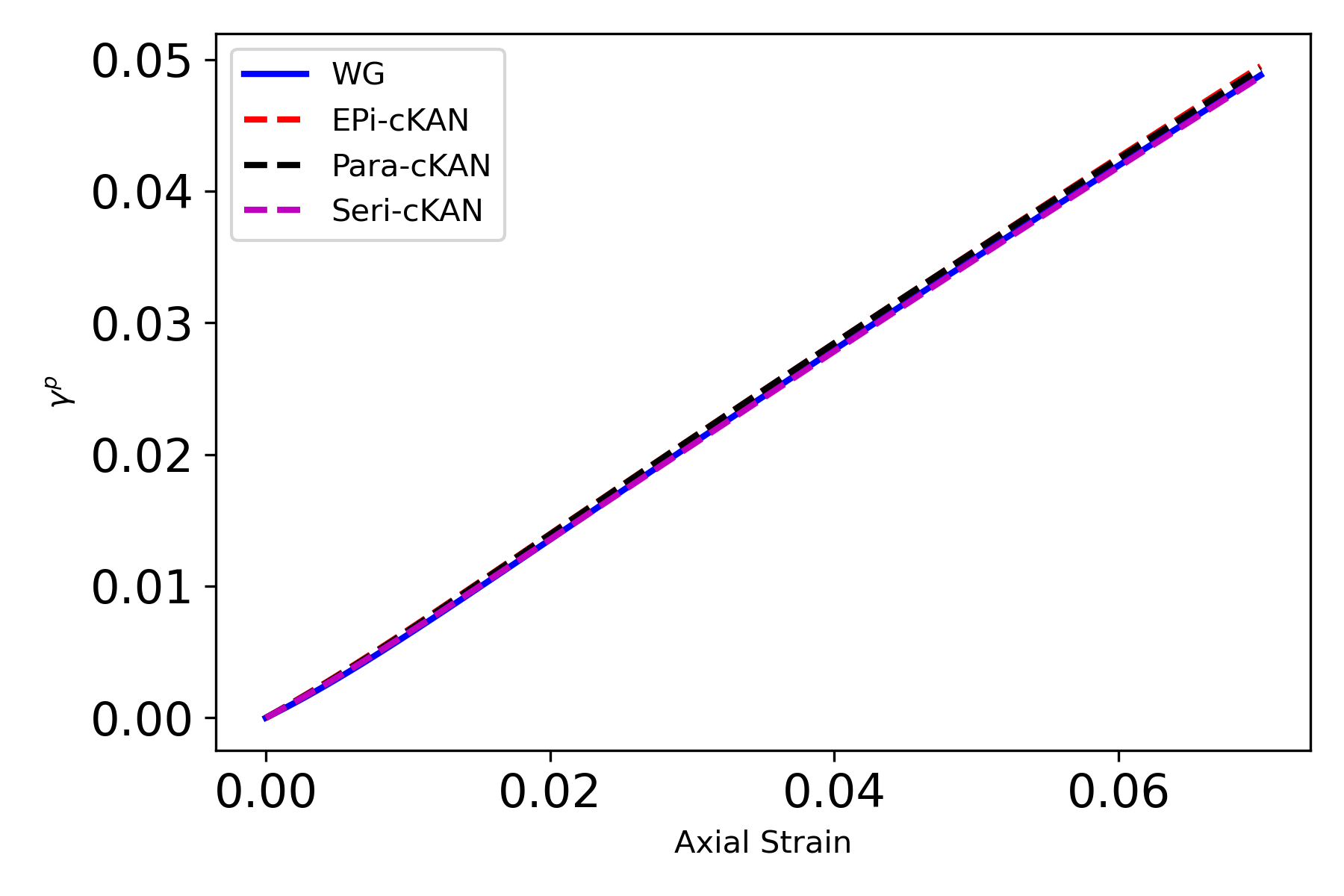}\hspace{.1cm}}
\subfigure[\(e\) vs \(\varepsilon_{33}\)]
{ \label{Pic.Fig24_cKAN_f}\includegraphics[width=.30\textwidth]{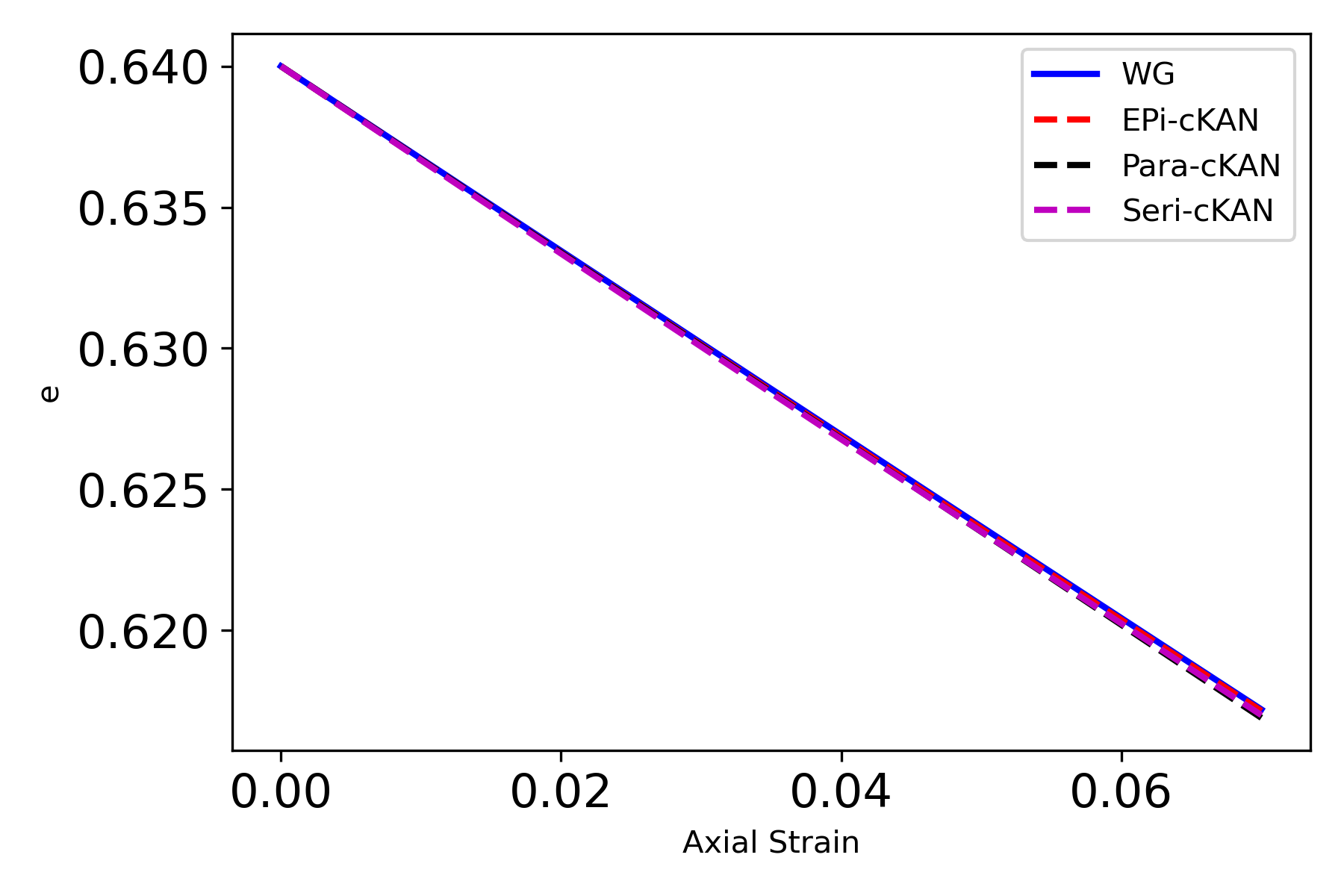}\hspace{.1cm}}

\subfigure[\(\vert\varepsilon^{v,p, *}-\varepsilon^{v,p}\vert\) vs \(\varepsilon_{33}\)]
{ \label{Pic.Fig24_cKAN_aERROR}\includegraphics[width=.30\textwidth]{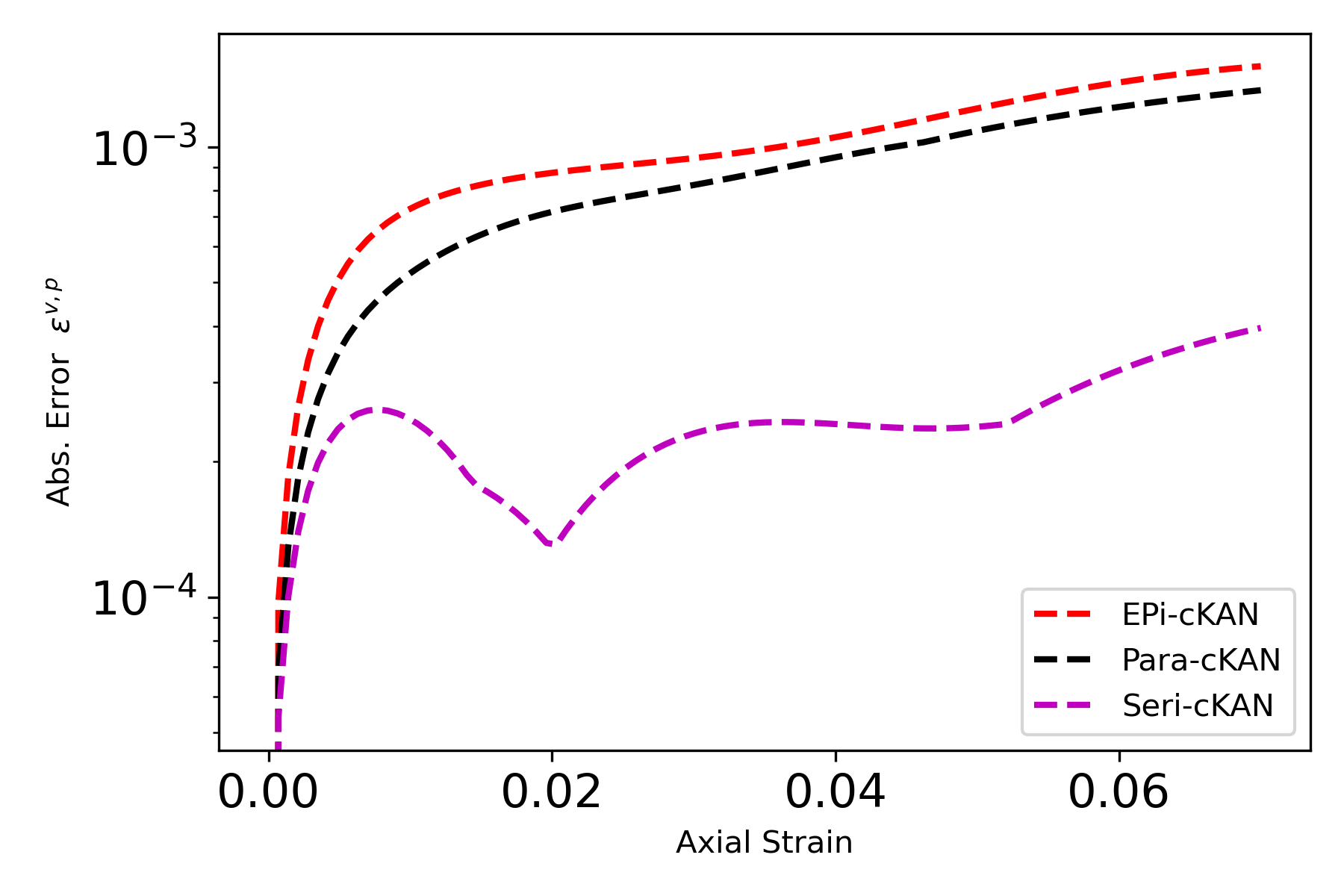}\hspace{.1cm}}
\subfigure[\(\vert\gamma^{p, *}-\gamma^{p}\vert\) vs \(\varepsilon_{33}\)]
{ \label{Pic.Fig24_cKAN_bERROR}\includegraphics[width=.30\textwidth]{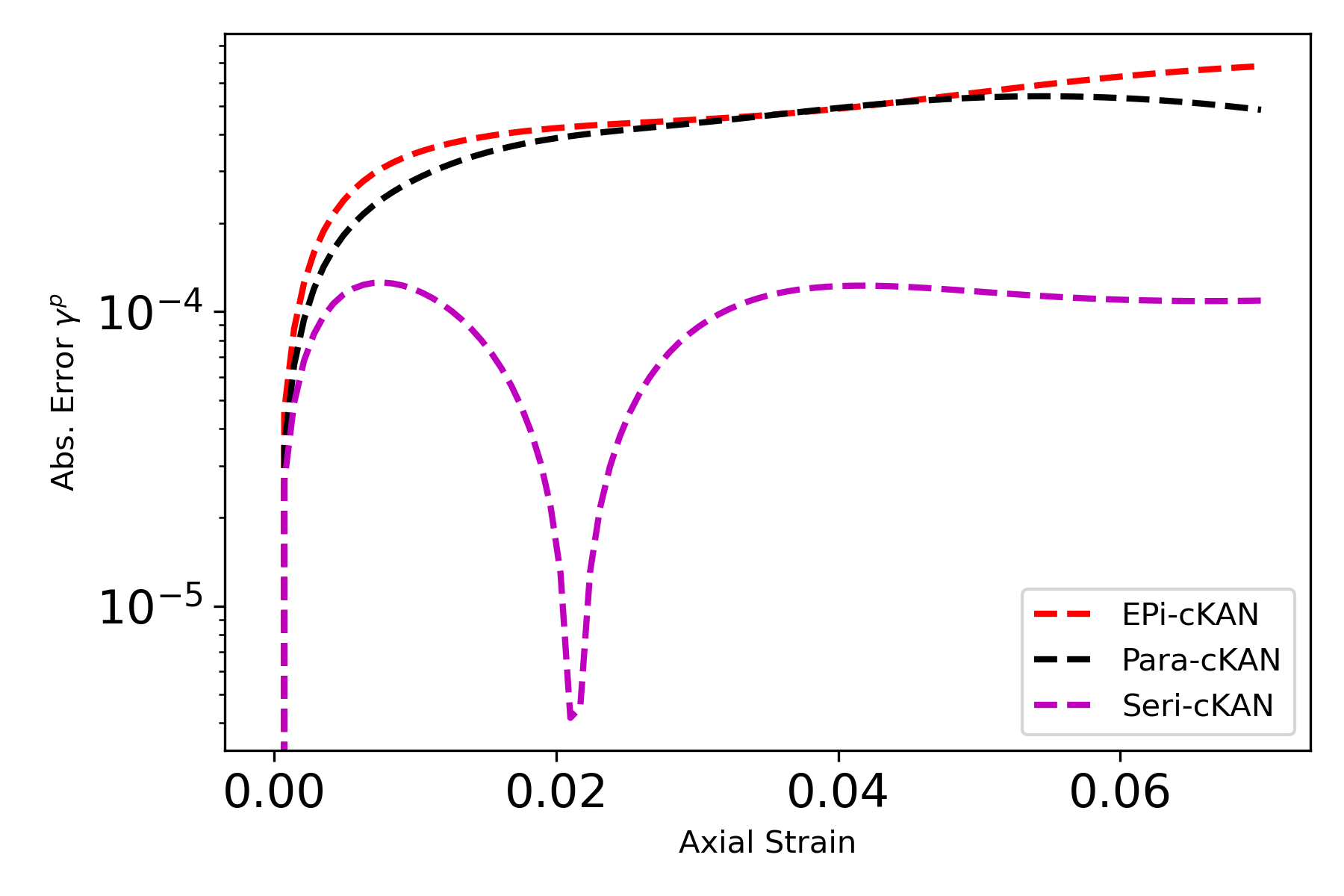}\hspace{.1cm}}
\subfigure[\(\vert e^{*}-e\vert\) vs \(\varepsilon_{33}\)]
{ \label{Pic.Fig24_cKAN_fERROR}\includegraphics[width=.30\textwidth]{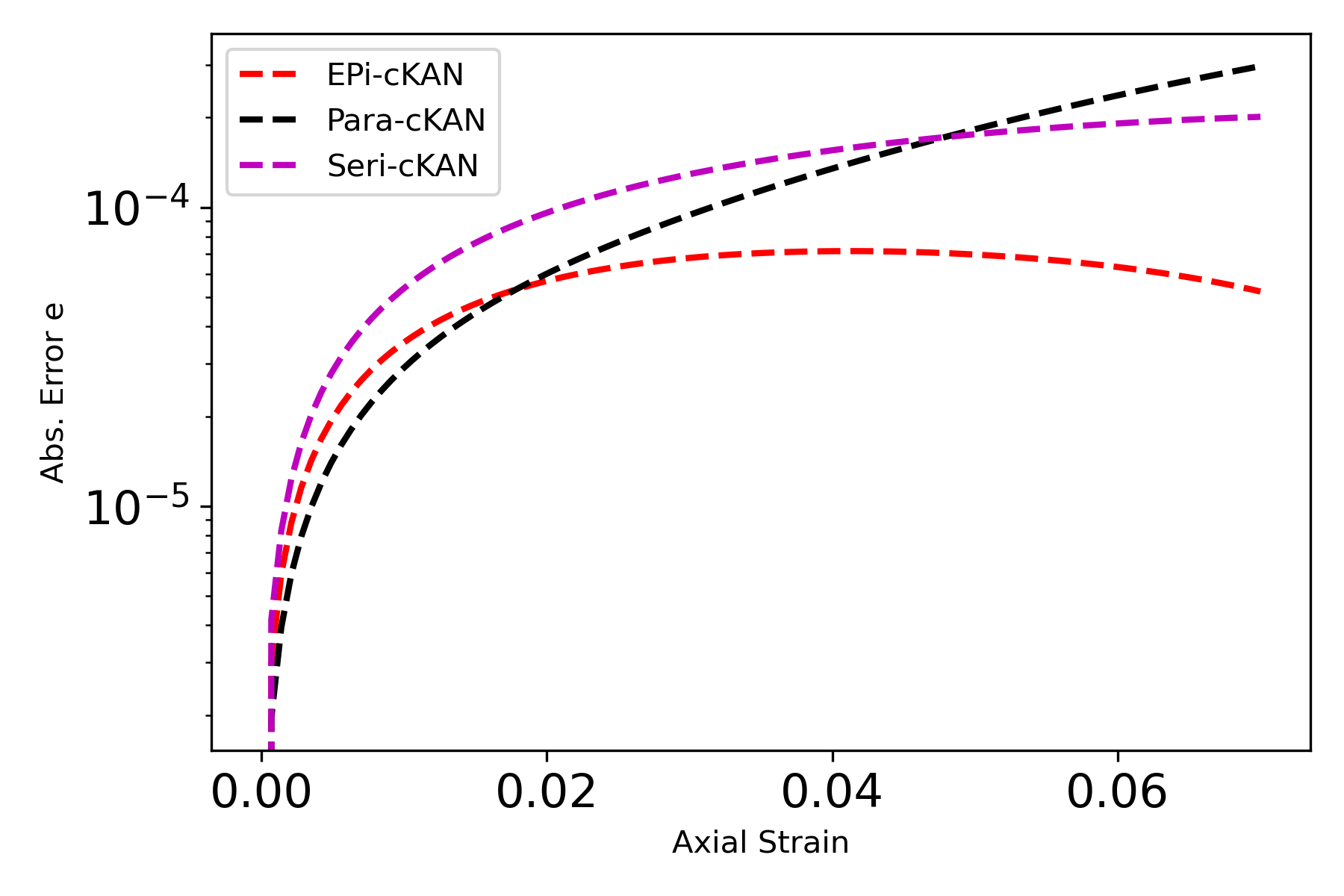}\hspace{.1cm}}

\subfigure[\(p\) (kPa) vs \(\varepsilon_{33}\)]
{ \label{Pic.Fig24_cKAN_c}\includegraphics[width=.30\textwidth]{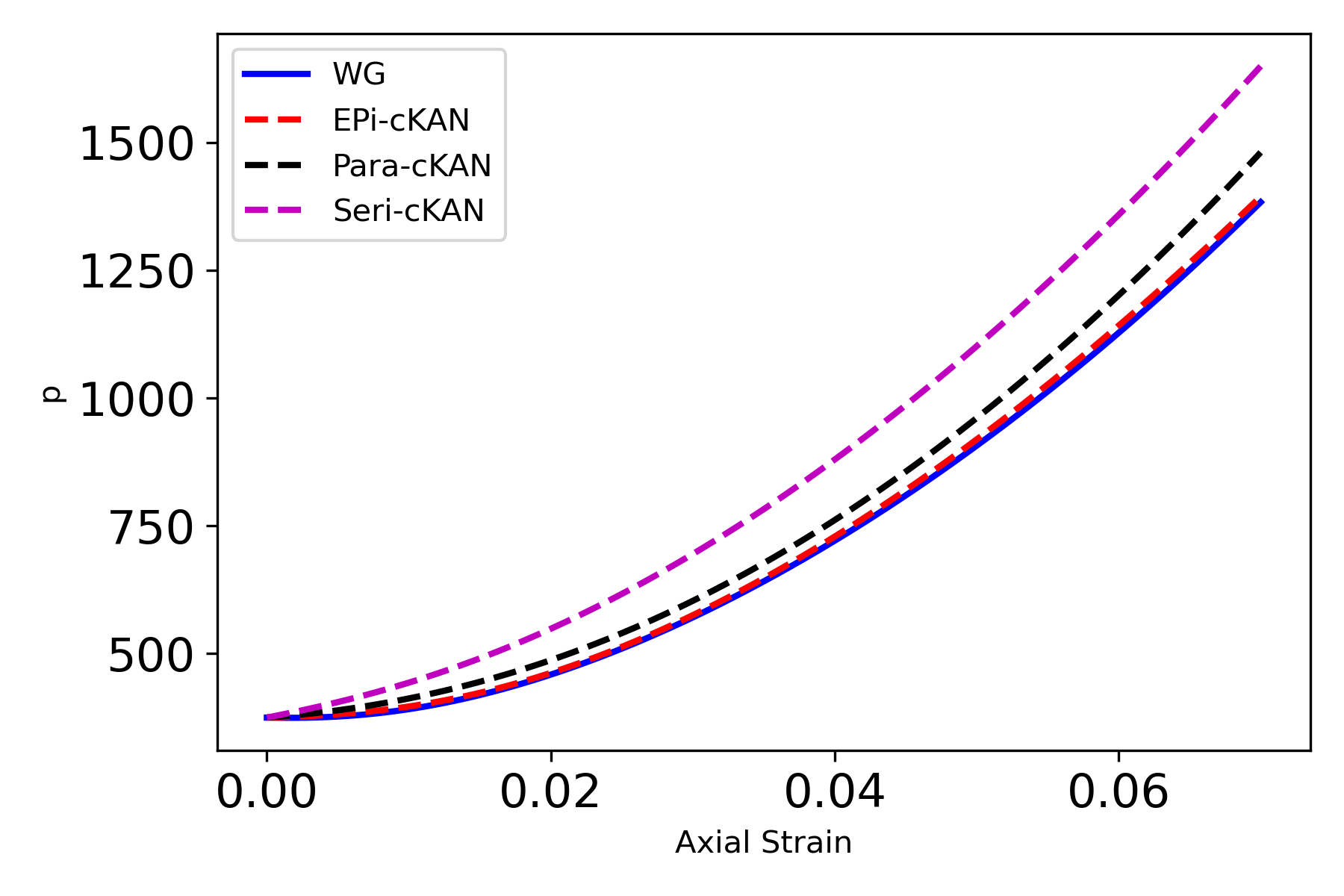}\hspace{.1cm}}
\subfigure[\(q\) (kPa) vs \(\varepsilon_{33}\)]
{ \label{Pic.Fig24_cKAN_d}\includegraphics[width=.30\textwidth]{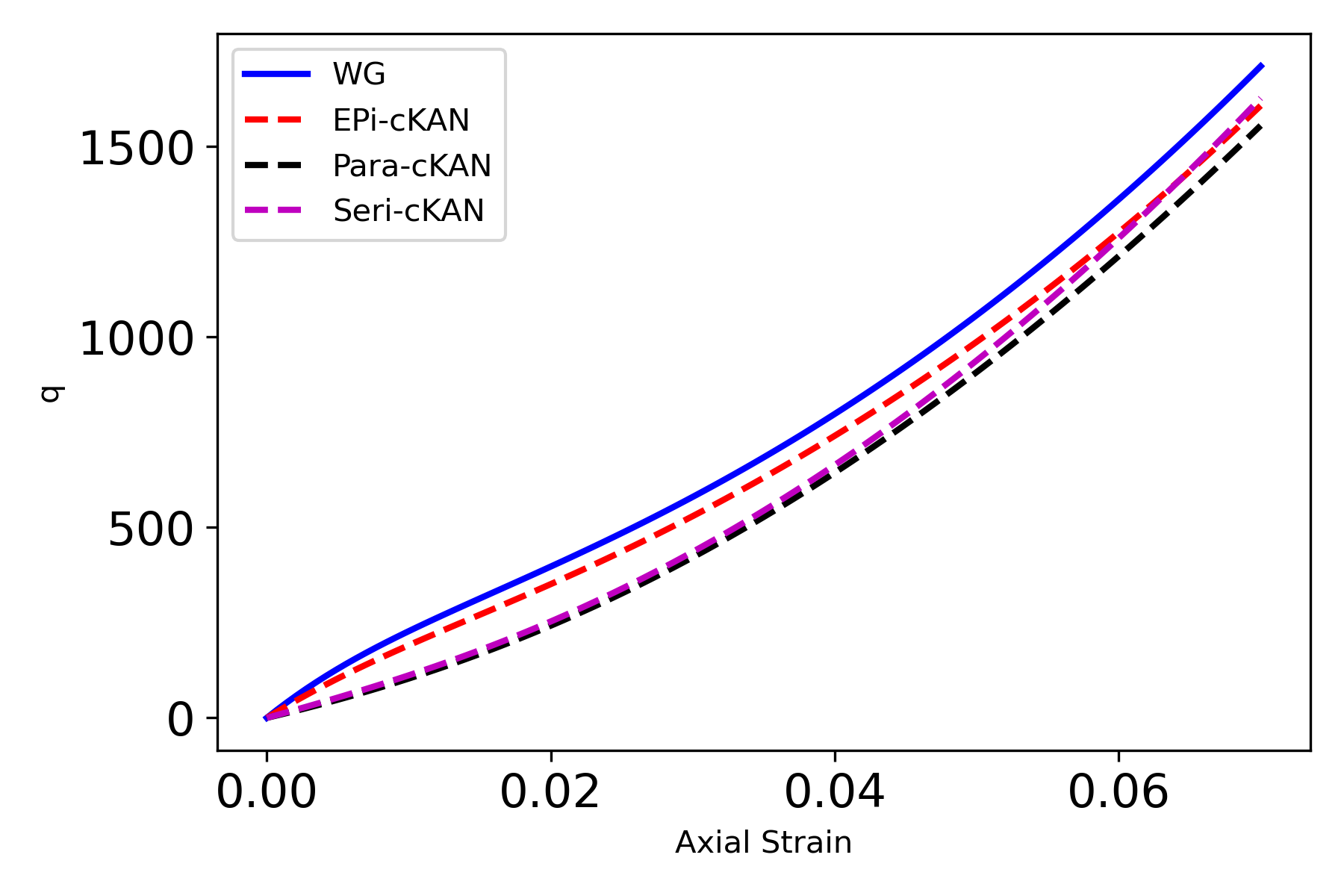}\hspace{.1cm}}
\subfigure[\(p\) (kPa) vs \(q\) (kPa)]
{ \label{Pic.Fig24_cKAN_e}\includegraphics[width=.30\textwidth]{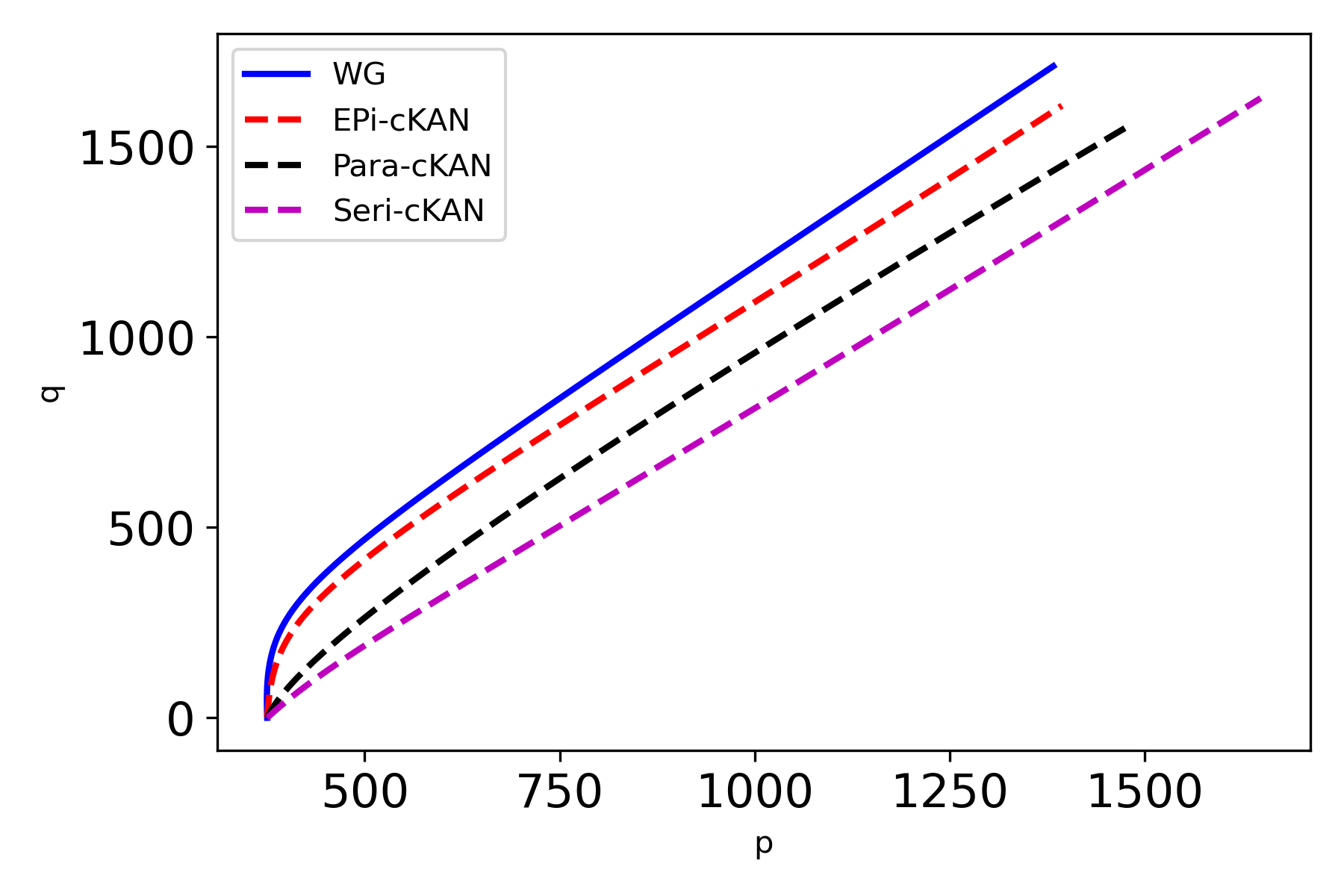}\hspace{.1cm}}

\subfigure[\(\vert p^{*}-p\vert\) vs \(\varepsilon_{33}\)]
{ \label{Pic.Fig24_cKAN_cERROR}\includegraphics[width=.30\textwidth]{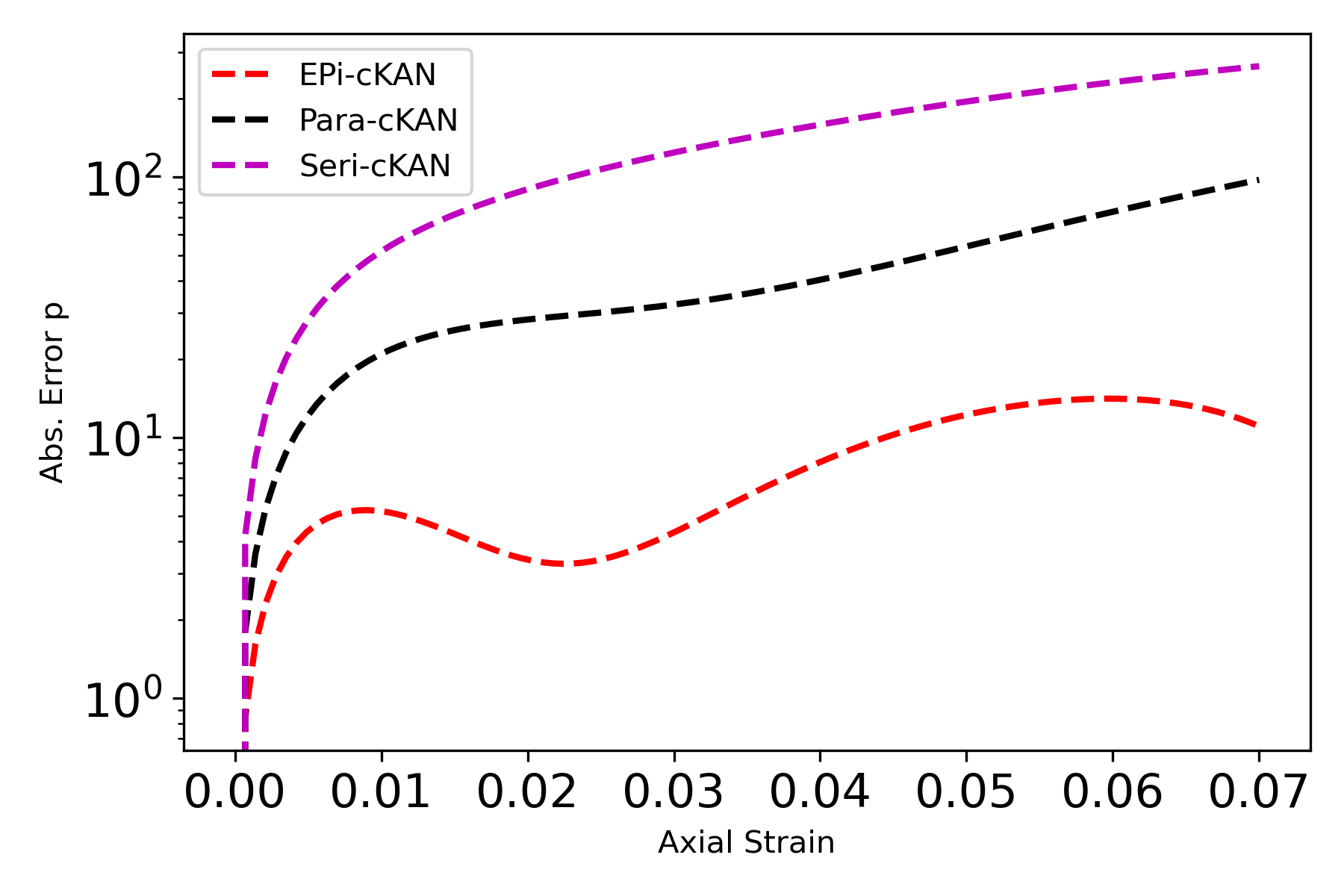}\hspace{.1cm}}
\subfigure[\(\vert q^{*}-q\vert\) vs \(\varepsilon_{33}\)]
{ \label{Pic.Fig24_cKAN_dERROR}\includegraphics[width=.30\textwidth]{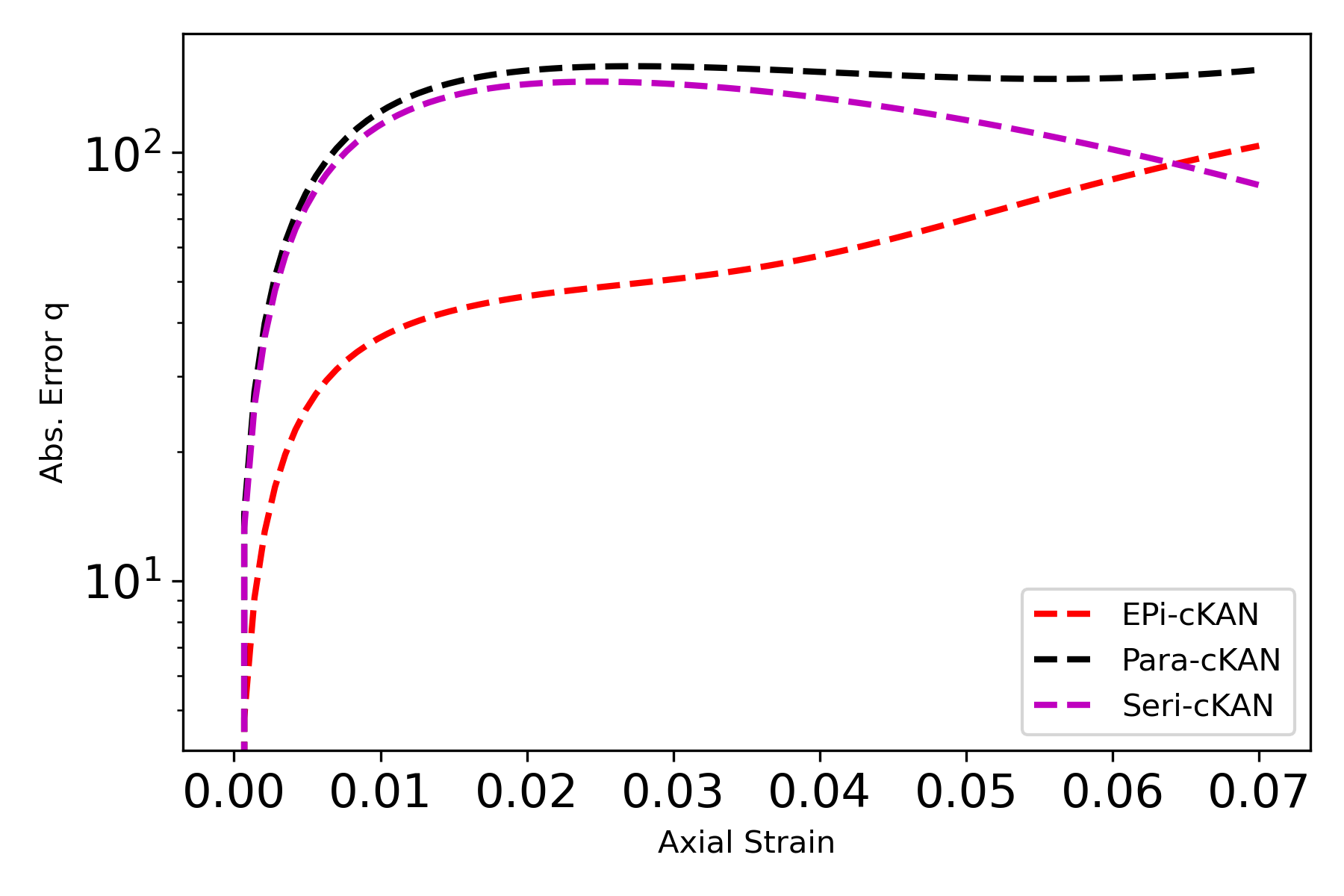}\hspace{.1cm}}

\caption{Comparison between the predictions of cKAN architectures (trained for \(2e4\) epochs) and the ground truth (numerical integration results), in the undrained test with \(\xi = -2.5\), \(p^{in}=375\) kPa and \(e^{in}=0.64\). Models used here are  trained on 31000 training data with the network structures reported in Table \ref{Tab.Settings}.} \label{Pic.Fig24_cKAN}
\end{figure}
\begin{figure}[!h]
\centering
\subfigure[\(\varepsilon^{v,p}\) vs \(\varepsilon_{33}\)]
{ \label{Pic.Fig24_EP_a}\includegraphics[width=.30\textwidth]{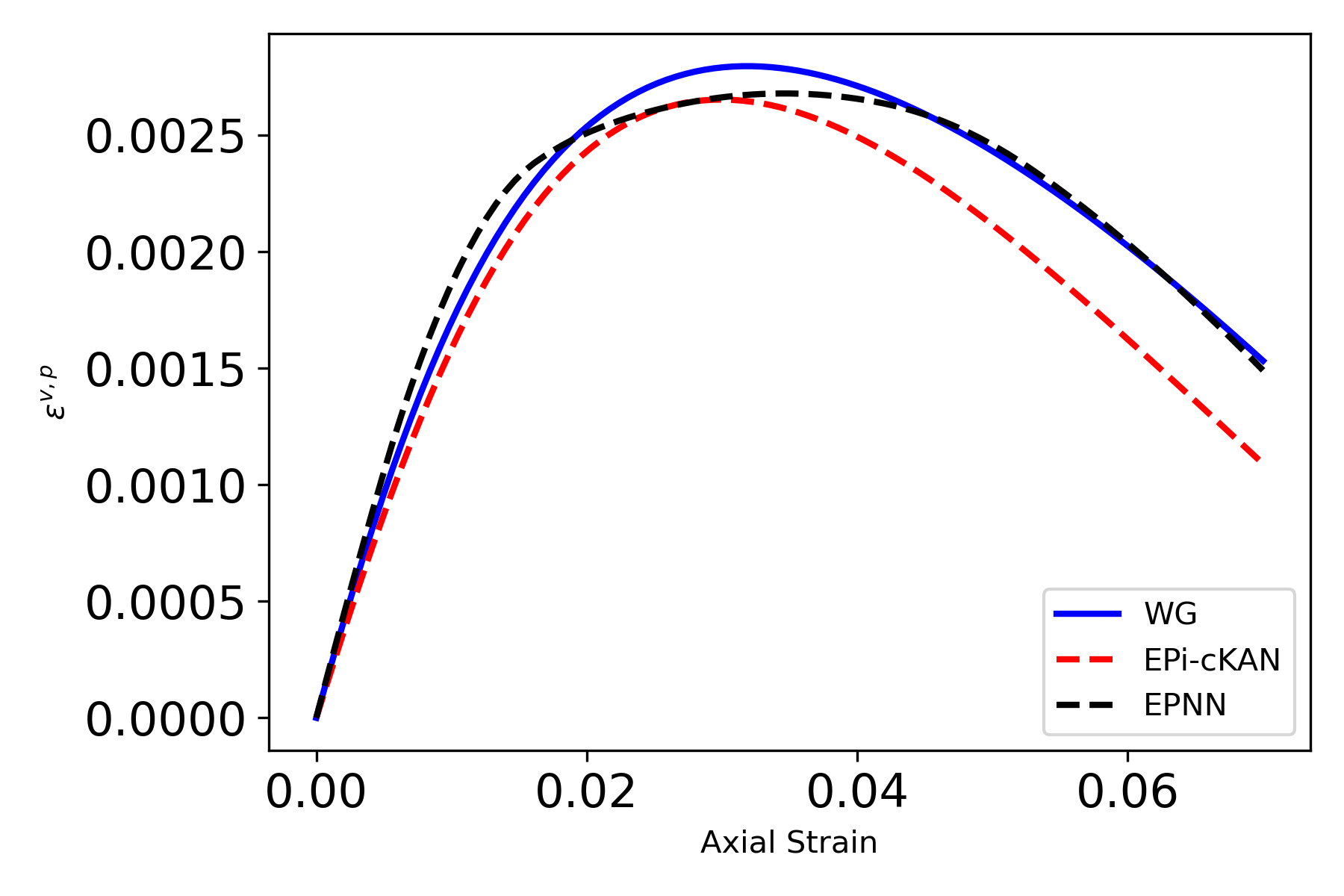}\hspace{.1cm}}
\subfigure[\(\gamma^{p}\) vs \(\varepsilon_{33}\)]
{ \label{Pic.Fig24_EP_b}\includegraphics[width=.30\textwidth]{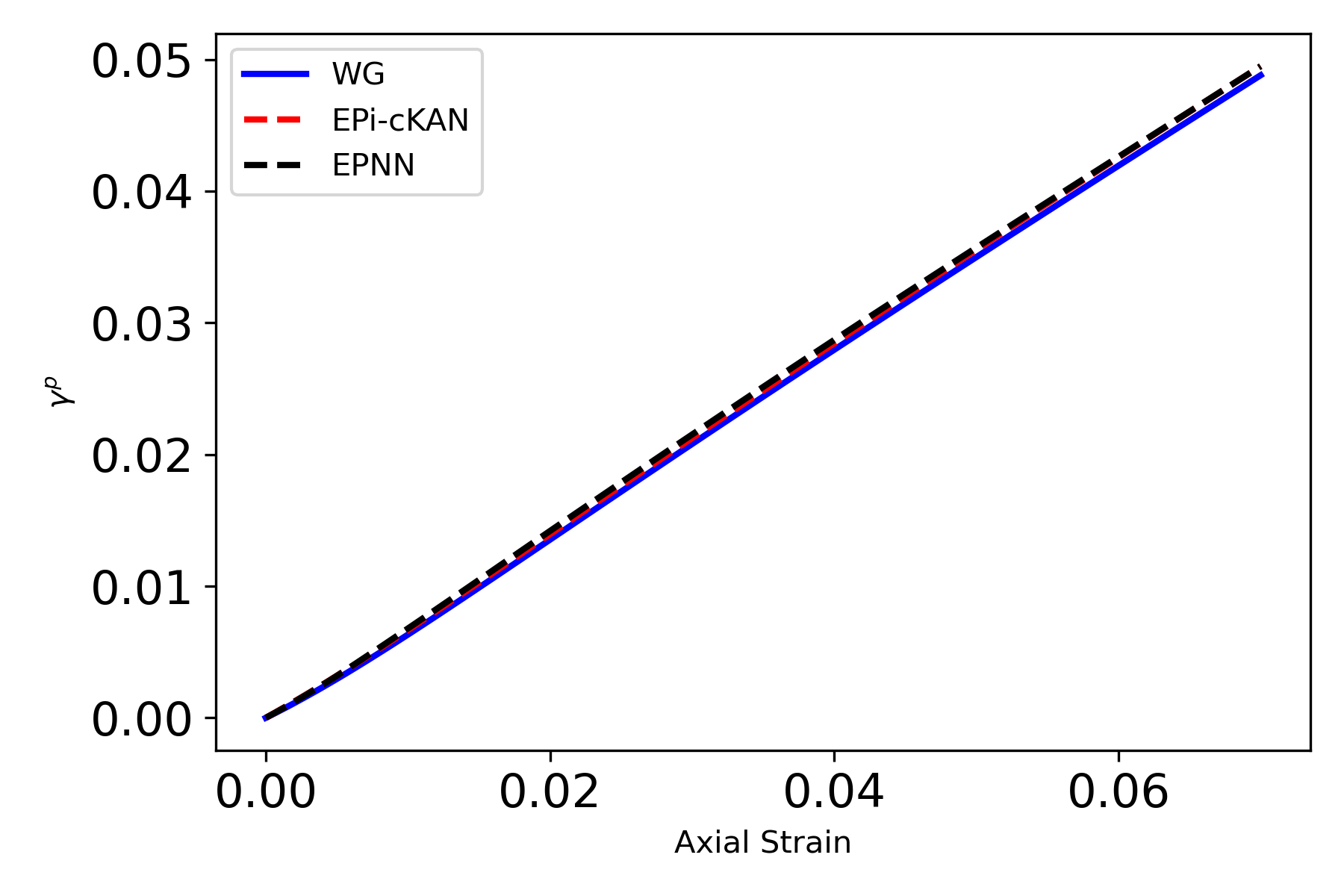}\hspace{.1cm}}
\subfigure[\(e\) vs \(\varepsilon_{33}\)]
{ \label{Pic.Fig24_EP_f}\includegraphics[width=.30\textwidth]{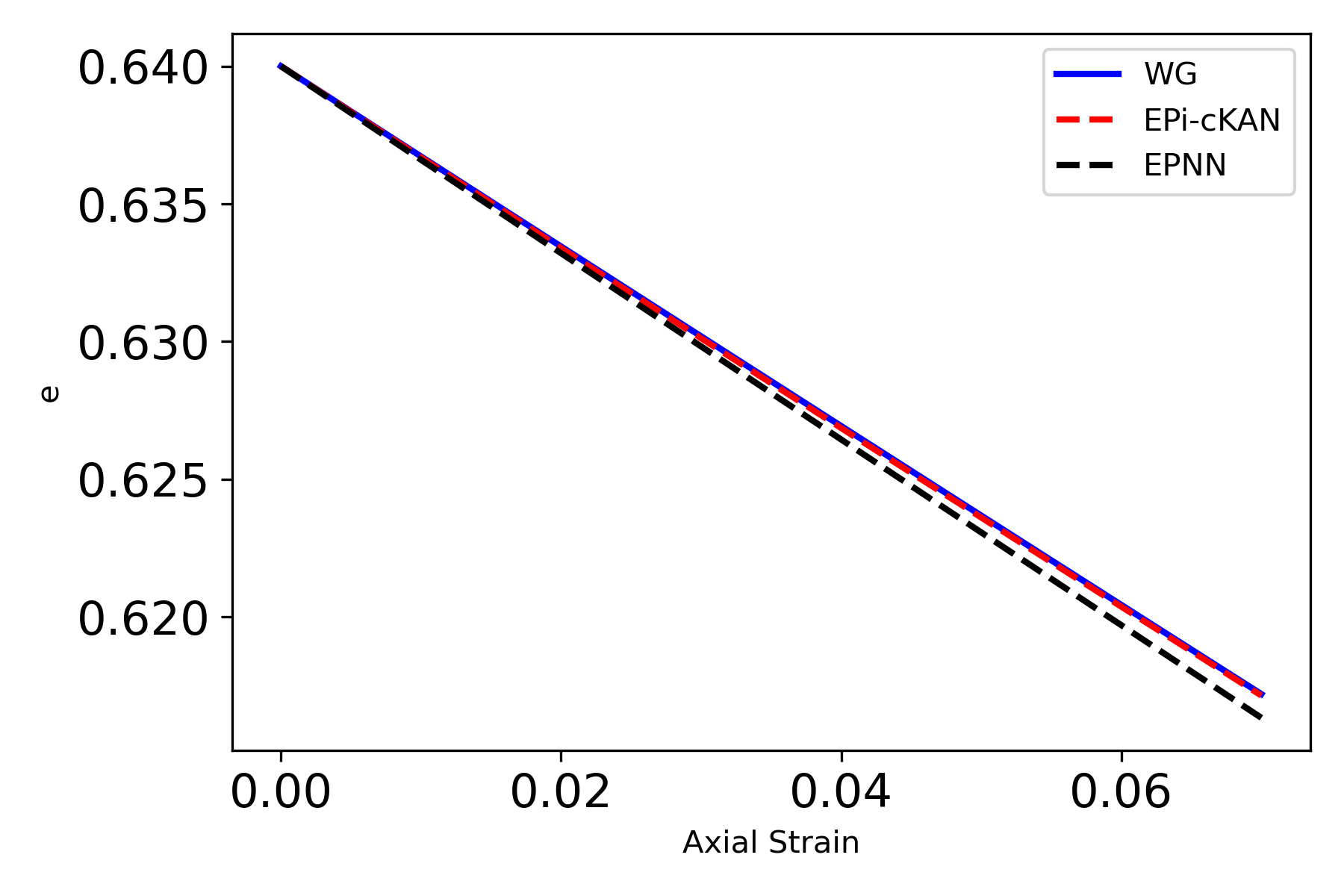}\hspace{.1cm}}

\subfigure[\(\vert\varepsilon^{v,p, *}-\varepsilon^{v,p}\vert\) vs \(\varepsilon_{33}\)]
{ \label{Pic.Fig24_EP_aERROR}\includegraphics[width=.30\textwidth]{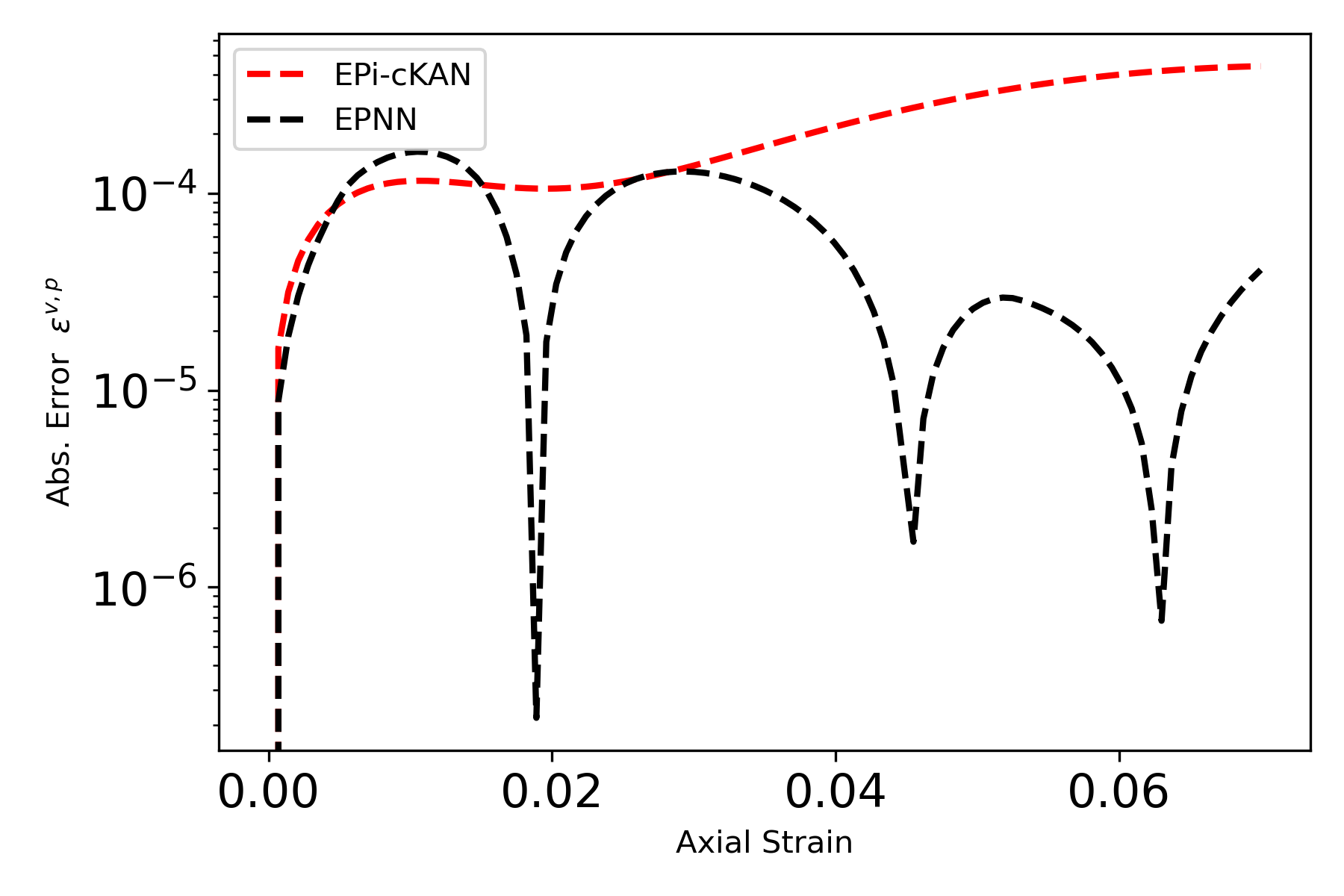}\hspace{.1cm}}
\subfigure[\(\vert\gamma^{p, *}-\gamma^{p}\vert\) vs \(\varepsilon_{33}\)]
{ \label{Pic.Fig24_EP_bERROR}\includegraphics[width=.30\textwidth]{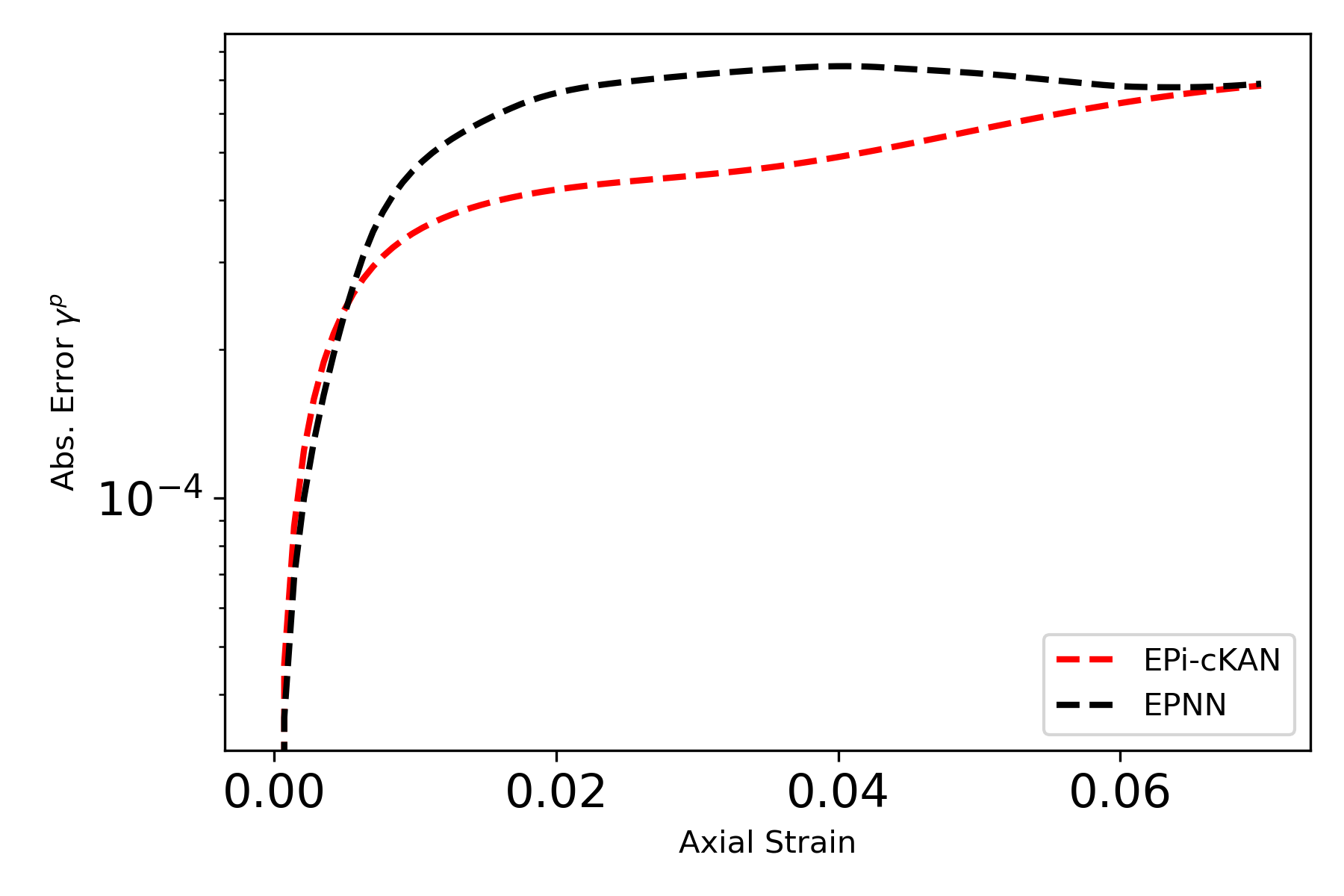}\hspace{.1cm}}
\subfigure[\(\vert e^{*}-e\vert\) vs \(\varepsilon_{33}\)]
{ \label{Pic.Fig24_EP_fERROR}\includegraphics[width=.30\textwidth]{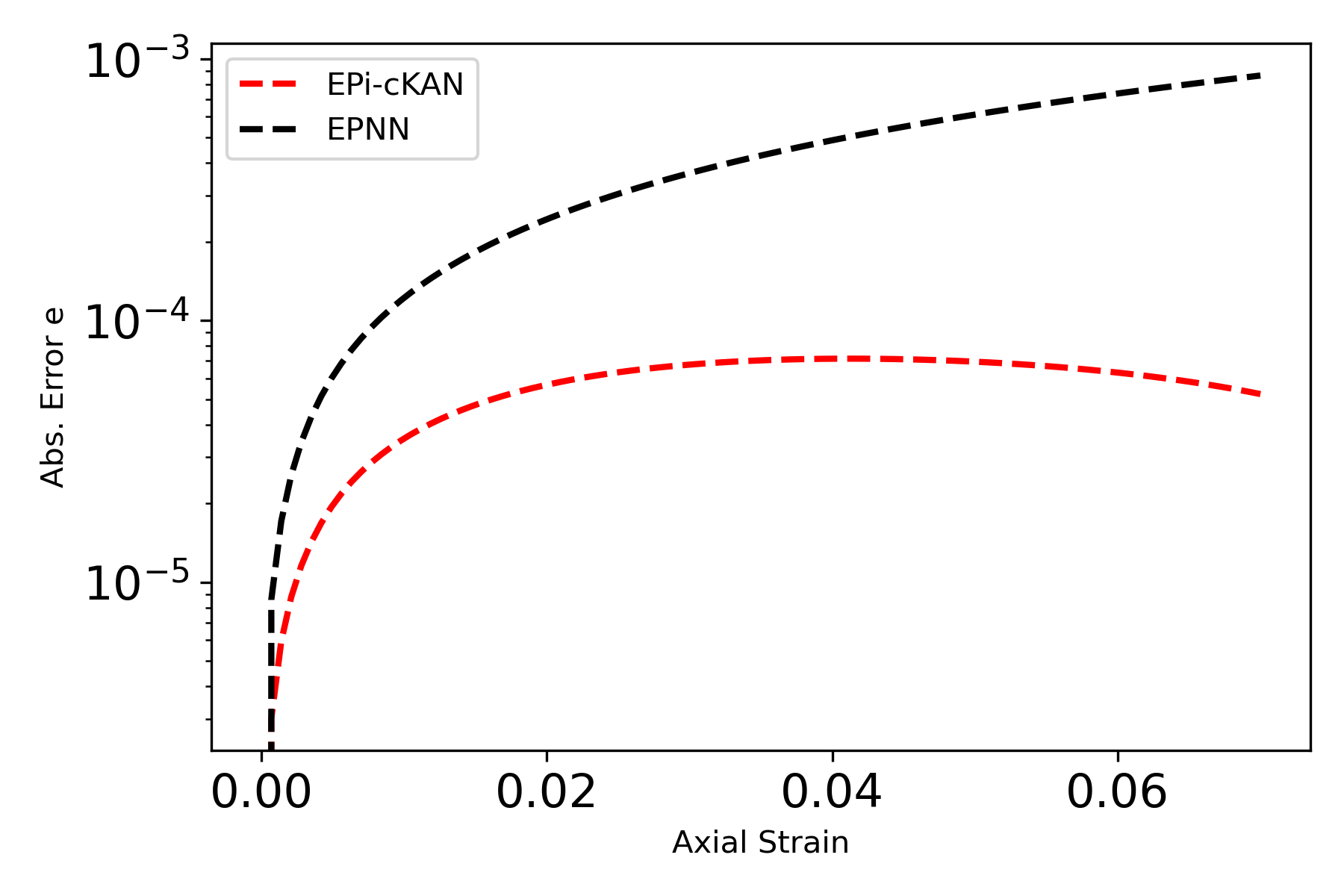}\hspace{.1cm}}

\subfigure[\(p\) (kPa) vs \(\varepsilon_{33}\)]
{ \label{Pic.Fig24_EP_c}\includegraphics[width=.30\textwidth]{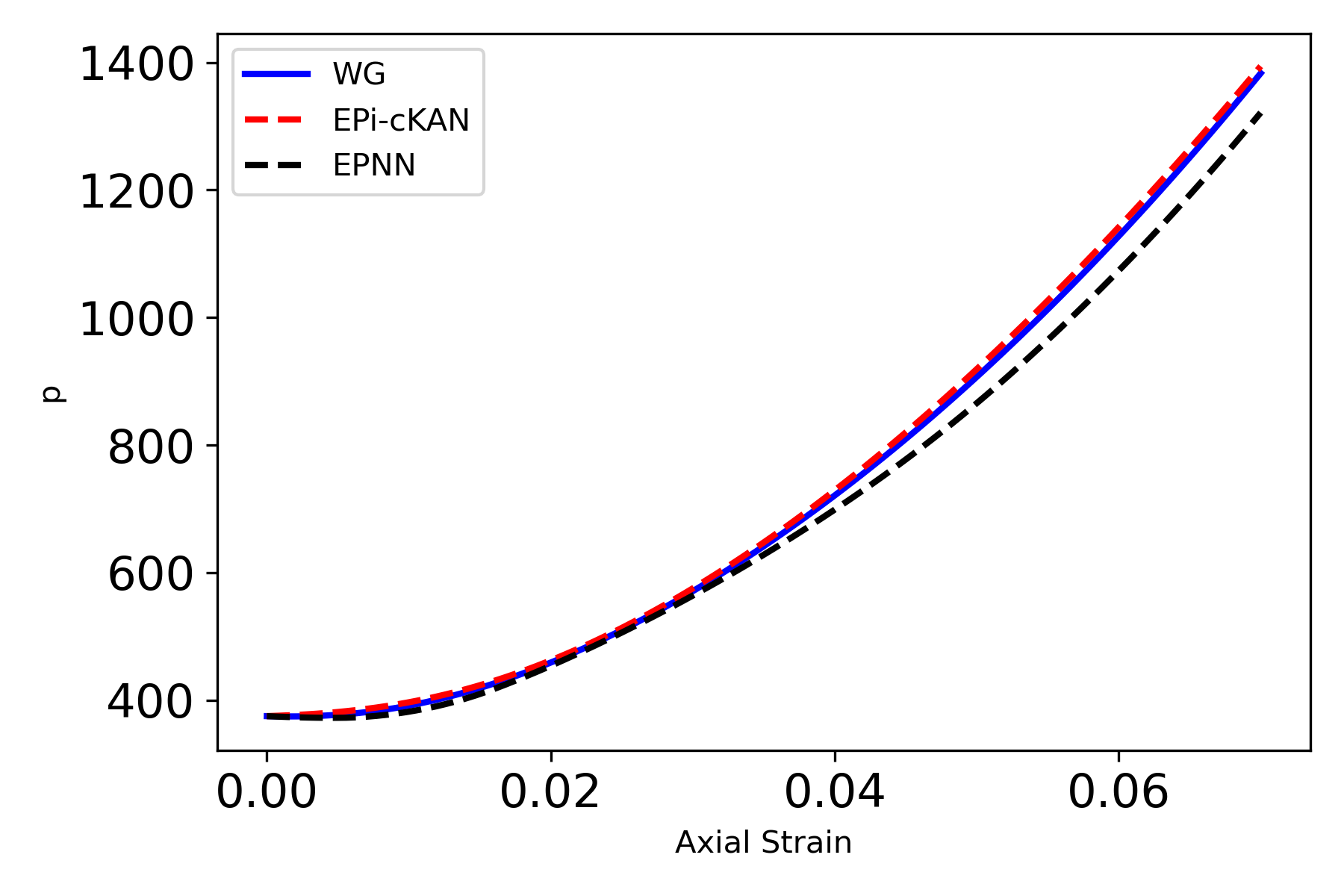}\hspace{.1cm}}
\subfigure[\(q\) (kPa) vs \(\varepsilon_{33}\)]
{ \label{Pic.Fig24_EP_d}\includegraphics[width=.30\textwidth]{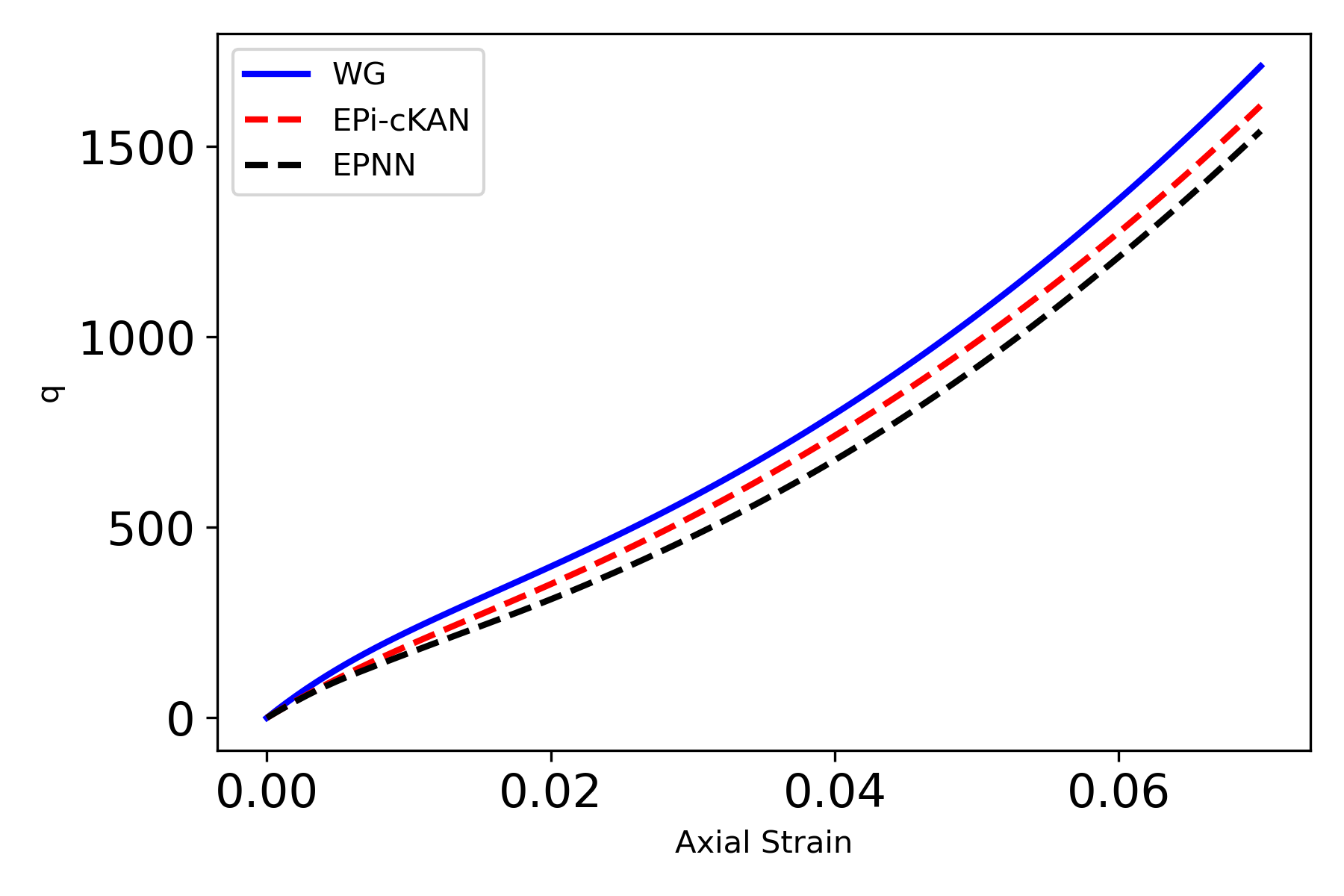}\hspace{.1cm}}
\subfigure[\(p\) (kPa) vs \(q\) (kPa)]
{ \label{Pic.Fig24_EP_e}\includegraphics[width=.30\textwidth]{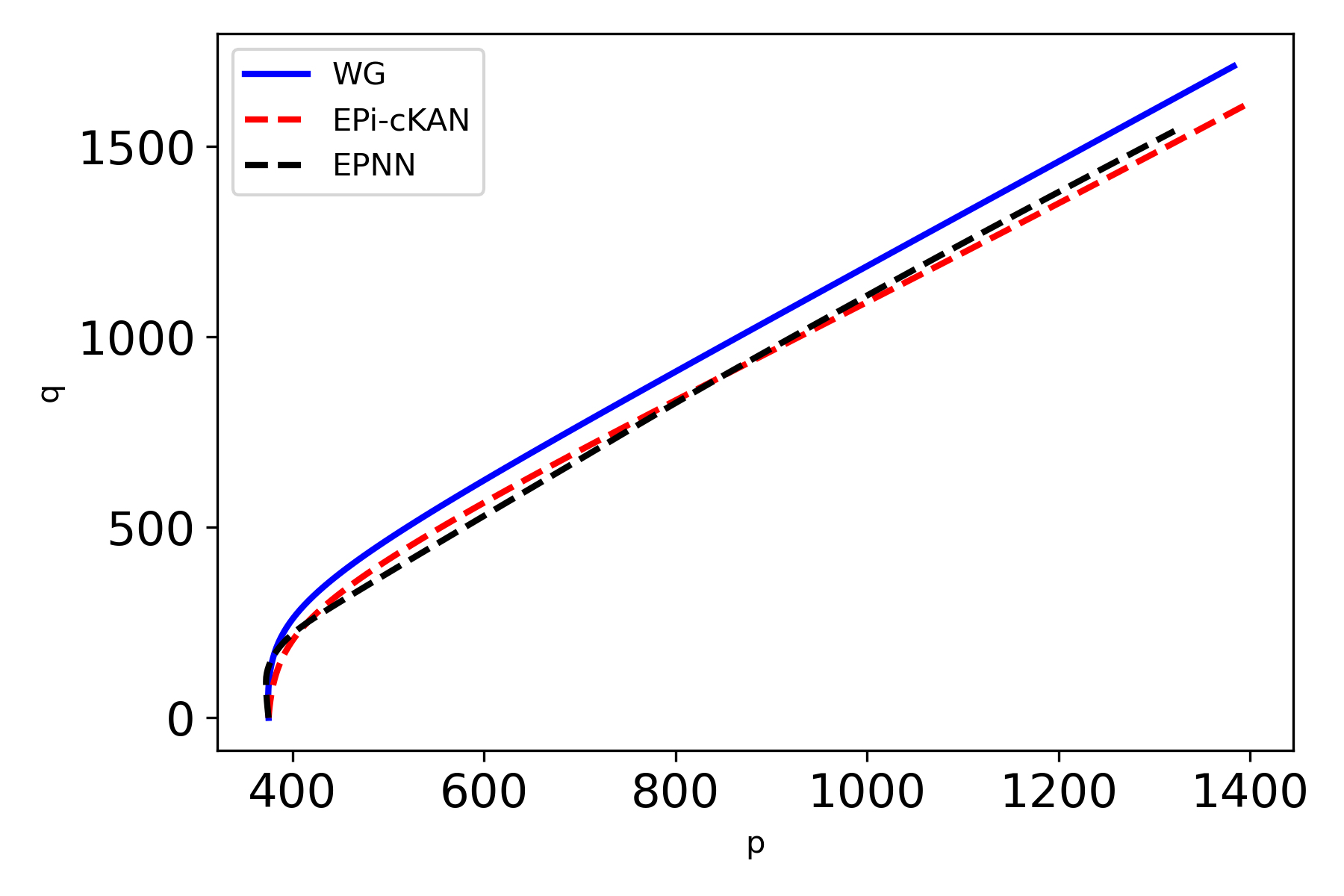}\hspace{.1cm}}

\subfigure[\(\vert p^{*}-p\vert\) vs \(\varepsilon_{33}\)]
{ \label{Pic.Fig24_EP_cERROR}\includegraphics[width=.30\textwidth]{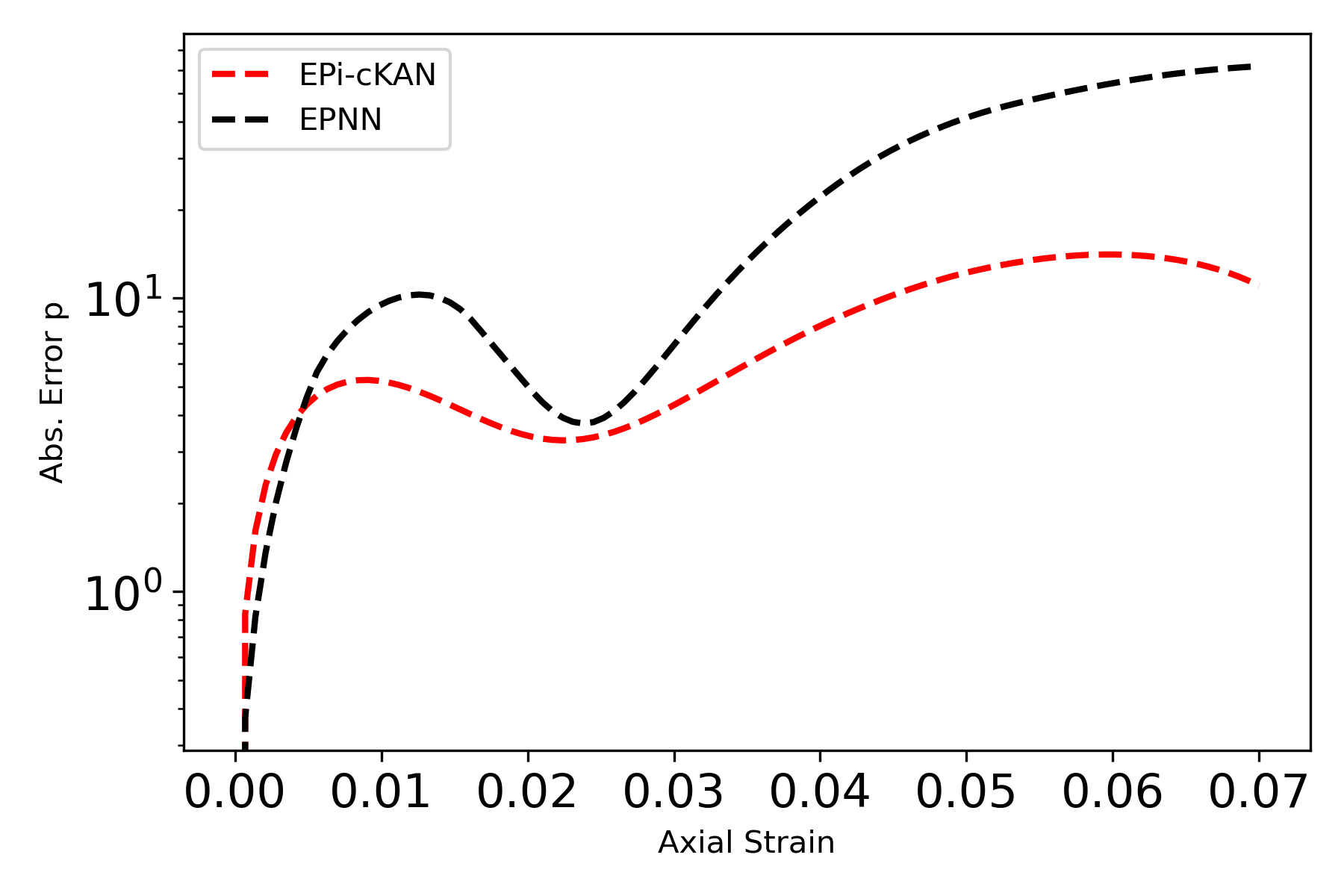}\hspace{.1cm}}
\subfigure[\(\vert q^{*}-q\vert\) vs \(\varepsilon_{33}\)]
{ \label{Pic.Fig24_EP_dERROR}\includegraphics[width=.30\textwidth]{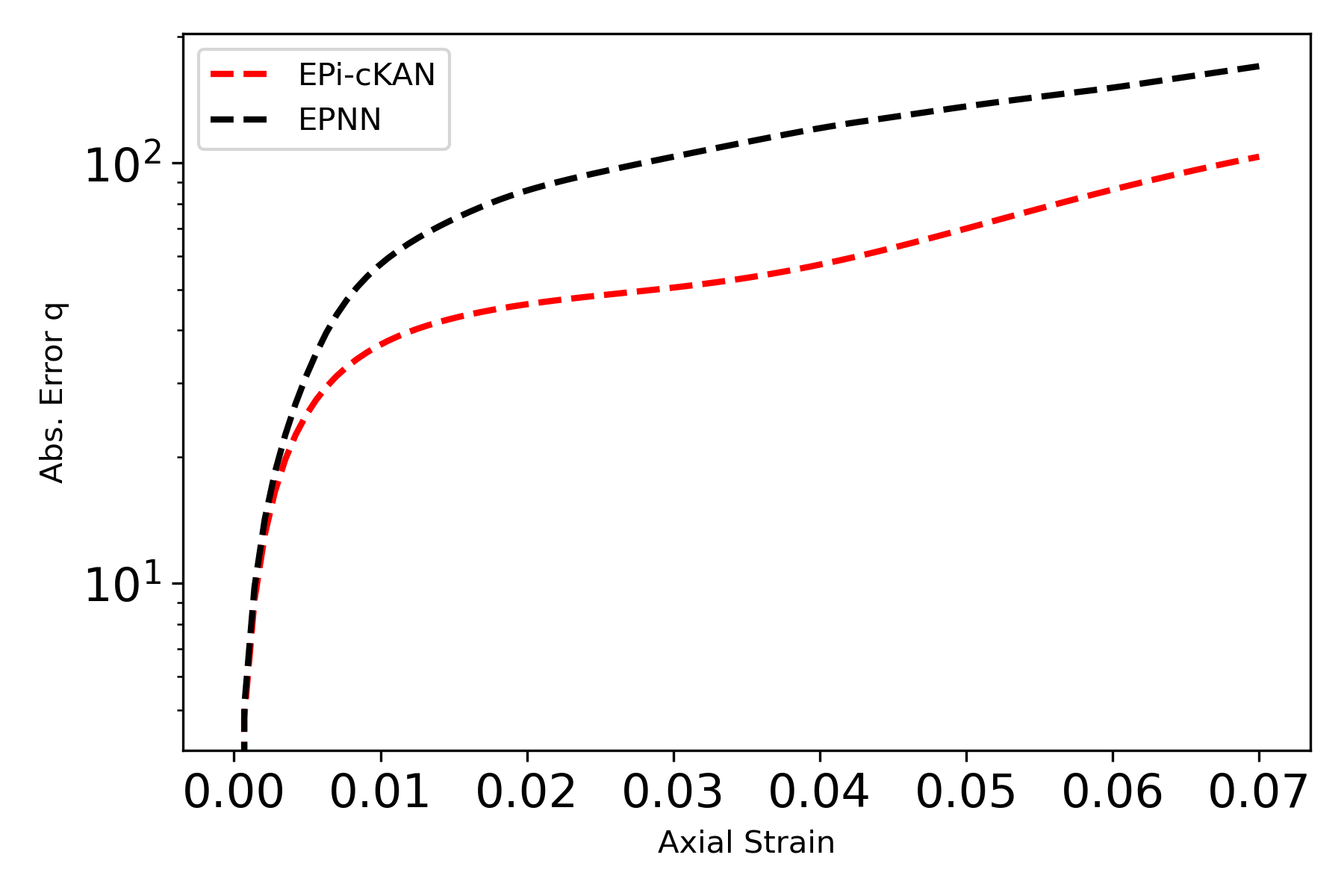}\hspace{.1cm}}

\caption{Comparison between the predictions of elasto-plastic based architectures (trained for \(2e4\) epochs) and the ground truth (numerical integration results), in the undrained test with \(\xi = -2.5\), \(p^{in}=375\) kPa and \(e^{in}=0.64\). Models used here are  trained on 31000 training data with the network structures reported in Table \ref{Tab.Settings}.} \label{Pic.Fig24_EP}
\end{figure}

\begin{figure}[!h]
\centering
\subfigure[\(\varepsilon^{v,p}\) vs \(\varepsilon_{33}\)]
{ \label{Pic.Fig25_cKAN_a}\includegraphics[width=.30\textwidth]{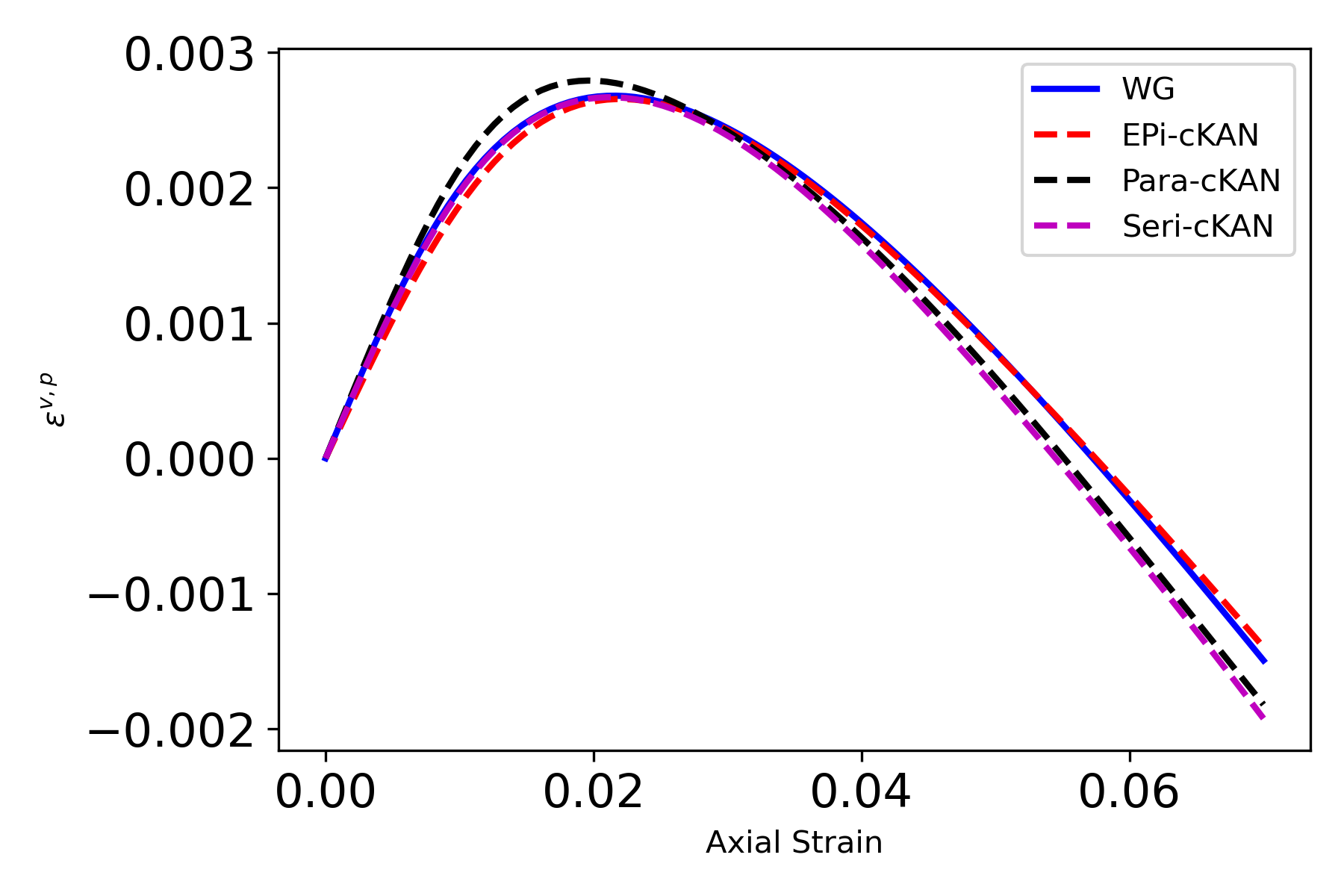}\hspace{.1cm}}
\subfigure[\(\gamma^{p}\) vs \(\varepsilon_{33}\)]
{ \label{Pic.Fig25_cKAN_b}\includegraphics[width=.30\textwidth]{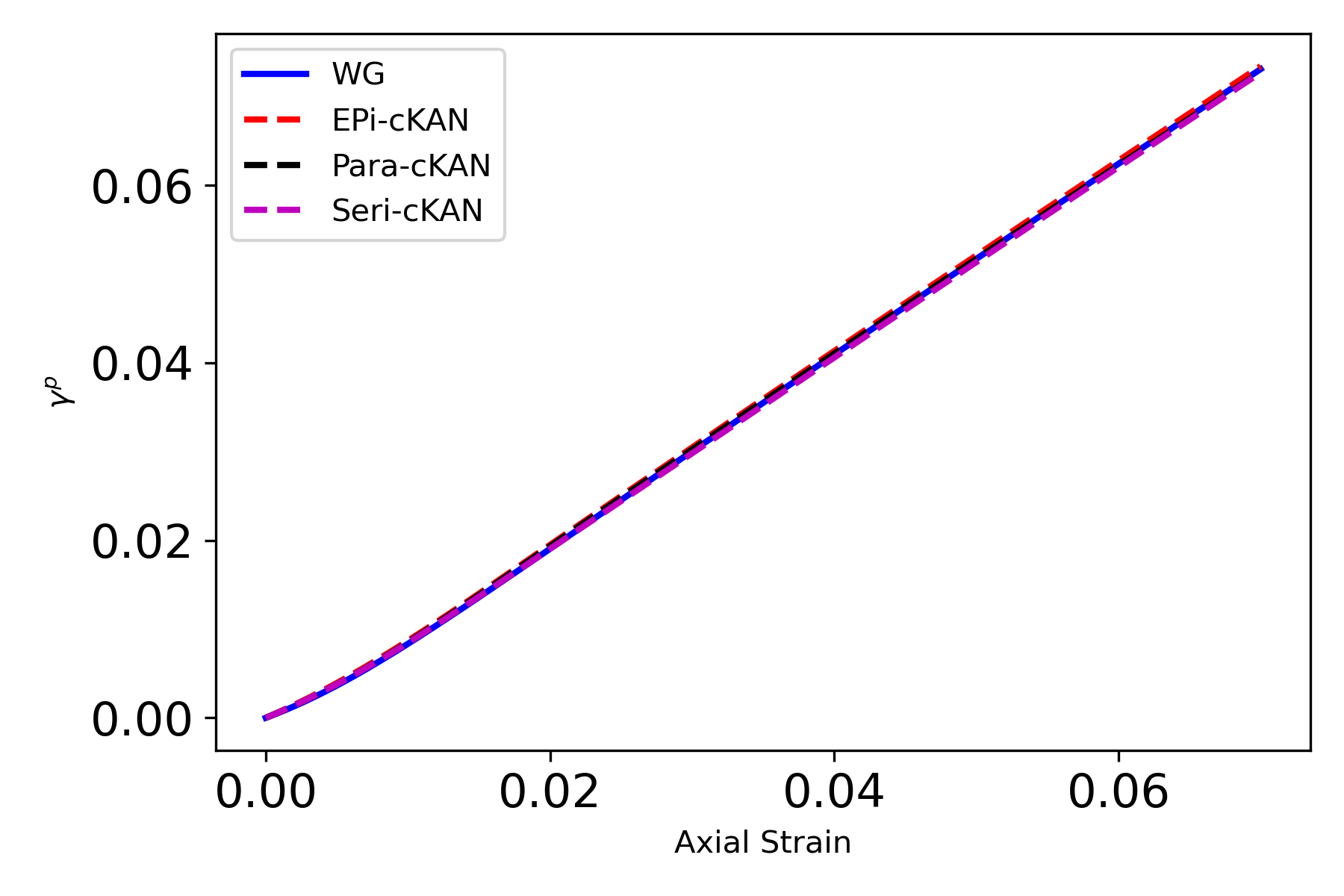}\hspace{.1cm}}
\subfigure[\(e\) vs \(\varepsilon_{33}\)]
{ \label{Pic.Fig25_cKAN_f}\includegraphics[width=.30\textwidth]{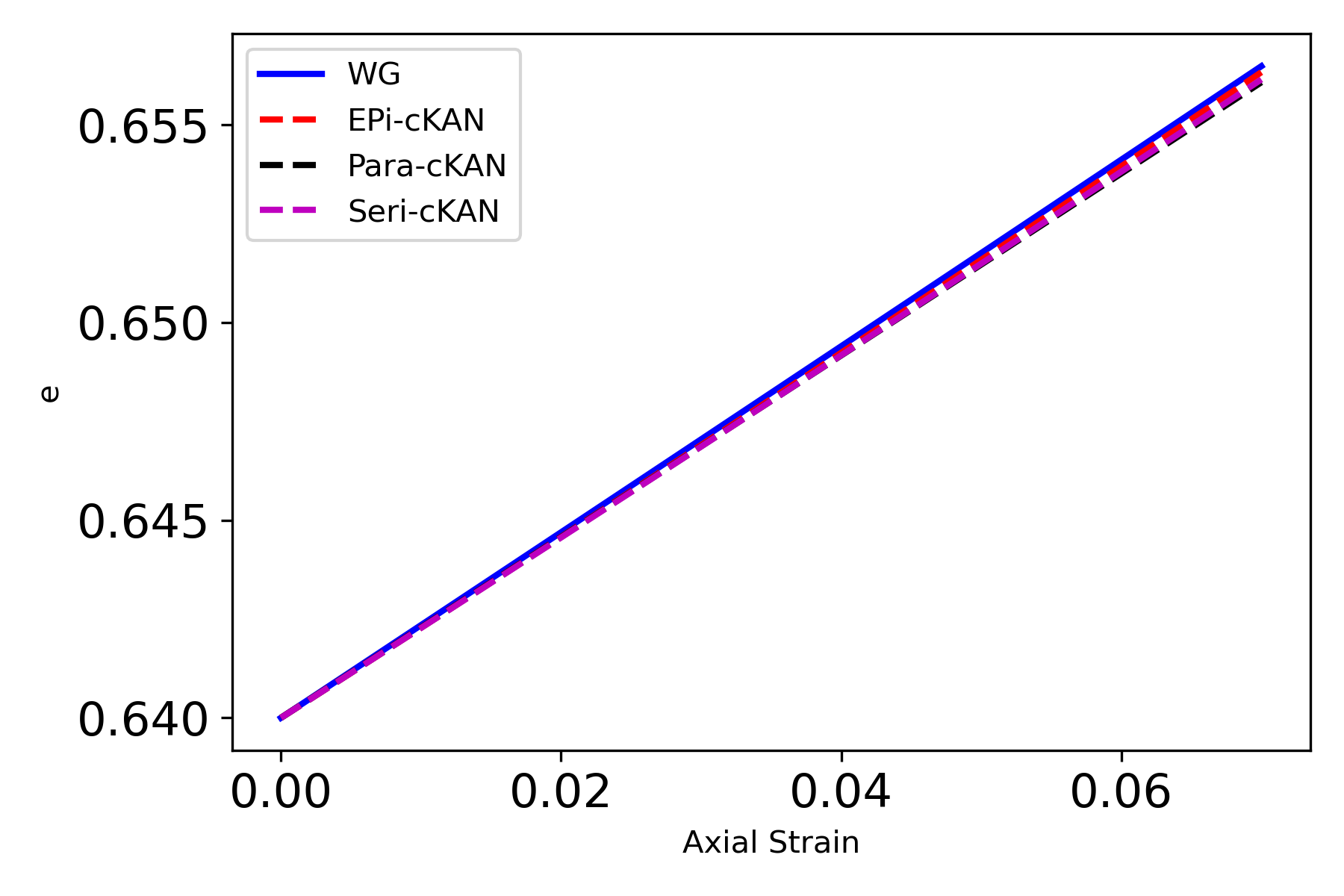}\hspace{.1cm}}

\subfigure[\(\vert\varepsilon^{v,p, *}-\varepsilon^{v,p}\vert\) vs \(\varepsilon_{33}\)]
{ \label{Pic.Fig25_cKAN_aERROR}\includegraphics[width=.30\textwidth]{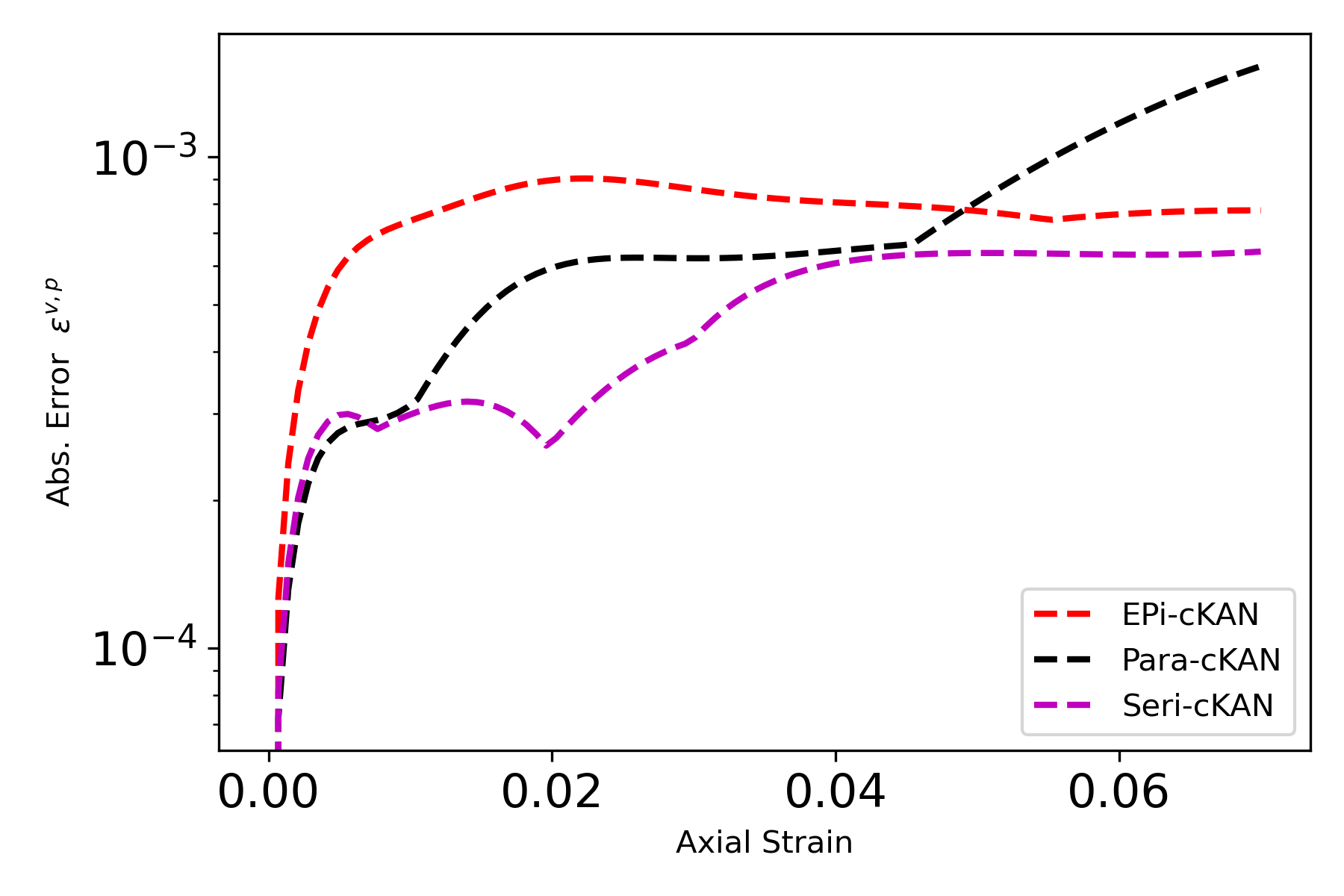}\hspace{.1cm}}
\subfigure[\(\vert\gamma^{p, *}-\gamma^{p}\vert\) vs \(\varepsilon_{33}\)]
{ \label{Pic.Fig25_cKAN_bERROR}\includegraphics[width=.30\textwidth]{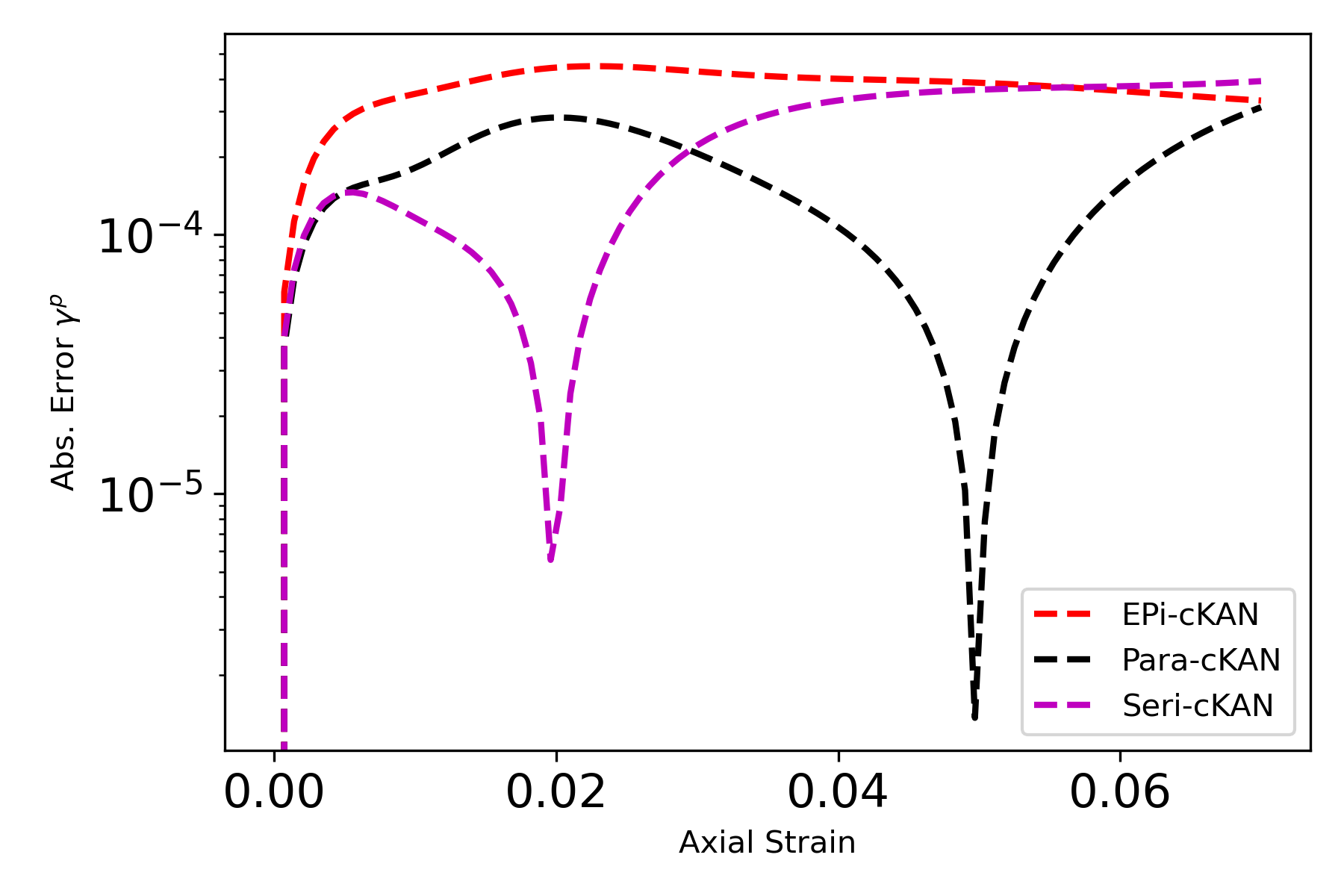}\hspace{.1cm}}
\subfigure[\(\vert e^{*}-e\vert\) vs \(\varepsilon_{33}\)]
{ \label{Pic.Fig25_cKAN_fERROR}\includegraphics[width=.30\textwidth]{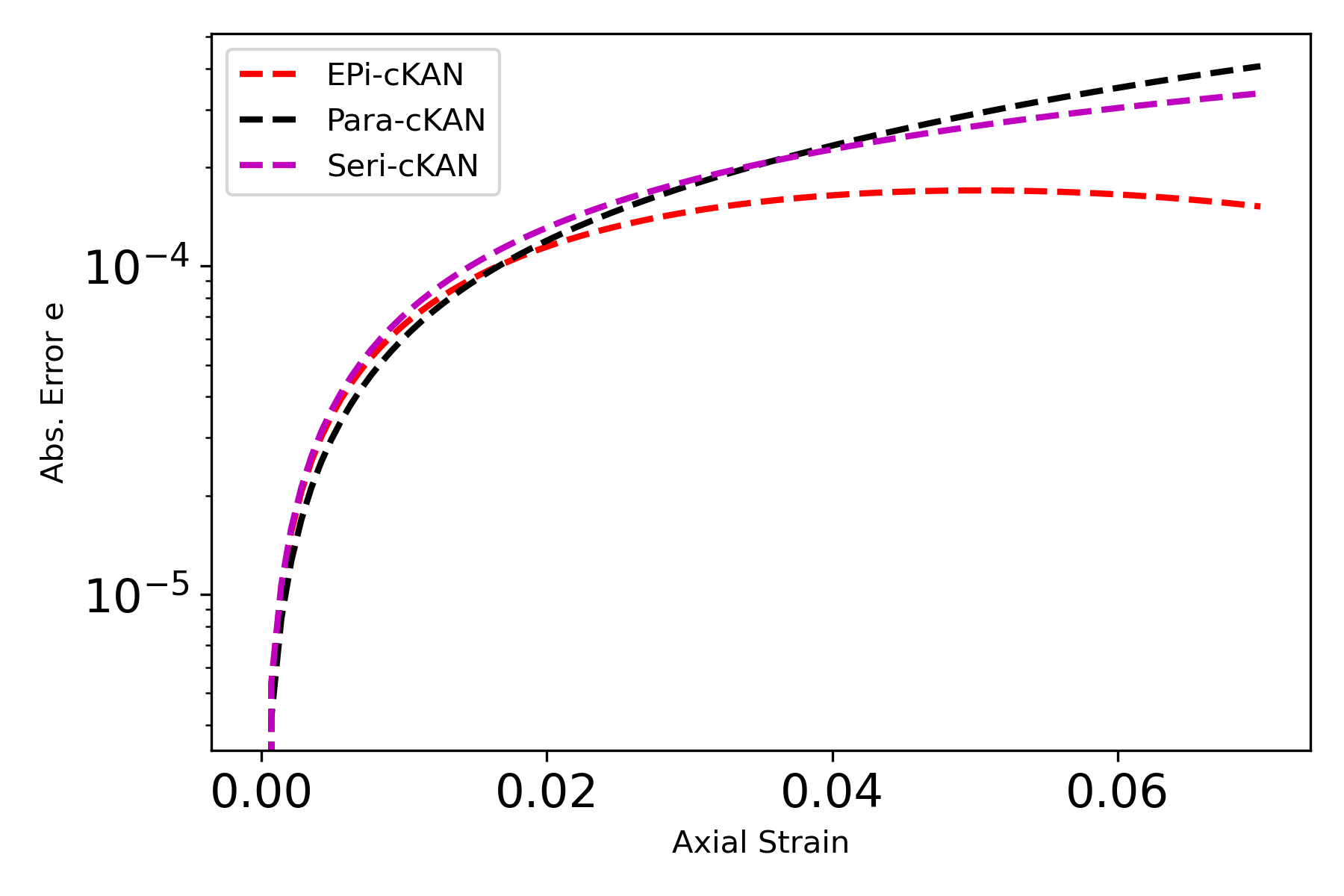}\hspace{.1cm}}

\subfigure[\(p\) (kPa) vs \(\varepsilon_{33}\)]
{ \label{Pic.Fig25_cKAN_c}\includegraphics[width=.30\textwidth]{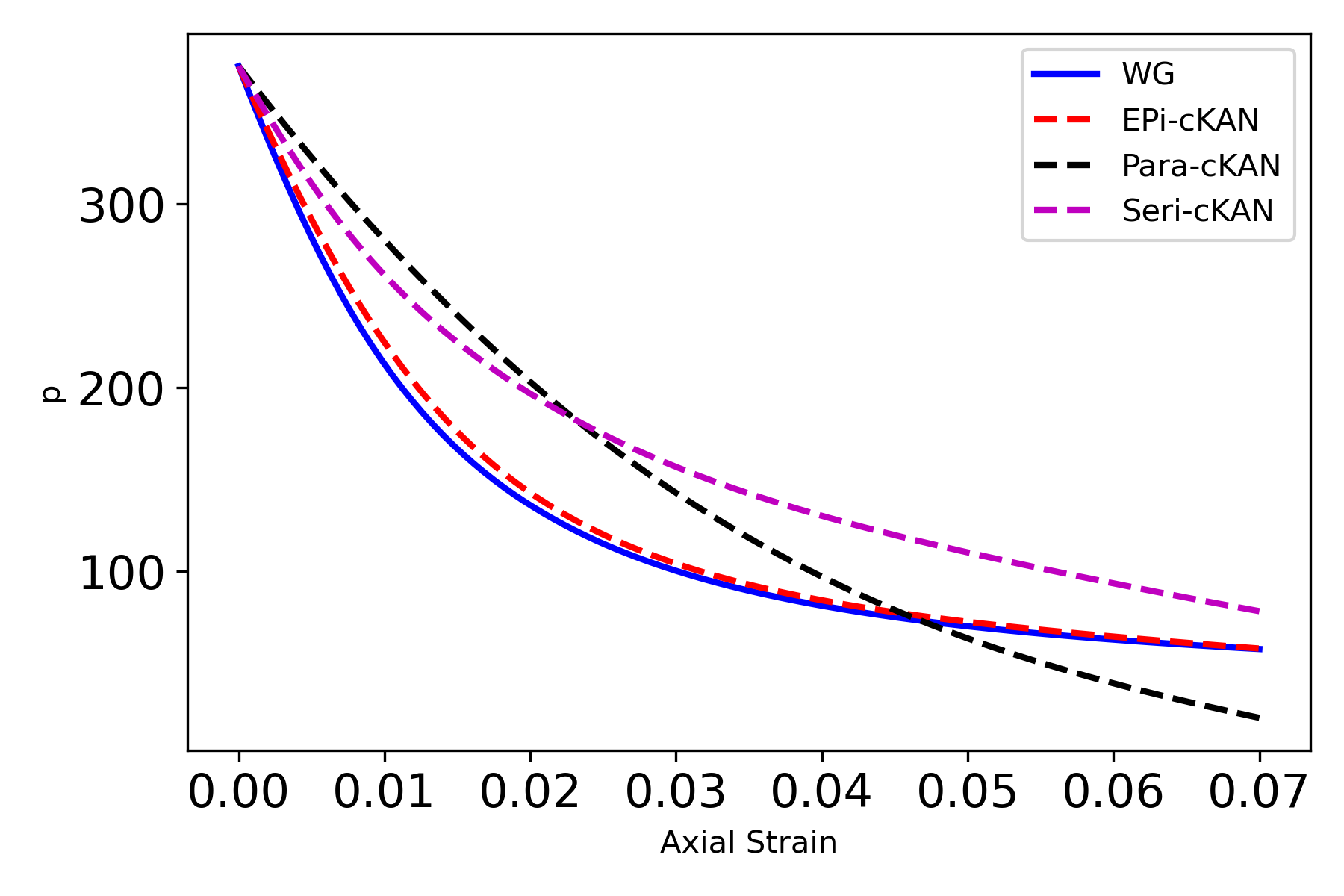}\hspace{.1cm}}
\subfigure[\(q\) (kPa) vs \(\varepsilon_{33}\)]
{ \label{Pic.Fig25_cKAN_d}\includegraphics[width=.30\textwidth]{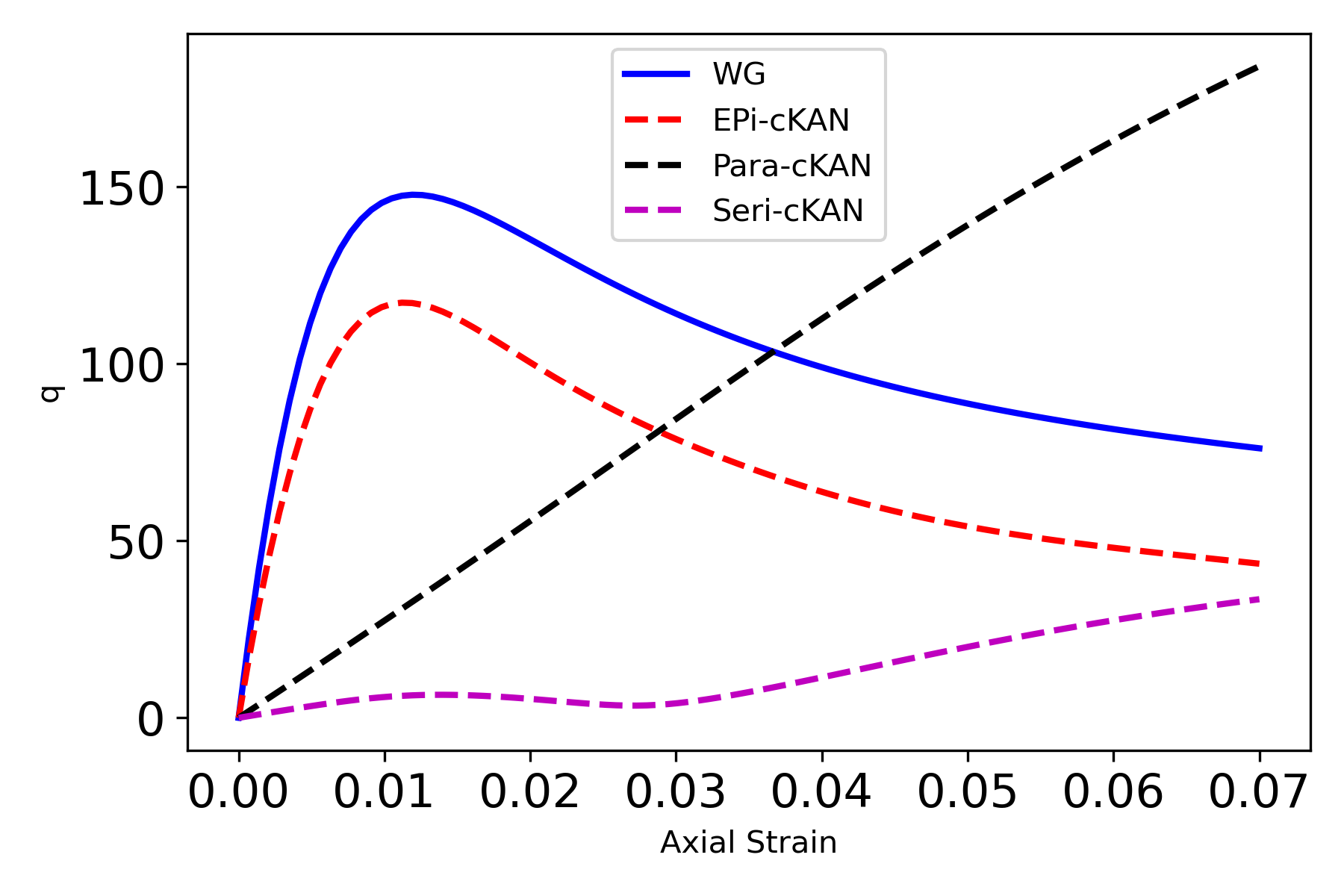}\hspace{.1cm}}
\subfigure[\(p\) (kPa) vs \(q\) (kPa)]
{ \label{Pic.Fig25_cKAN_e}\includegraphics[width=.30\textwidth]{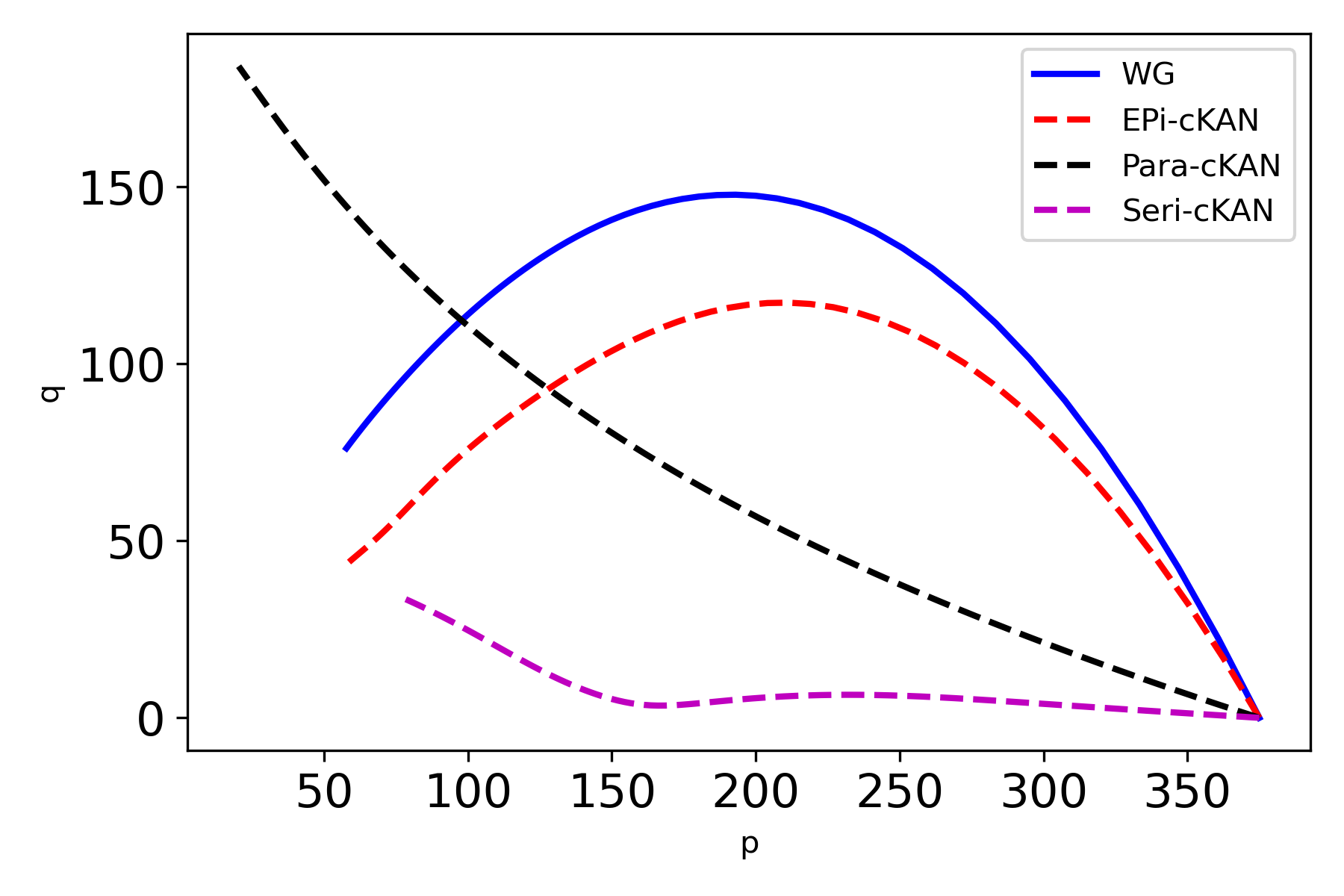}\hspace{.1cm}}

\subfigure[\(\vert p^{*}-p\vert\) vs \(\varepsilon_{33}\)]
{ \label{Pic.Fig25_cKAN_cERROR}\includegraphics[width=.30\textwidth]{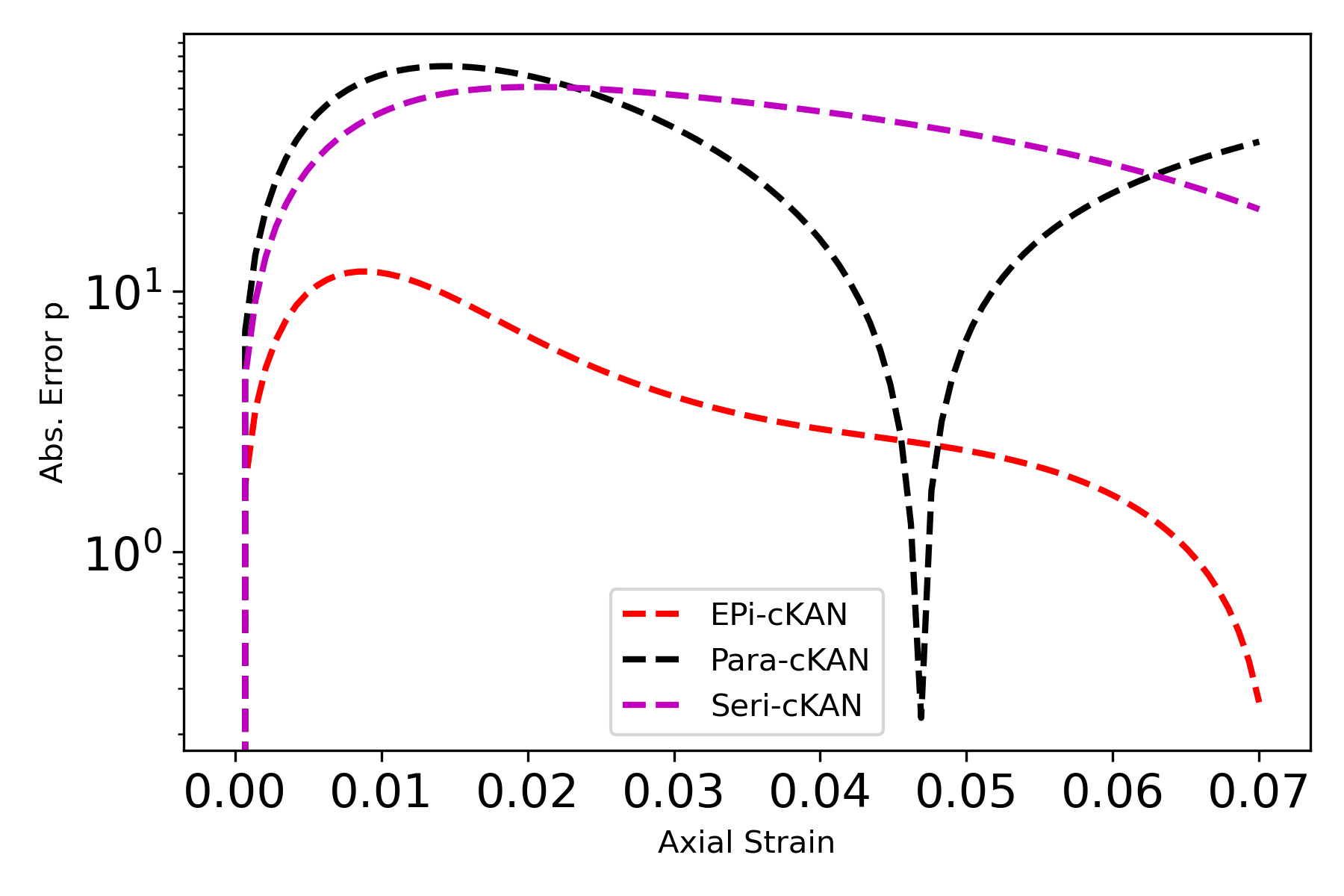}\hspace{.1cm}}
\subfigure[\(\vert q^{*}-q\vert\) vs \(\varepsilon_{33}\)]
{ \label{Pic.Fig25_cKAN_dERROR}\includegraphics[width=.30\textwidth]{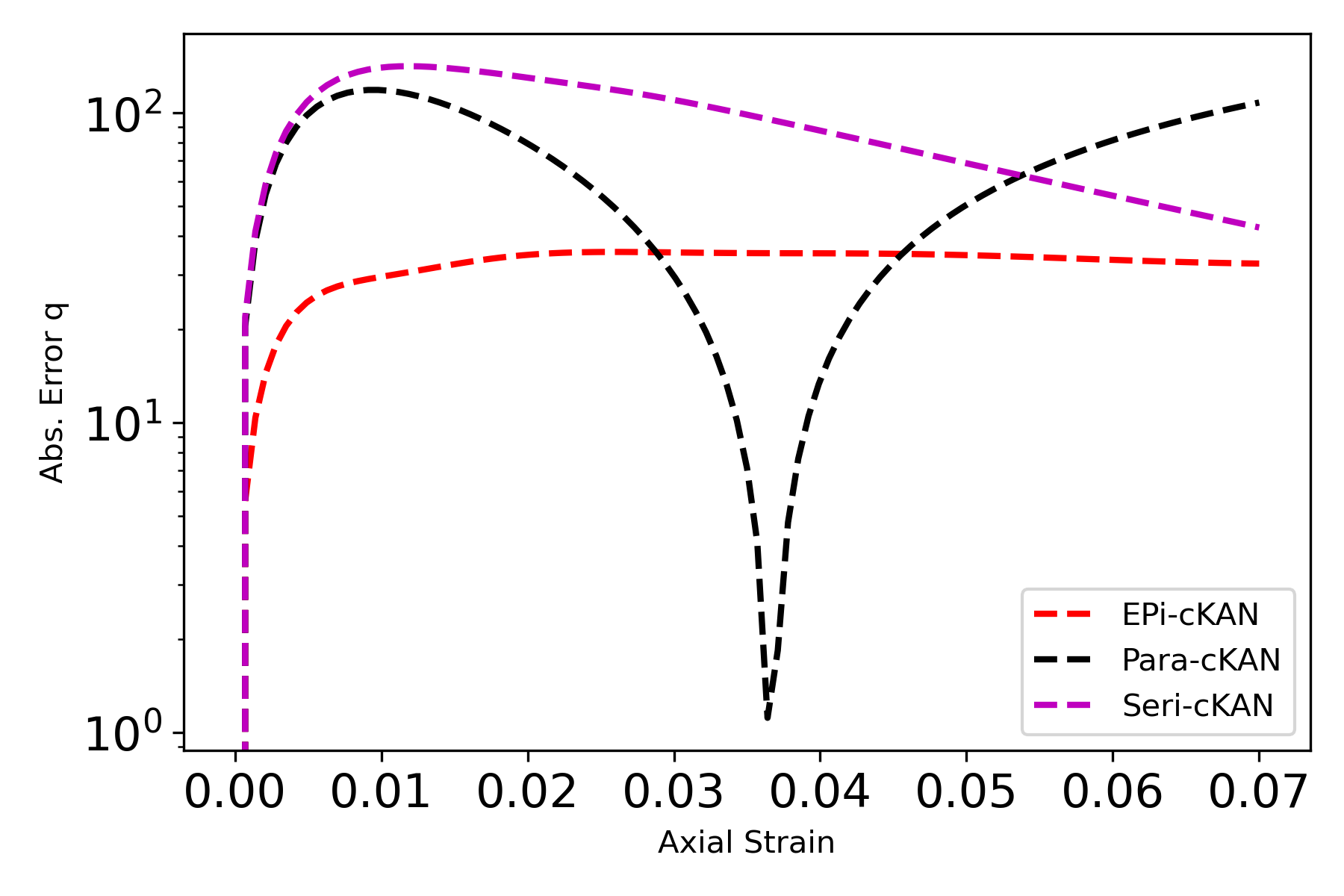}\hspace{.1cm}}

\caption{Comparison between the predictions of cKAN architectures (trained for \(2e4\) epochs) and the ground truth (numerical integration results), in the undrained test with \(\xi = -1.75\), \(p^{in}=375\) kPa and \(e^{in}=0.64\). Models used here are  trained on 31000 training data with the network structures reported in Table \ref{Tab.Settings}.} \label{Pic.Fig25_cKAN}
\end{figure}
\begin{figure}[!h]
\centering
\subfigure[\(\varepsilon^{v,p}\) vs \(\varepsilon_{33}\)]
{ \label{Pic.Fig25_EP_a}\includegraphics[width=.30\textwidth]{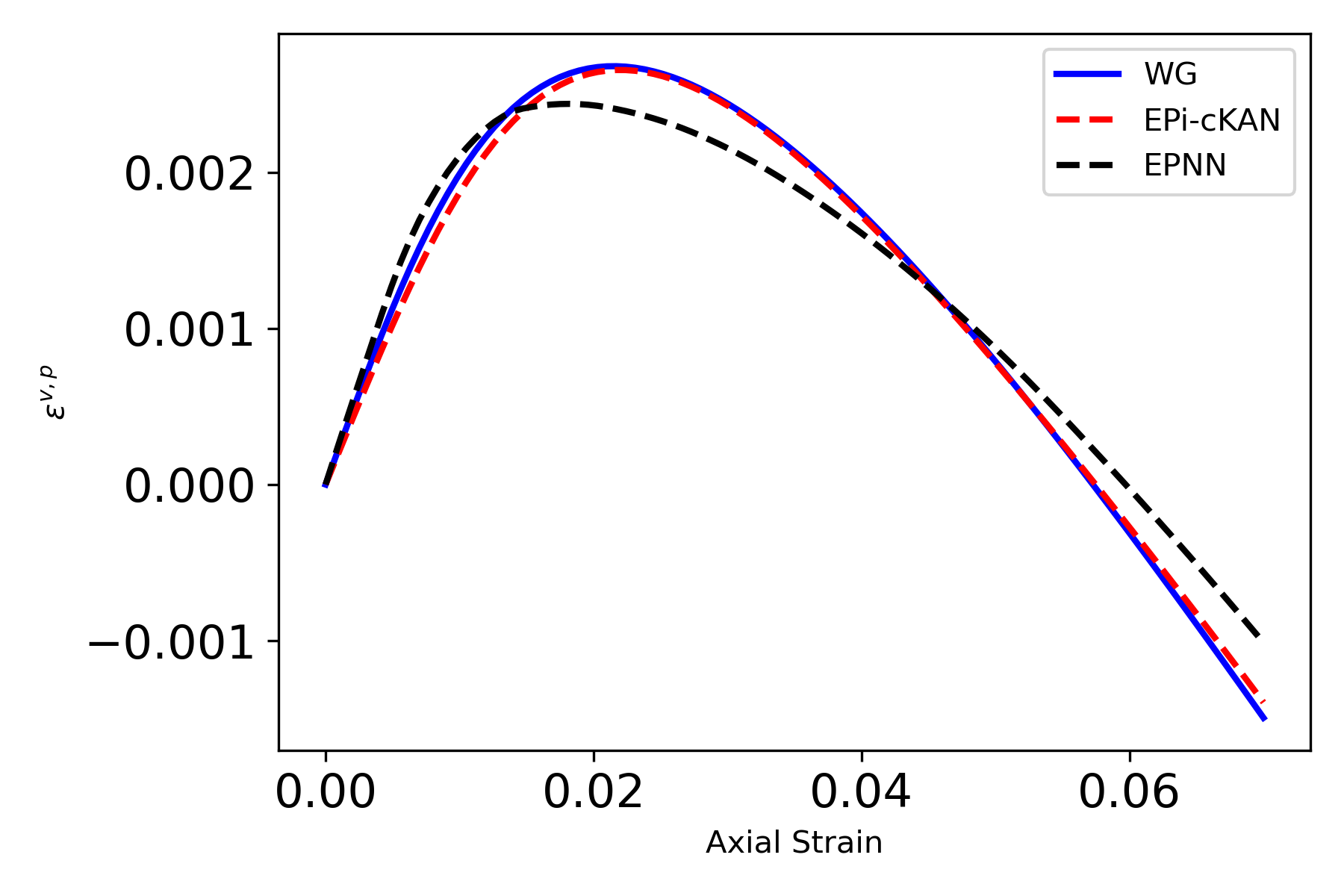}\hspace{.1cm}}
\subfigure[\(\gamma^{p}\) vs \(\varepsilon_{33}\)]
{ \label{Pic.Fig25_EP_b}\includegraphics[width=.30\textwidth]{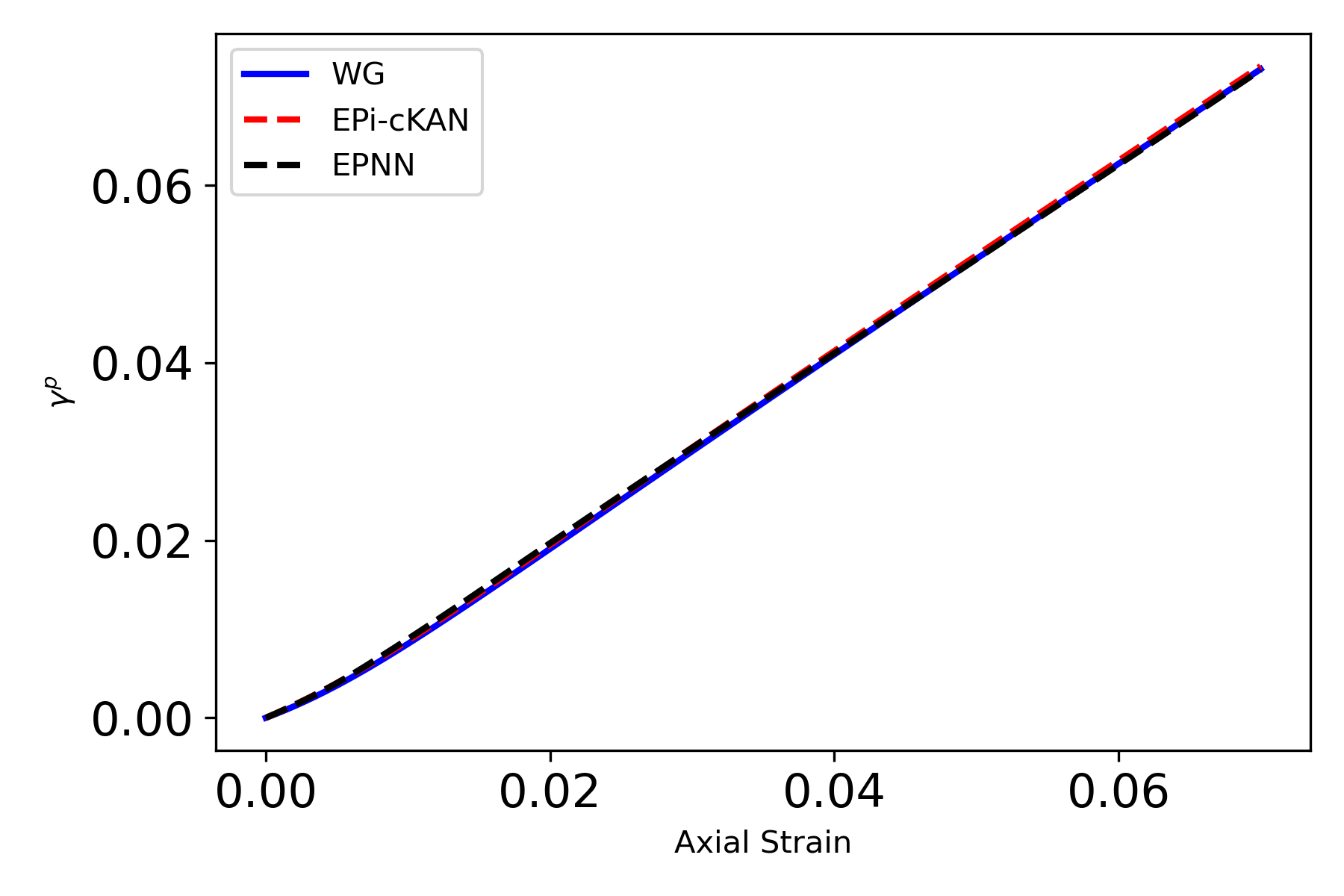}\hspace{.1cm}}
\subfigure[\(e\) vs \(\varepsilon_{33}\)]
{ \label{Pic.Fig25_EP_f}\includegraphics[width=.30\textwidth]{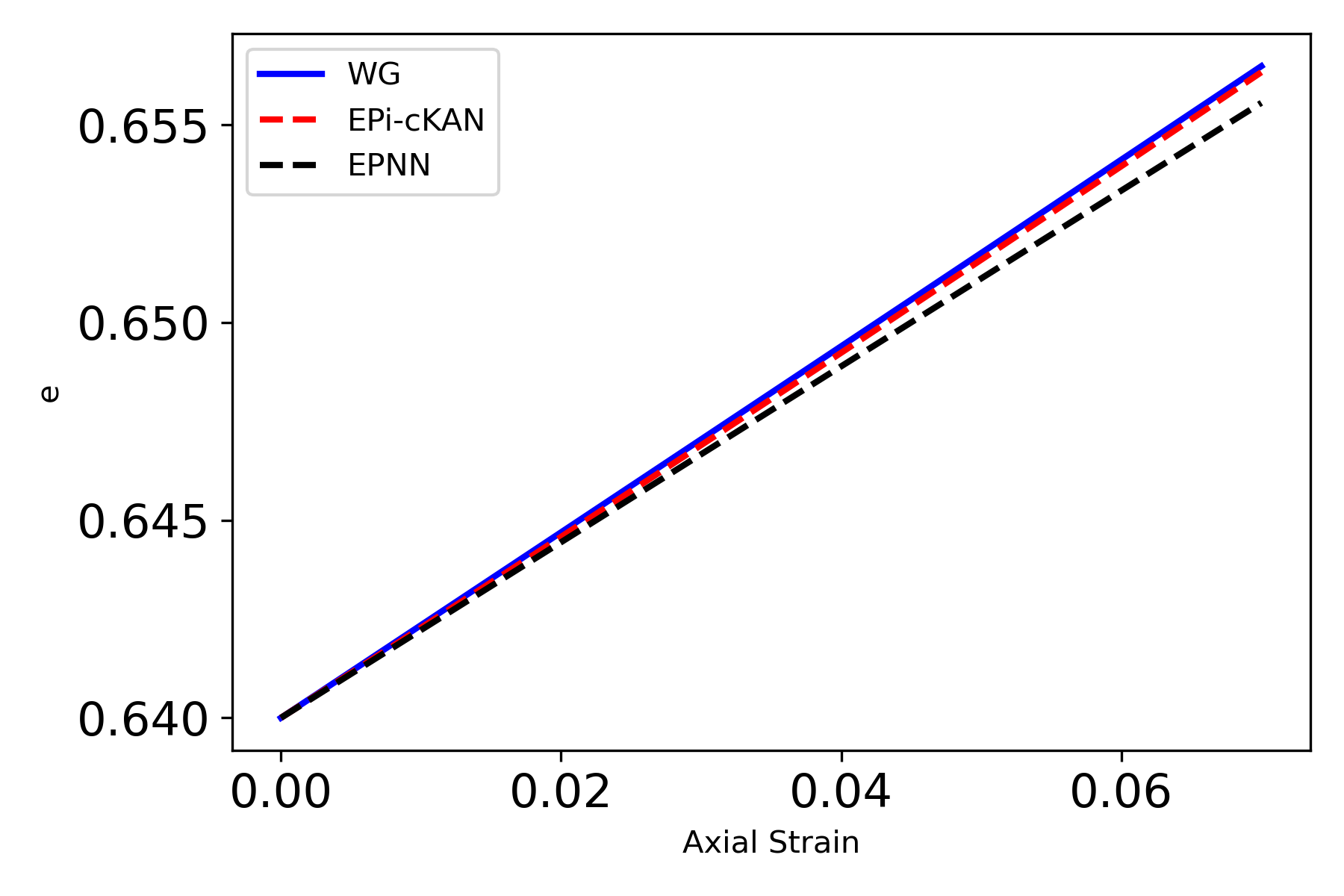}\hspace{.1cm}}

\subfigure[\(\vert\varepsilon^{v,p, *}-\varepsilon^{v,p}\vert\) vs \(\varepsilon_{33}\)]
{ \label{Pic.Fig25_EP_aERROR}\includegraphics[width=.30\textwidth]{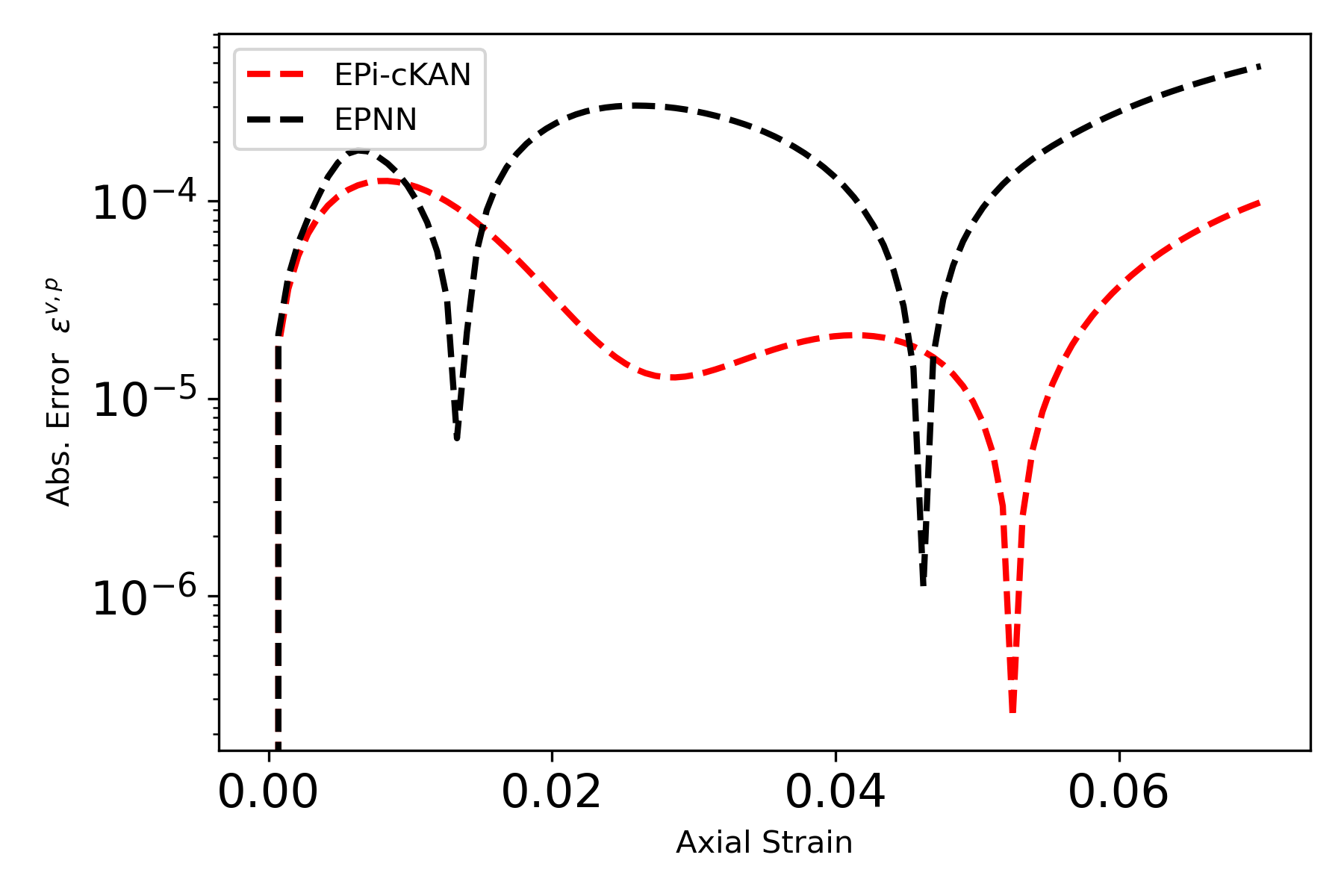}\hspace{.1cm}}
\subfigure[\(\vert\gamma^{p, *}-\gamma^{p}\vert\) vs \(\varepsilon_{33}\)]
{ \label{Pic.Fig25_EP_bERROR}\includegraphics[width=.30\textwidth]{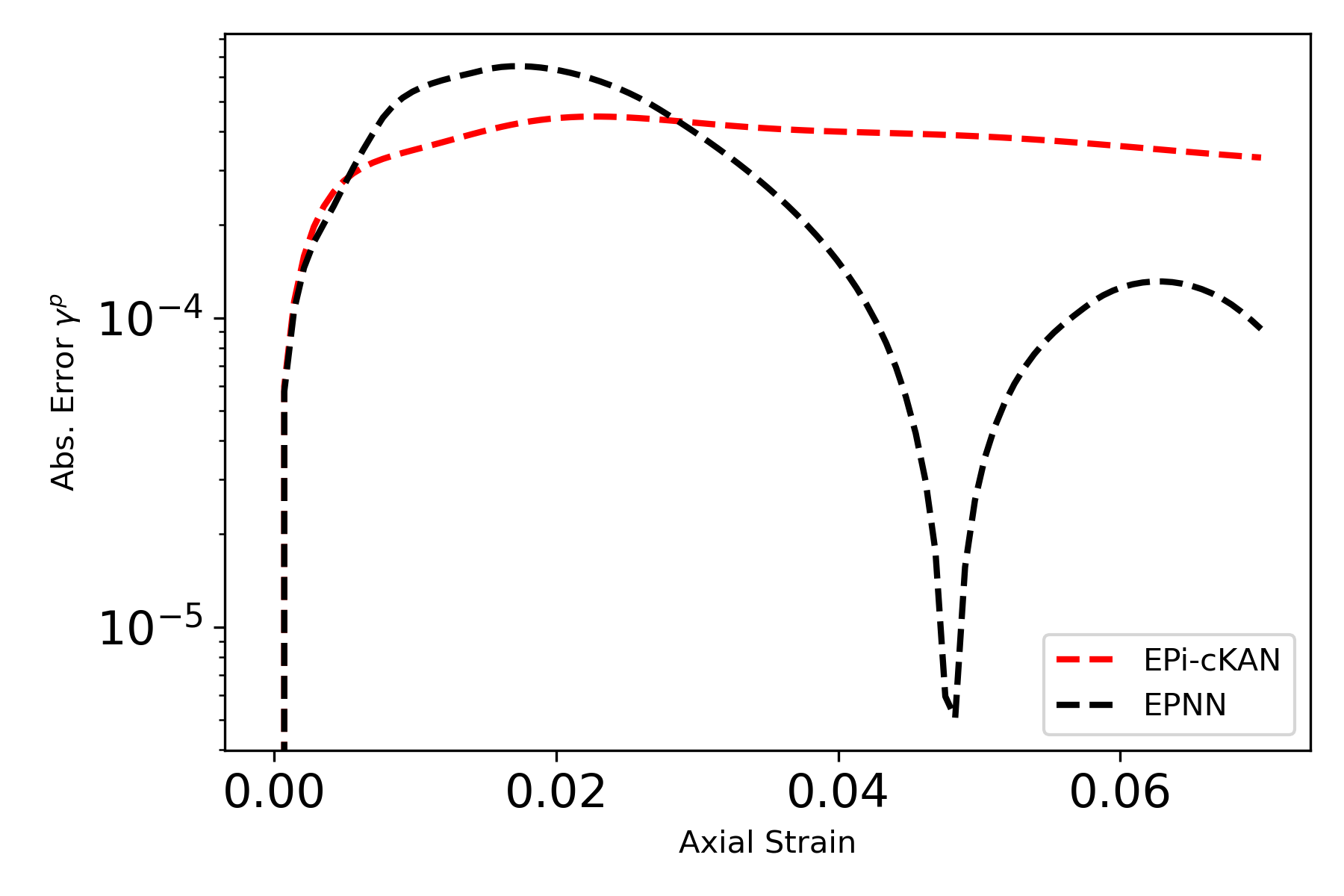}\hspace{.1cm}}
\subfigure[\(\vert e^{*}-e\vert\) vs \(\varepsilon_{33}\)]
{ \label{Pic.Fig25_EP_fERROR}\includegraphics[width=.30\textwidth]{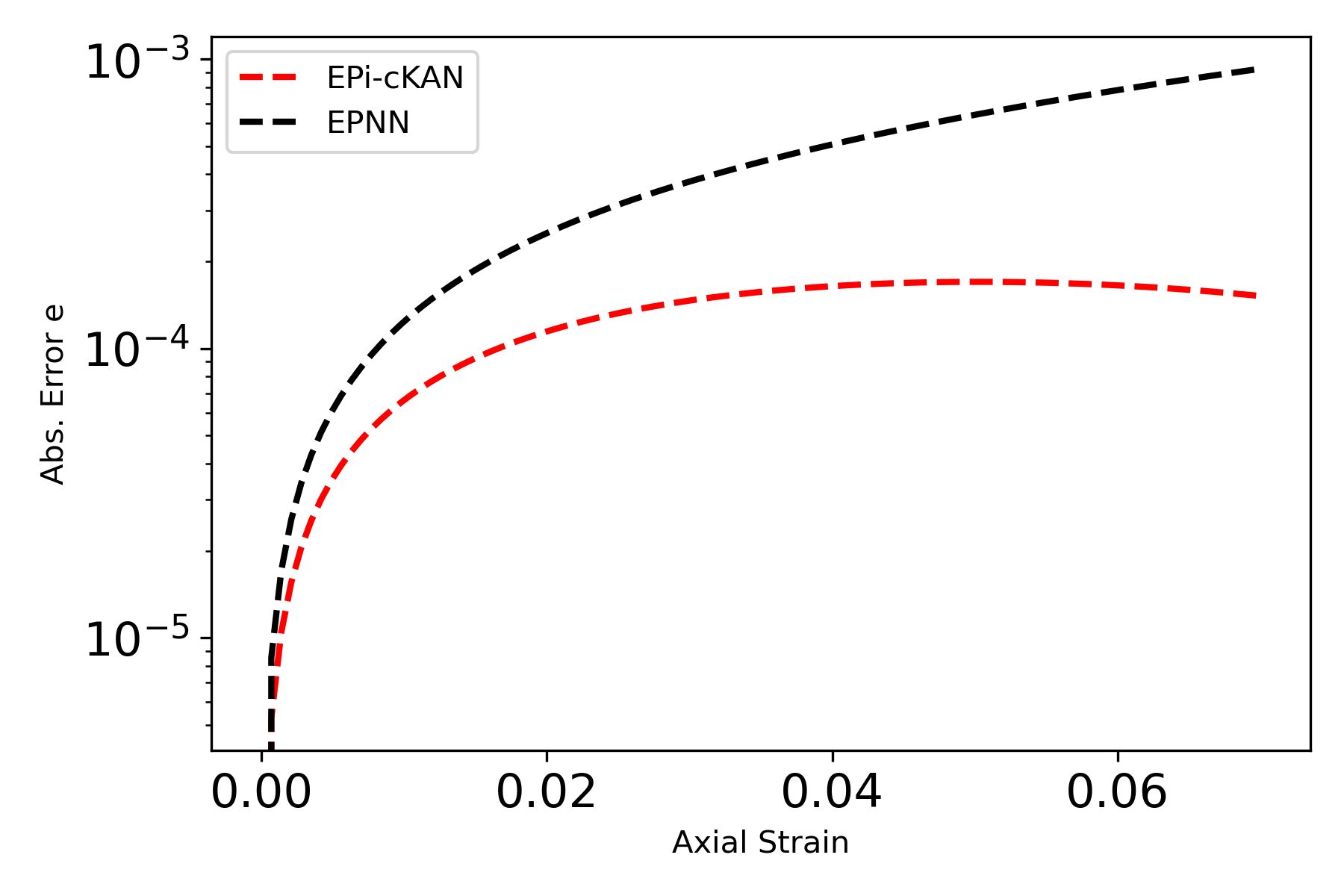}\hspace{.1cm}}

\subfigure[\(p\) (kPa) vs \(\varepsilon_{33}\)]
{ \label{Pic.Fig25_EP_c}\includegraphics[width=.30\textwidth]{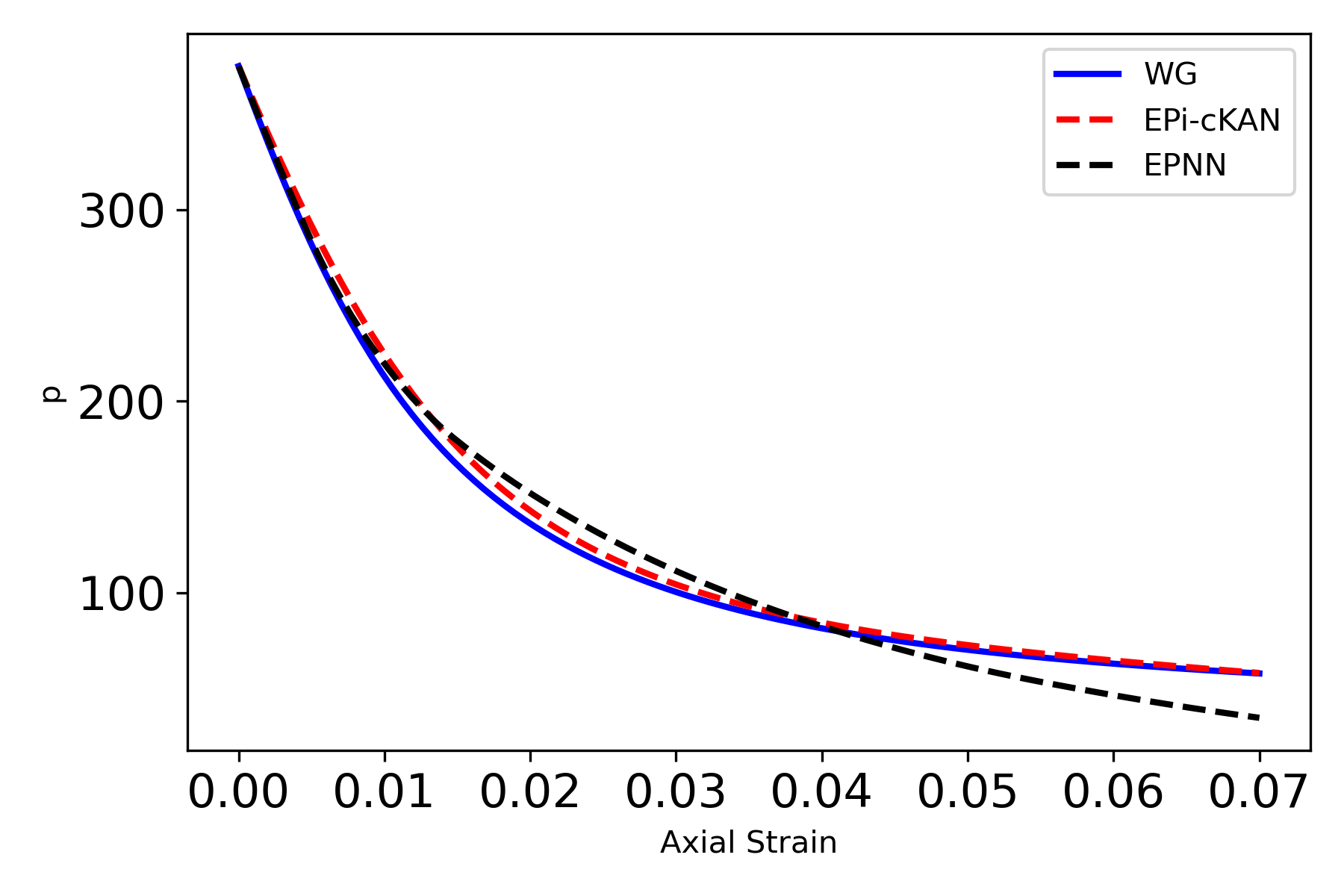}\hspace{.1cm}}
\subfigure[\(q\) (kPa) vs \(\varepsilon_{33}\)]
{ \label{Pic.Fig25_EP_d}\includegraphics[width=.30\textwidth]{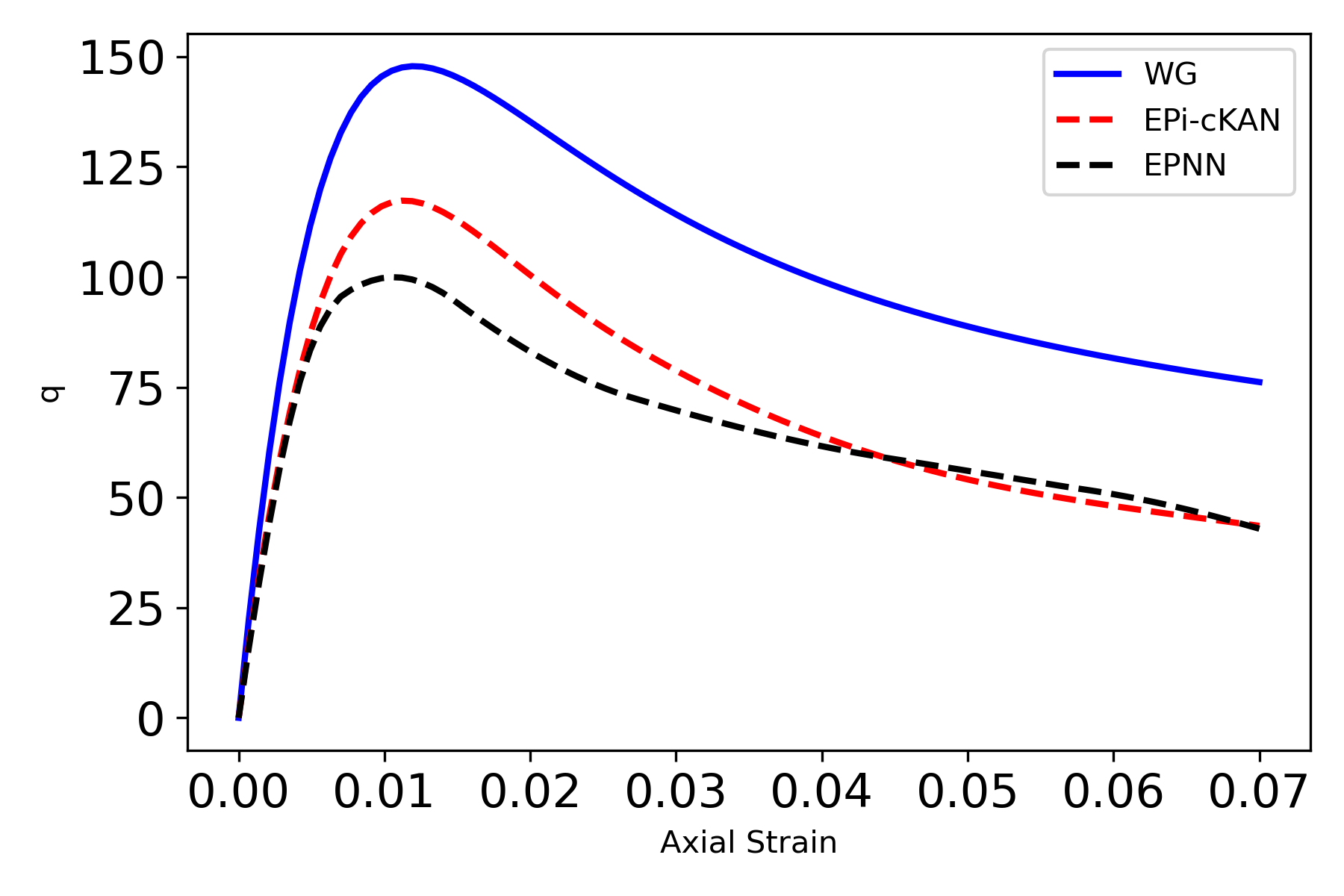}\hspace{.1cm}}
\subfigure[\(p\) (kPa) vs \(q\) (kPa)]
{ \label{Pic.Fig25_EP_e}\includegraphics[width=.30\textwidth]{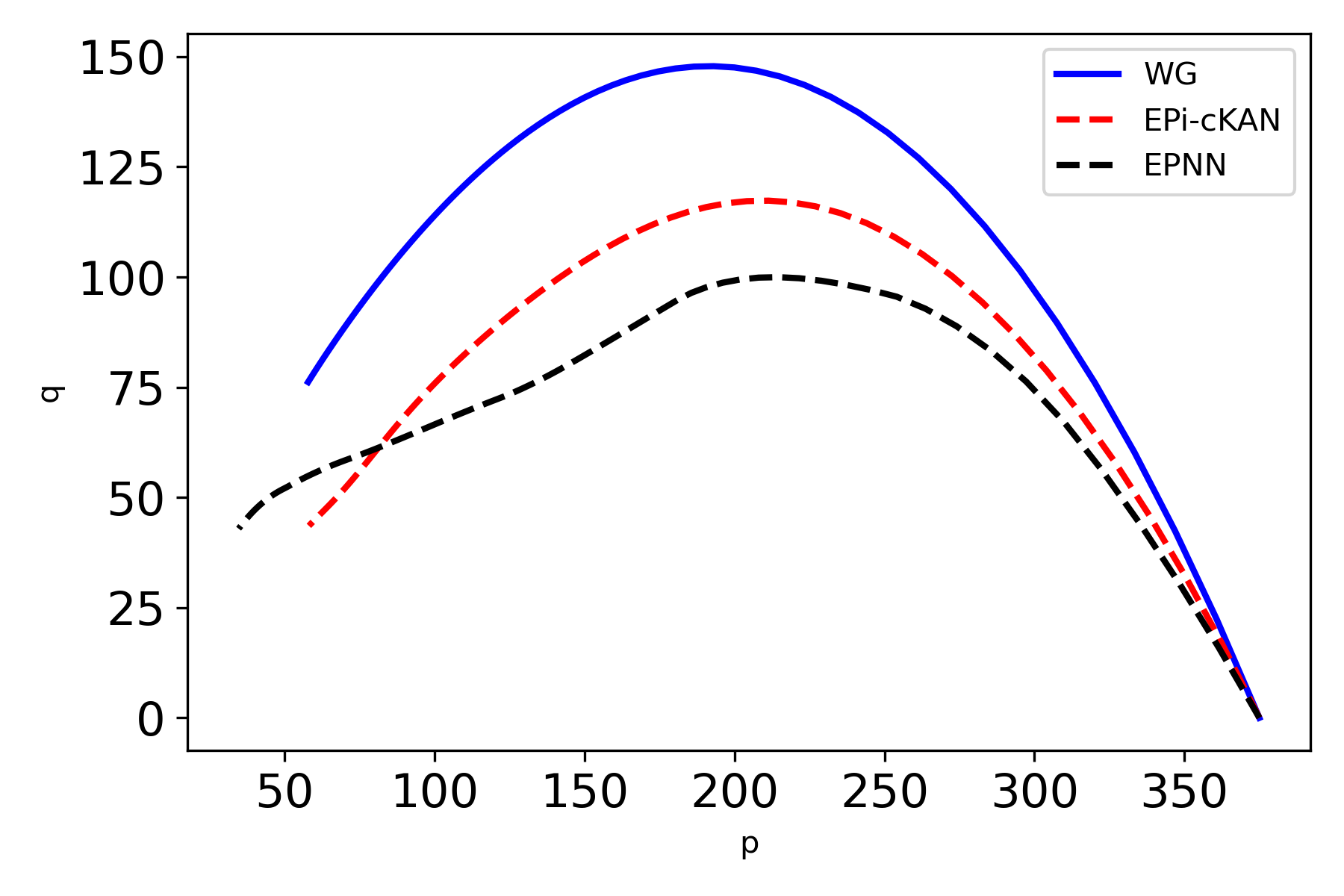}\hspace{.1cm}}

\subfigure[\(\vert p^{*}-p\vert\) vs \(\varepsilon_{33}\)]
{ \label{Pic.Fig25_EP_cERROR}\includegraphics[width=.30\textwidth]{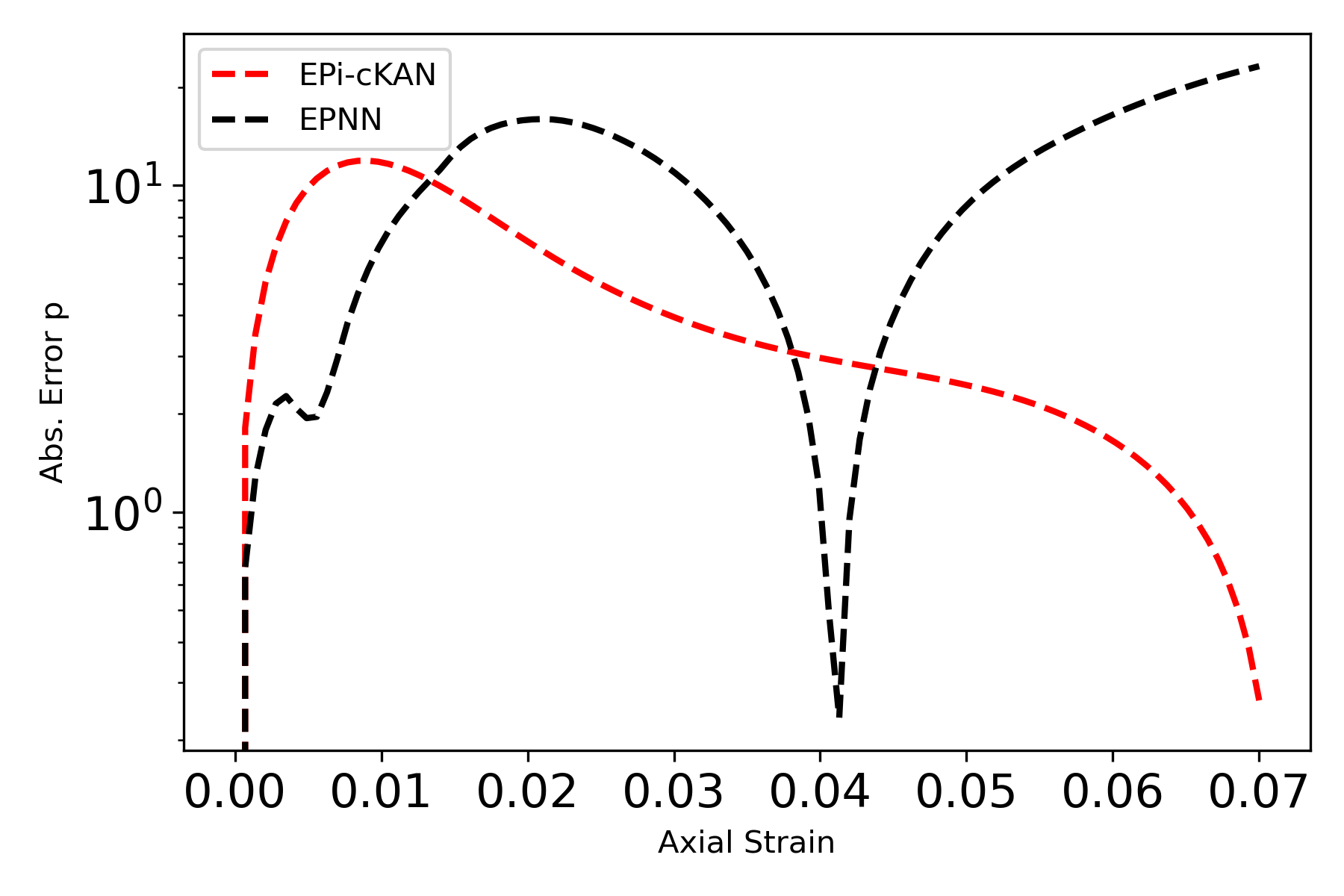}\hspace{.1cm}}
\subfigure[\(\vert q^{*}-q\vert\) vs \(\varepsilon_{33}\)]
{ \label{Pic.Fig25_EP_dERROR}\includegraphics[width=.30\textwidth]{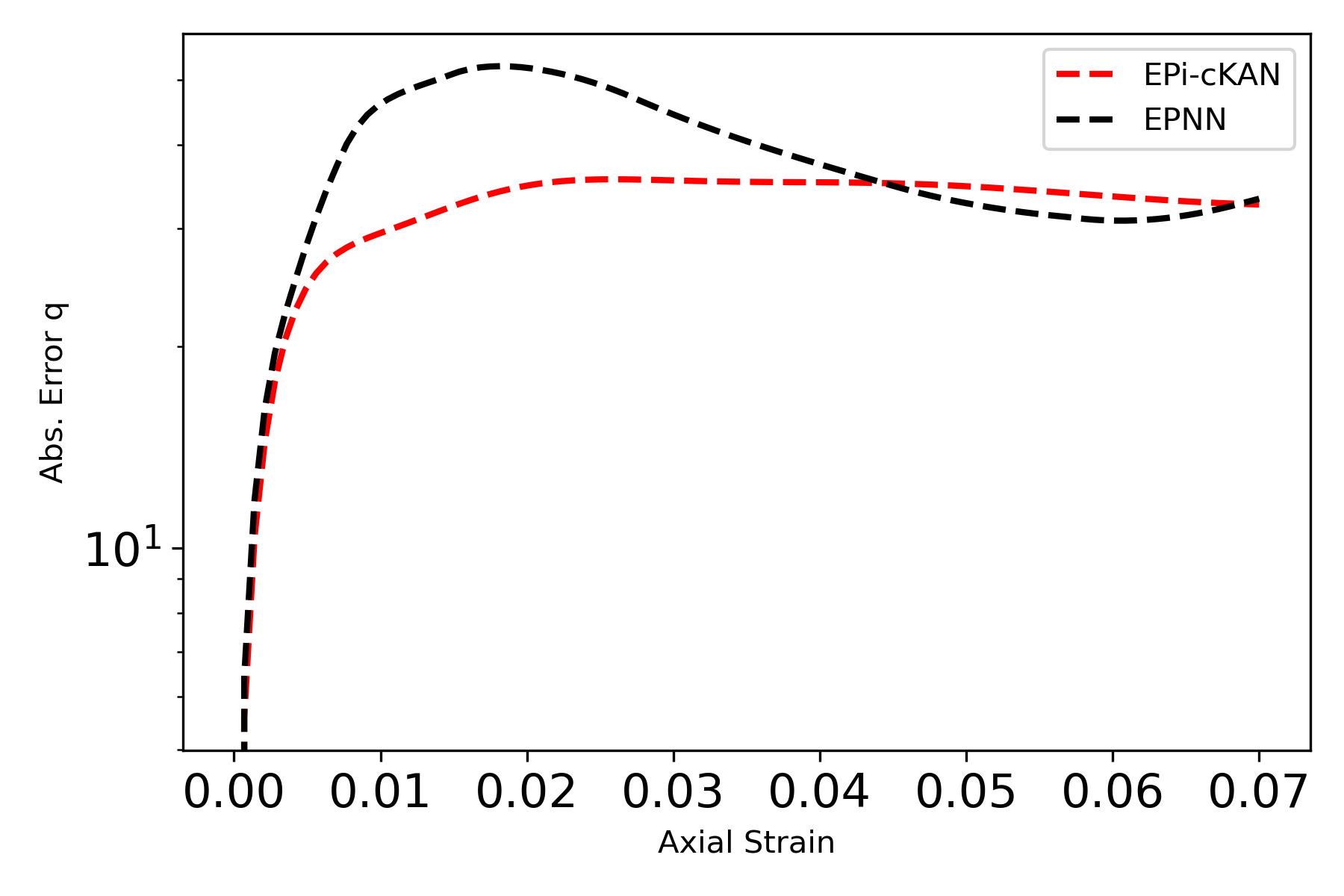}\hspace{.1cm}}

\caption{Comparison between the predictions of elasto-plastic based architectures (trained for \(2e4\) epochs) and the ground truth (numerical integration results), in the undrained test with \(\xi = -1.75\), \(p^{in}=375\) kPa and \(e^{in}=0.64\). Models used here are  trained on 31000 training data with the network structures reported in Table \ref{Tab.Settings}.} \label{Pic.Fig25_EP}
\end{figure}
\begin{figure}[!h]
\centering
\subfigure[\(\xi = -1.75\)]
{ \label{Pic.Size175}\includegraphics[width=.30\textwidth]{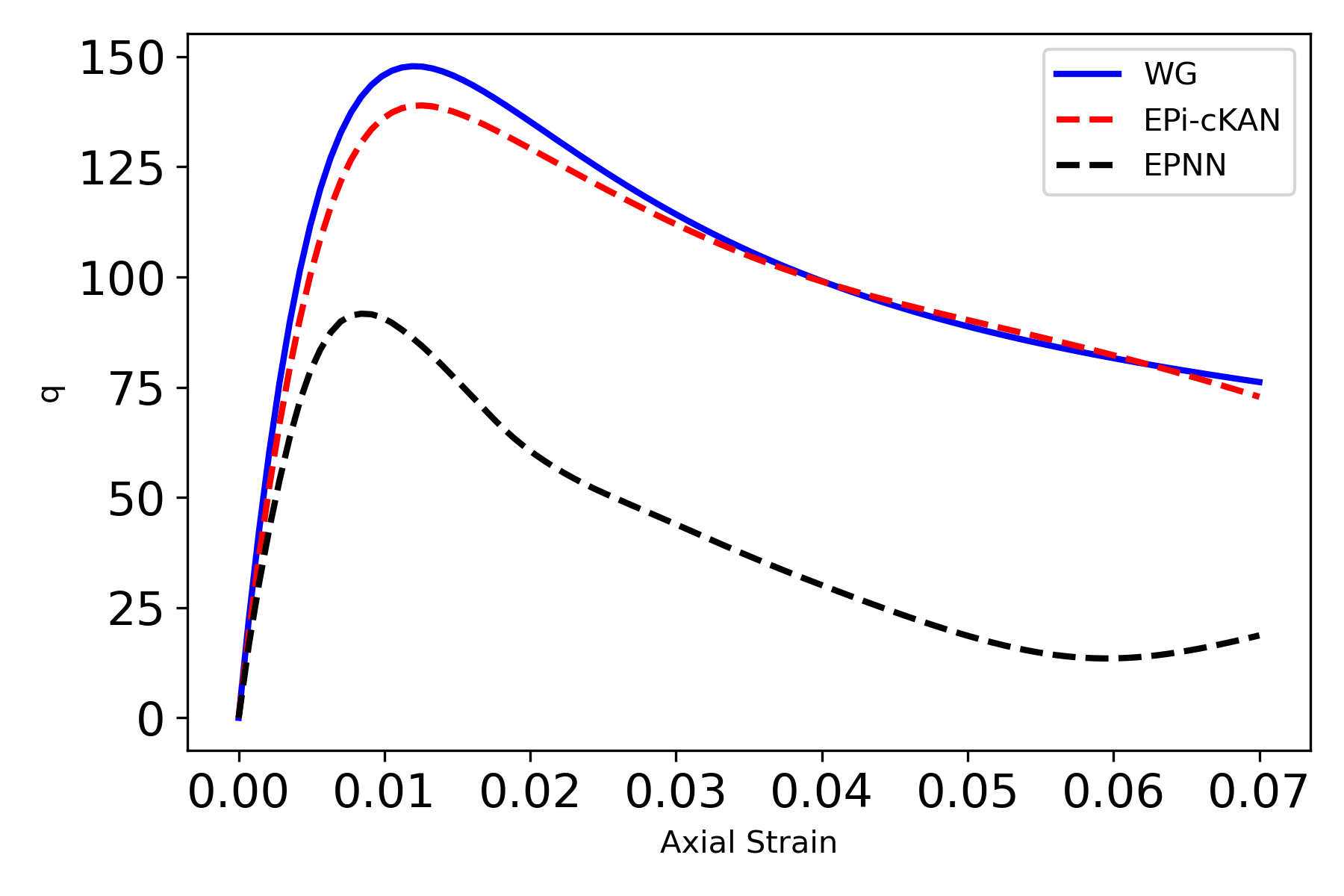}\hspace{.1cm}}
\subfigure[\(\xi = -2.0\)]
{ \label{Pic.Size20}\includegraphics[width=.30\textwidth]{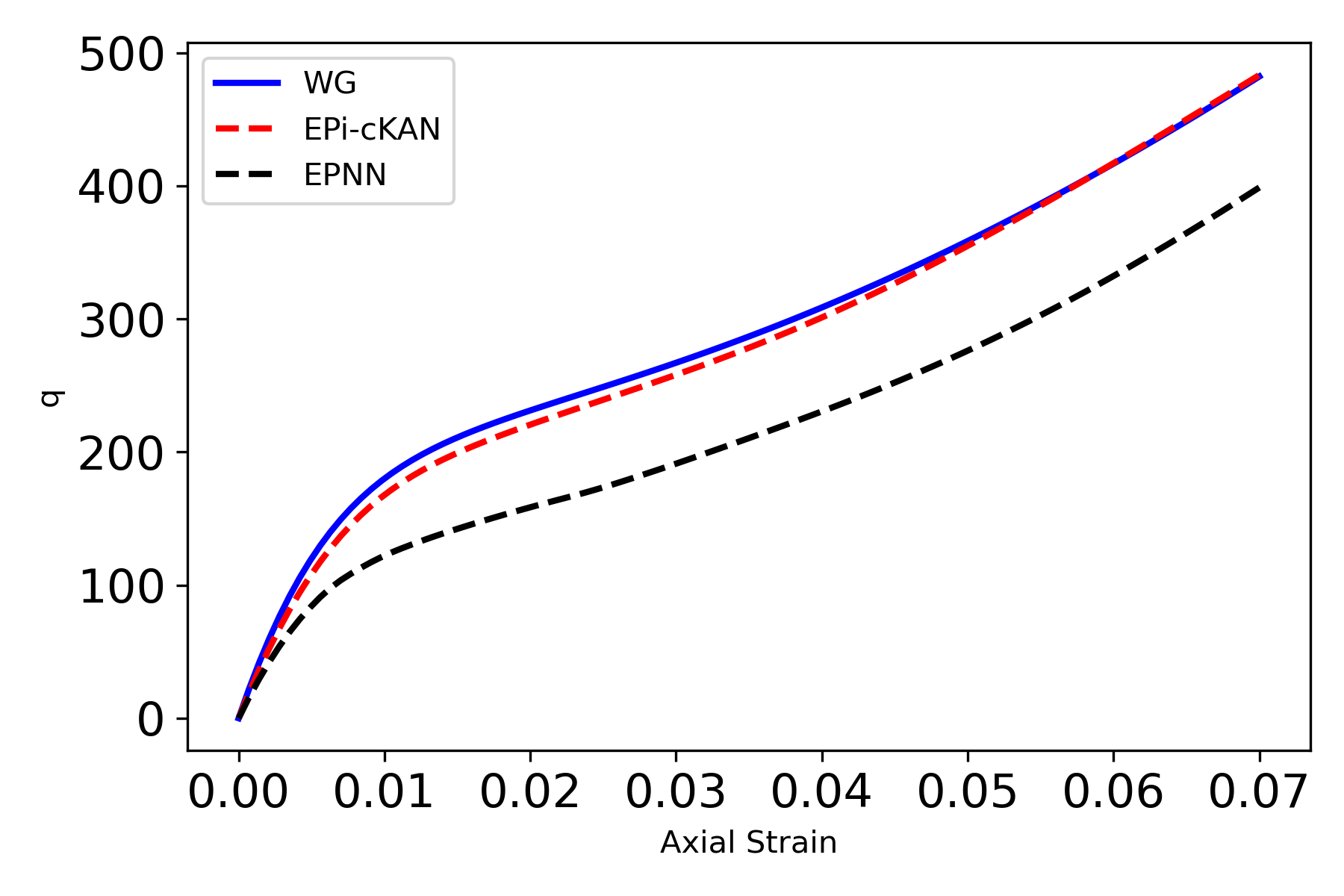}\hspace{.1cm}}
\subfigure[\(\xi = -2.5\)]
{ \label{Pic.Size25}\includegraphics[width=.30\textwidth]{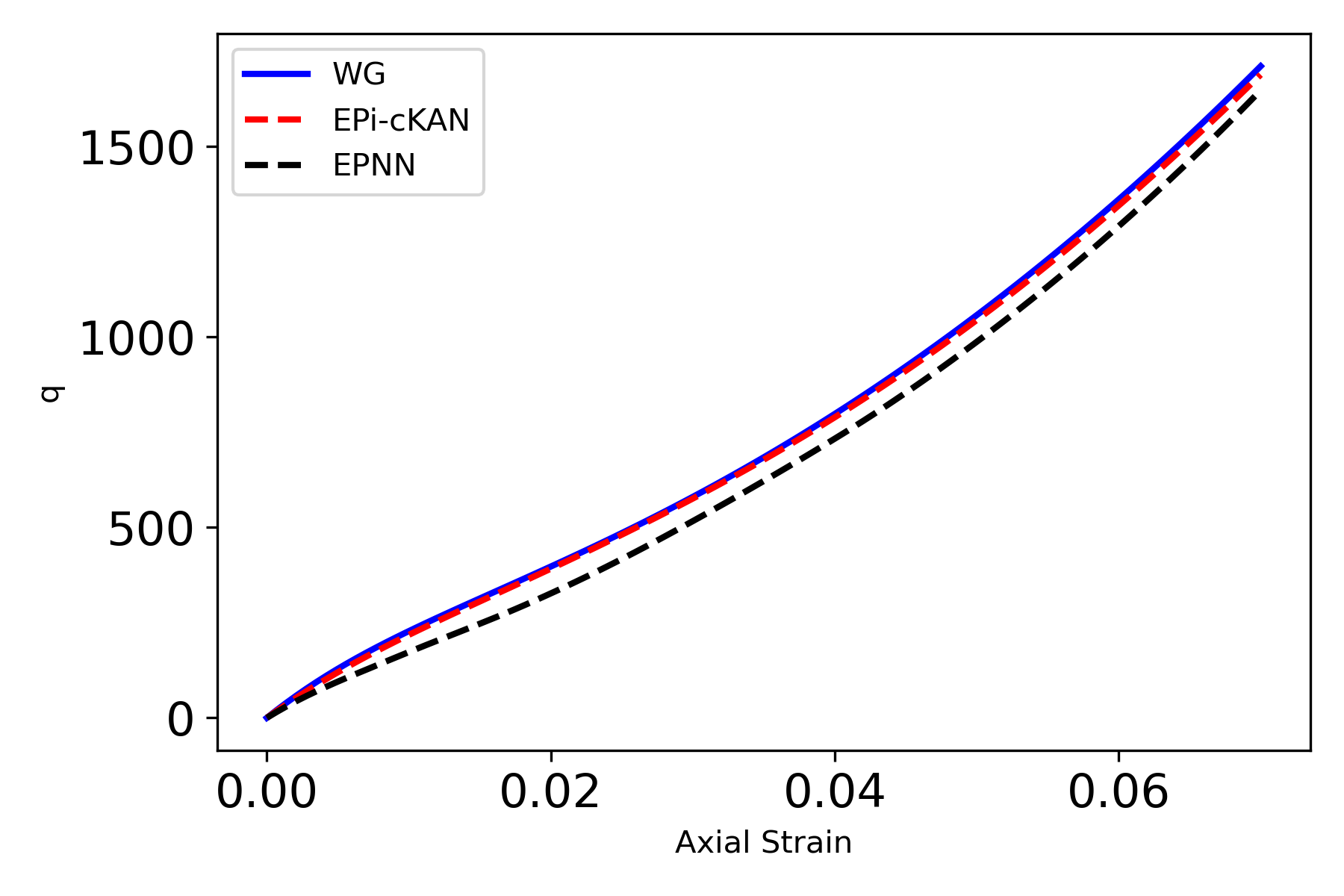}\hspace{.1cm}}

\subfigure[\(\xi = -1.75\)]
{ \label{Pic.Size175ERROR}\includegraphics[width=.30\textwidth]{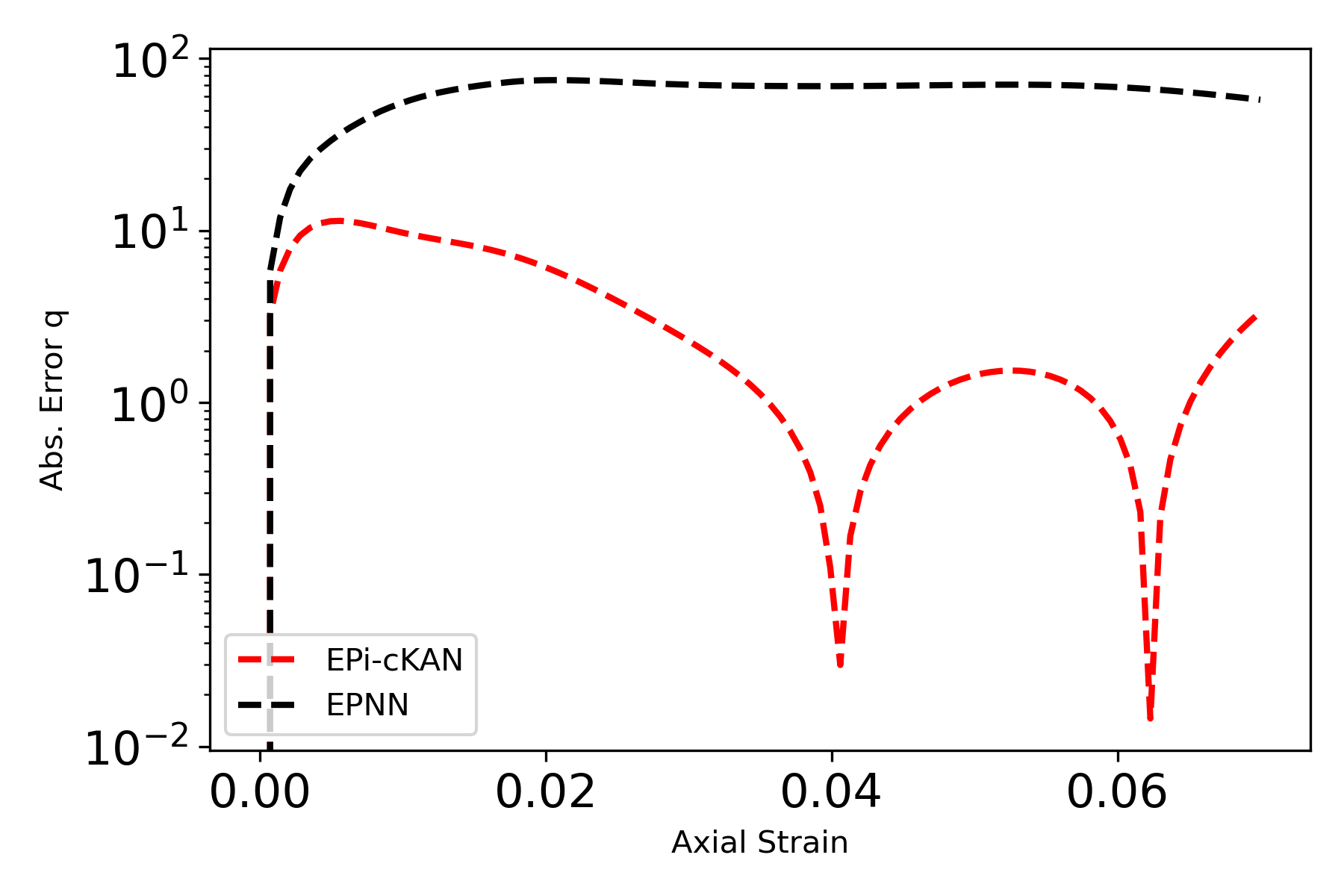}\hspace{.1cm}}
\subfigure[\(\xi = -2.0\)]
{ \label{Pic.Size2ERROR}\includegraphics[width=.30\textwidth]{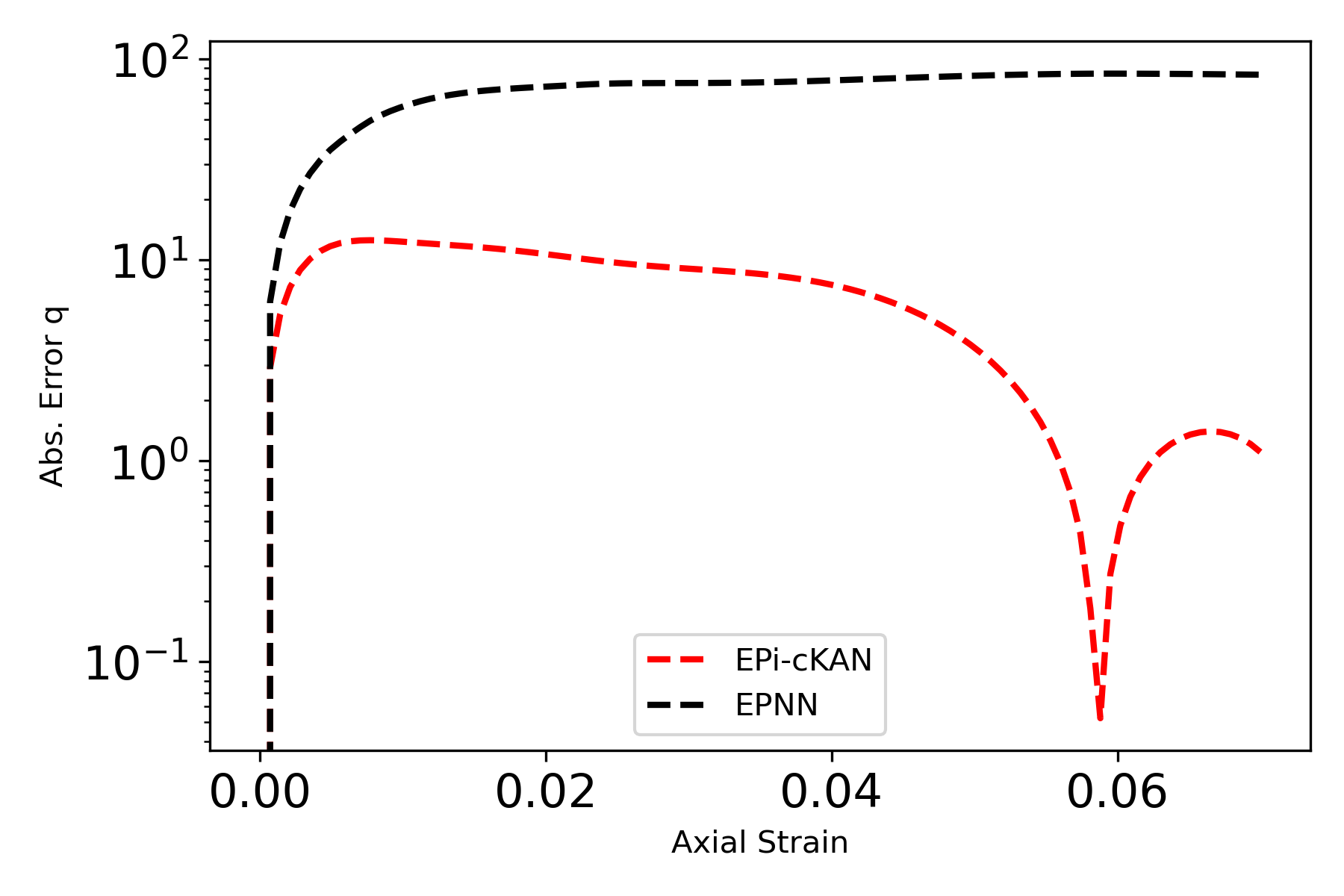}\hspace{.1cm}}
\subfigure[\(\xi = -2.5\)]
{ \label{Pic.Size25ERROR}\includegraphics[width=.30\textwidth]{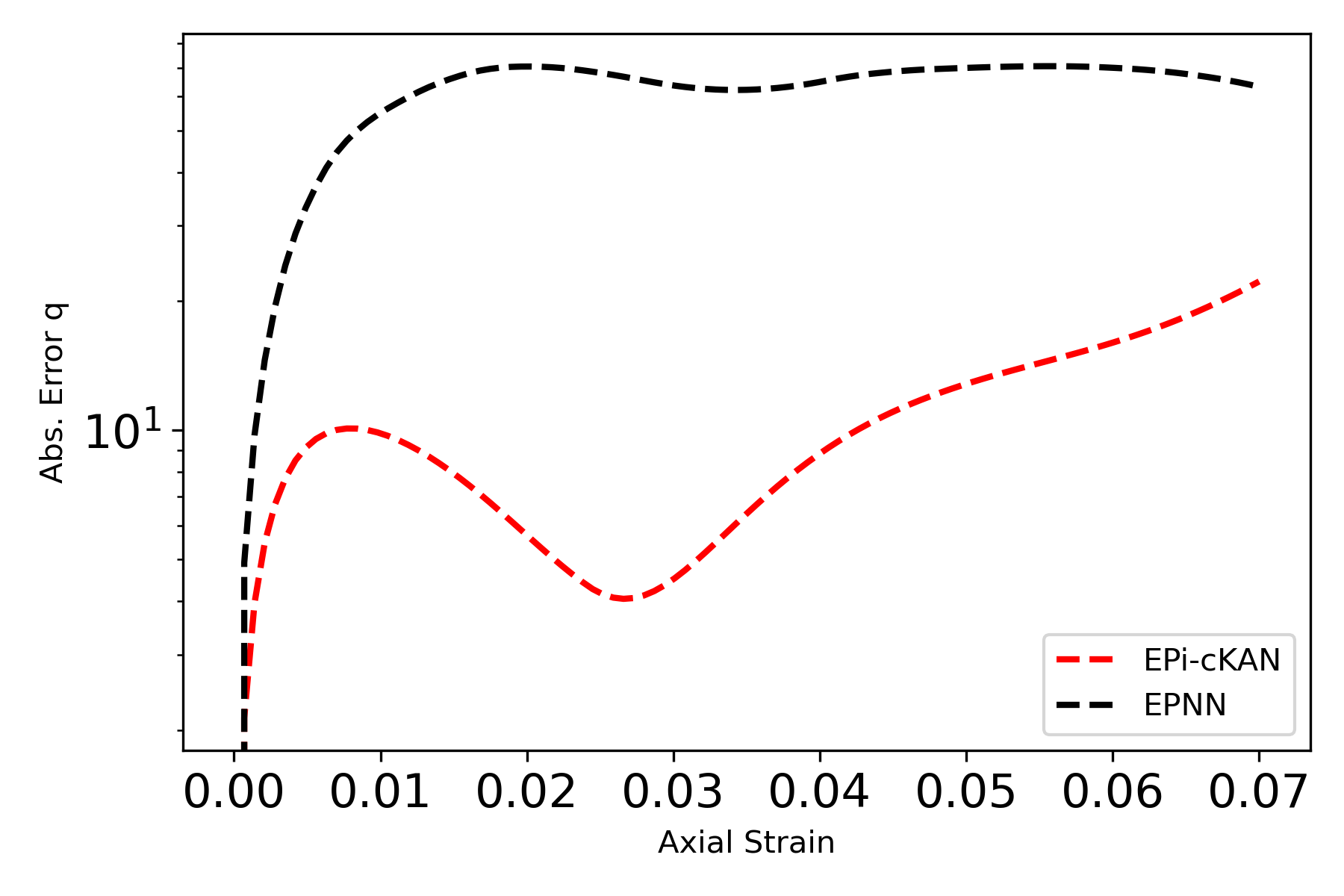}\hspace{.1cm}}

\subfigure[\(\xi = -4.0\)]
{ \label{Pic.Size4}\includegraphics[width=.30\textwidth]{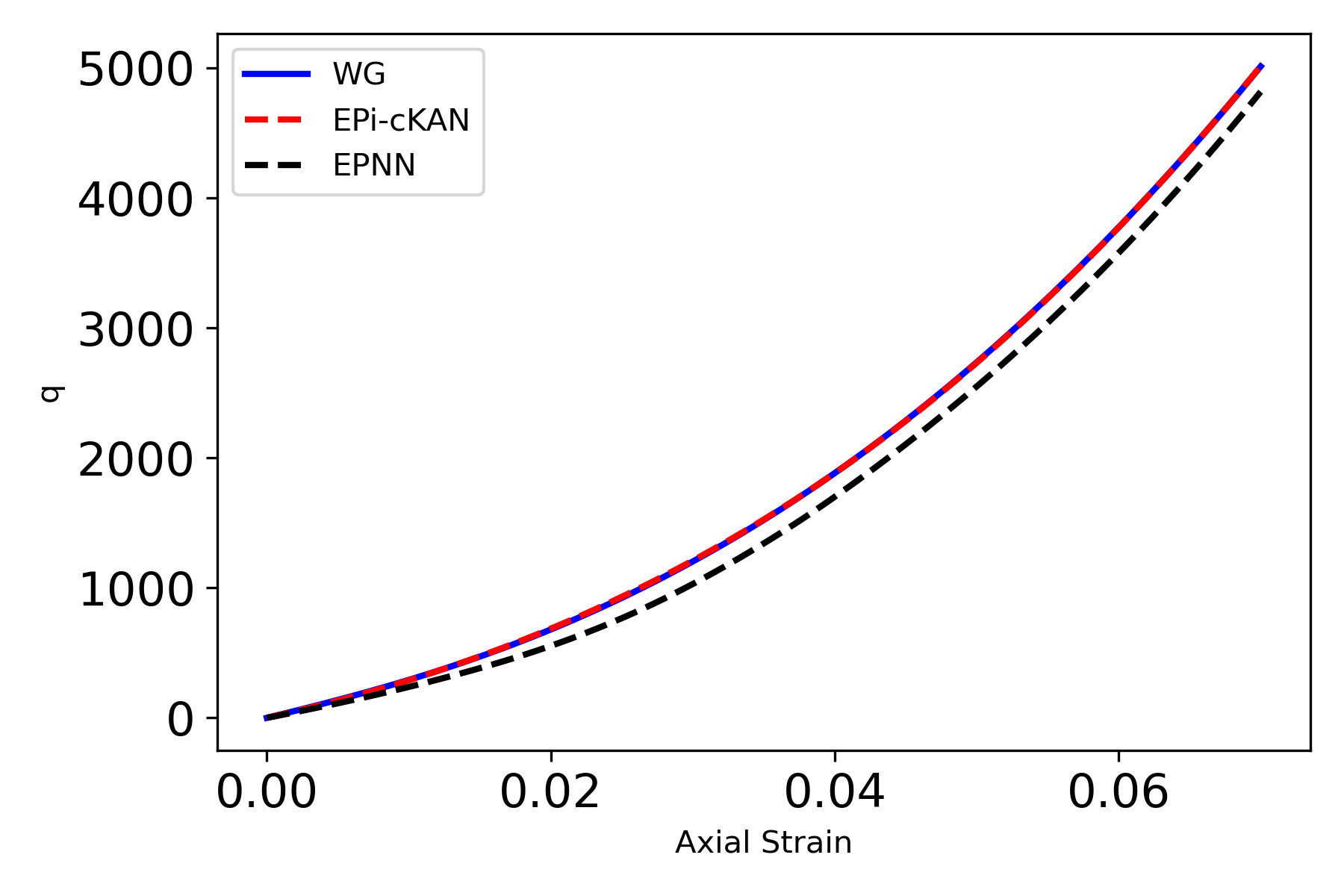}\hspace{.1cm}}
\subfigure[\(\xi = -10.0\)]
{ \label{Pic.Size10}\includegraphics[width=.30\textwidth]{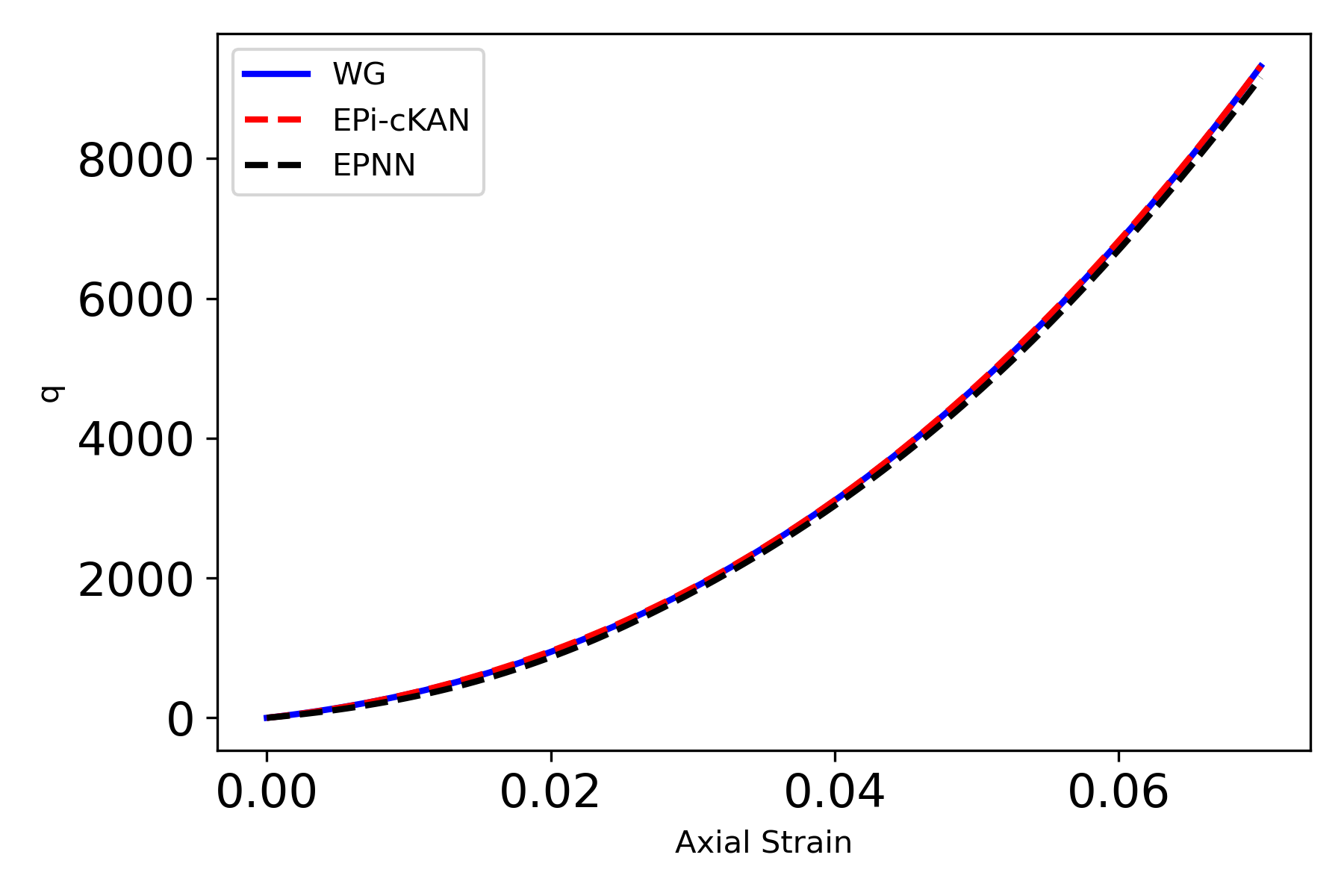}\hspace{.1cm}}
\subfigure[\(\xi = -1000.0\)]
{ \label{Pic.Size1000}\includegraphics[width=.30\textwidth]{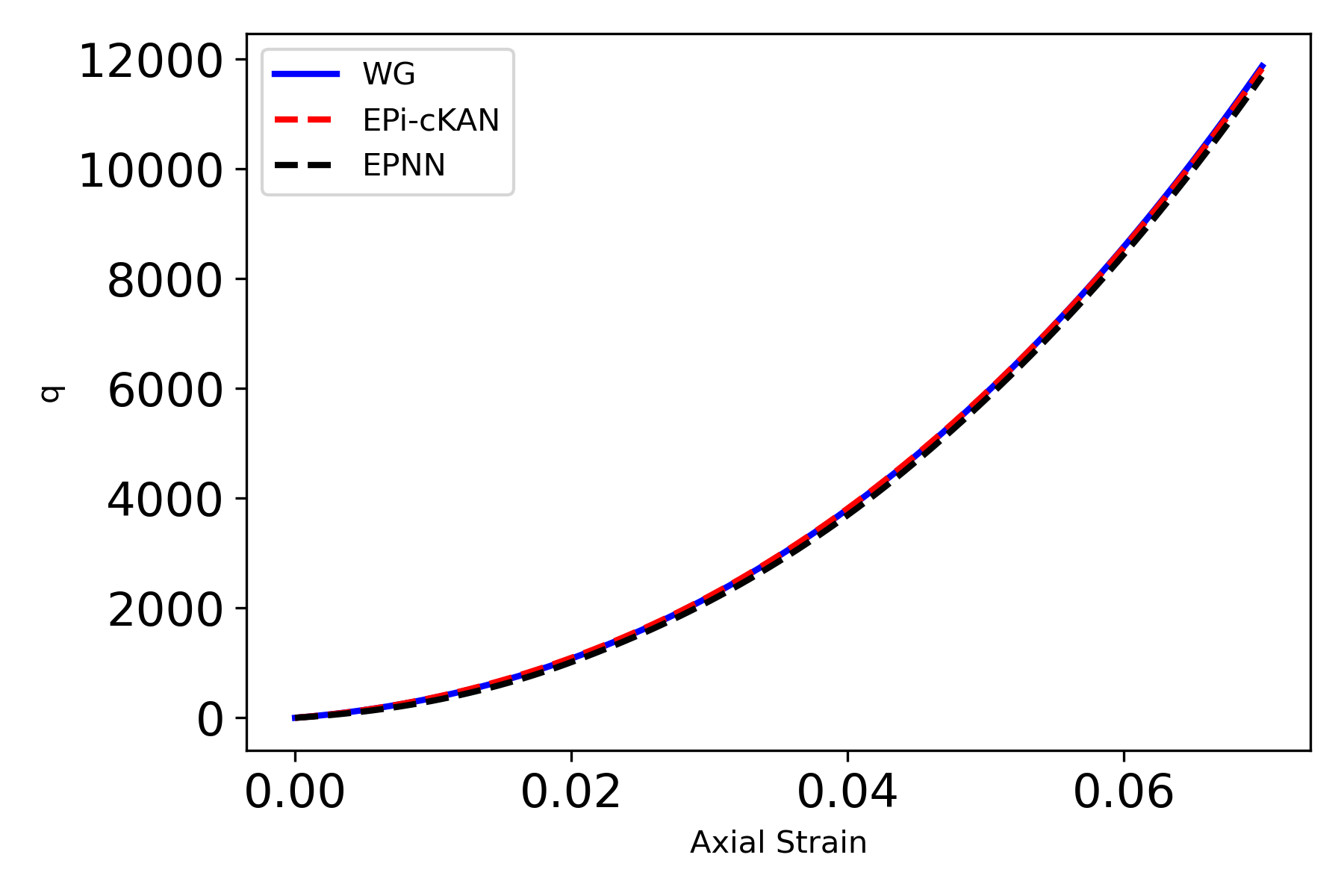}\hspace{.1cm}}

\subfigure[\(\xi = -4.0\)]
{ \label{Pic.Size4ERROR}\includegraphics[width=.30\textwidth]{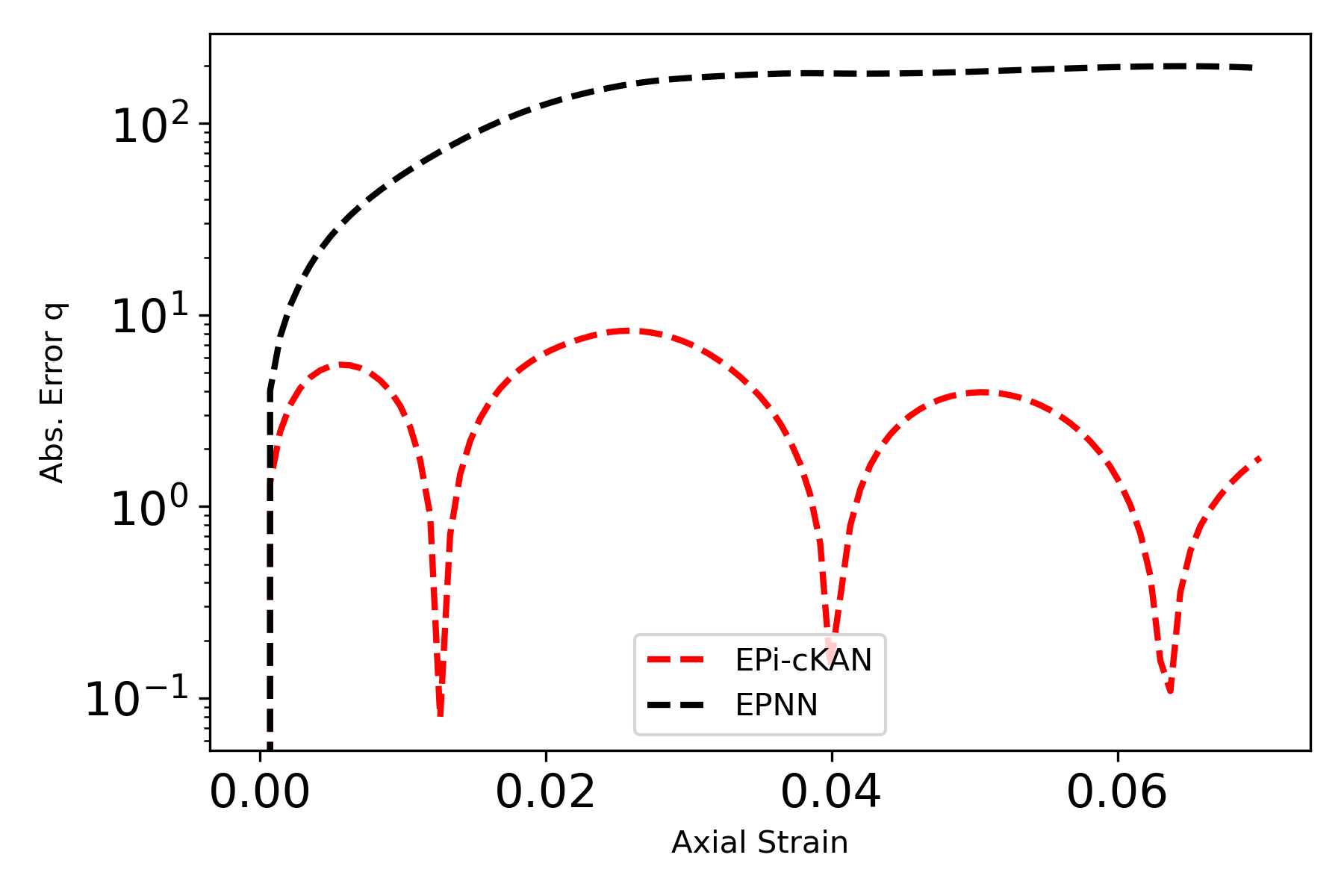}\hspace{.1cm}}
\subfigure[\(\xi = -10.0\)]
{ \label{Pic.Size10ERROR}\includegraphics[width=.30\textwidth]{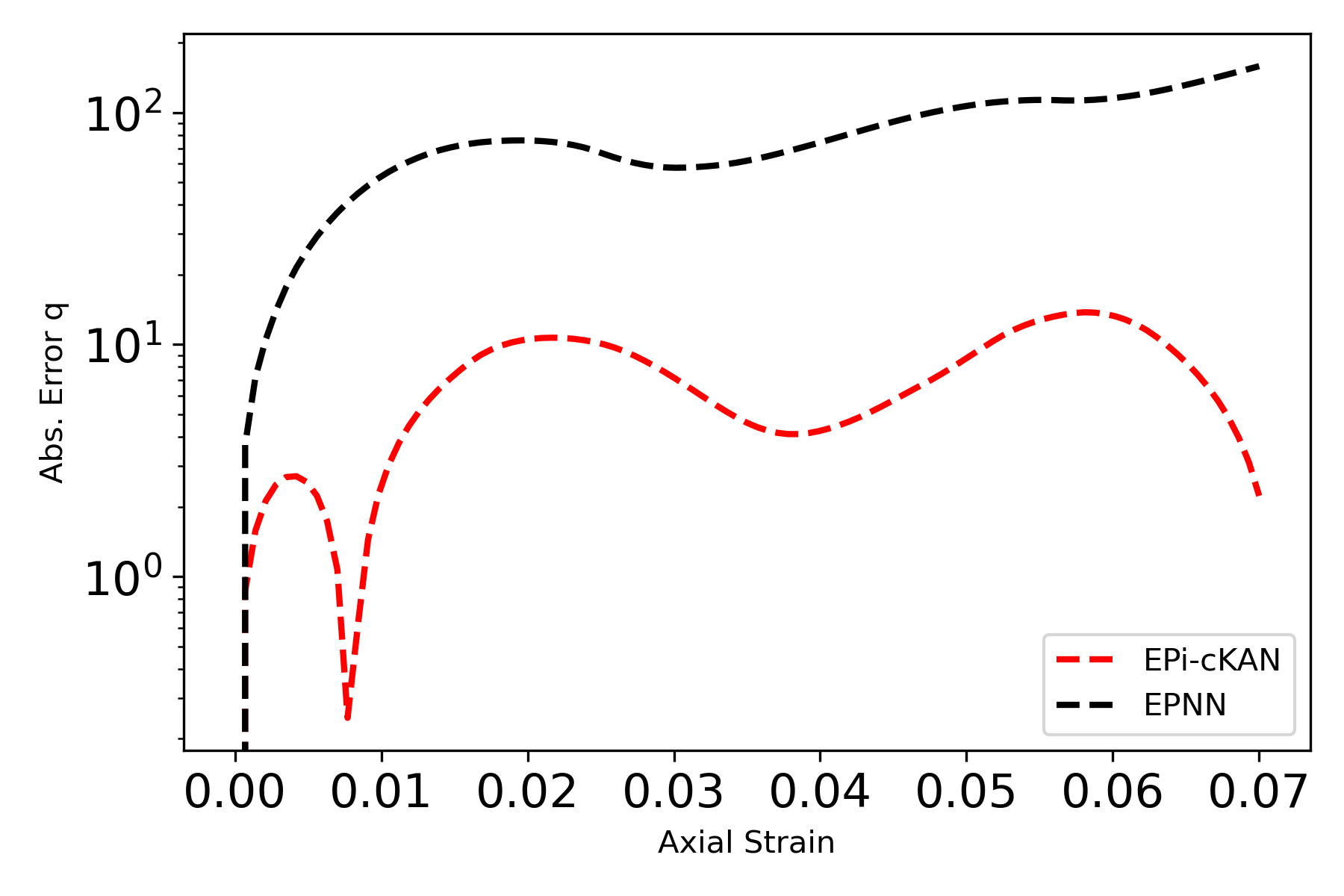}\hspace{.1cm}}
\subfigure[\(\xi = -1000.0\)]
{ \label{Pic.Size1000ERROR}\includegraphics[width=.30\textwidth]{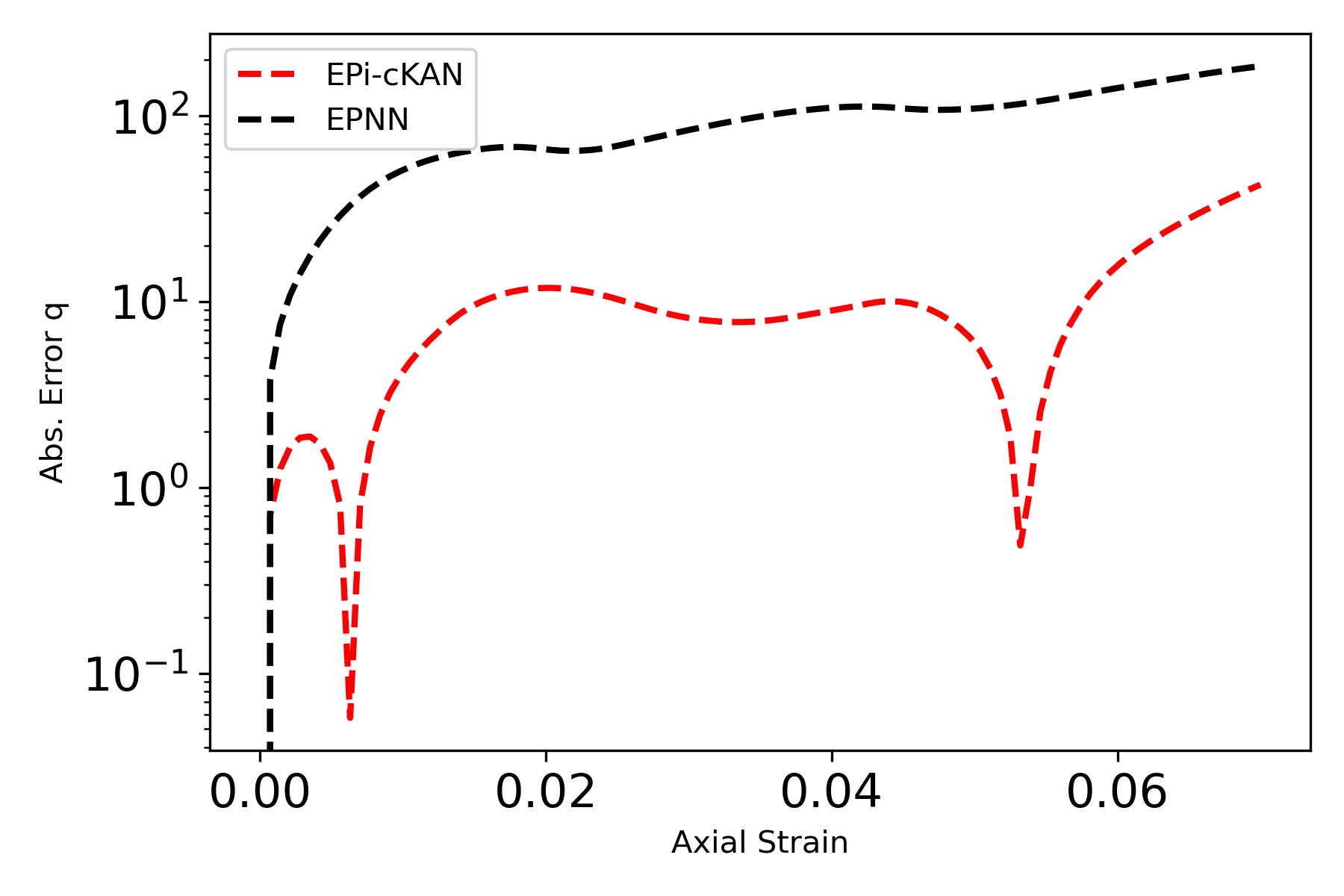}\hspace{.1cm}}

\caption{Comparison between the stress predictions of elasto-plastic based architectures (trained for \(2e4\) epochs) and the ground truth (numerical integration results), in the triaxial axisymmetric test with \(p^{in}=375\) kPa and \(e^{in}=0.64\). Models used here are trained on 187000 training data with the network structures reported in Table \ref{Tab.Settings}. First and third rows: \(q\) (kPa) vs \(\varepsilon_{33}\), second and forth rows: \(\vert q^{*}-q\vert\) vs \(\varepsilon_{33}\).} \label{Pic.Size60_EP}
\end{figure}

\subsection{Second  Phase: Blind Testing}\label{Sec.secondphase}
This section deploys trained networks on elasto-plastic behavior of Ottawa sand to predict the behavior of material under blind strain-controlled loading path. Trained networks are applied repeatedly, and during each step, the void ratio as well as the stress and strain tensor components are updated based on the outputs from the sub-networks. These updated values serve as feedback input for the next step. A feedback test is crucial for evaluating the accuracy of proposed data-driven constitutive models as it helps to identify if minor errors can build up over time. Should these errors increase progressively, they might lead the models’ predictions to diverge from the actual behavior. Conducting this test allows us to assess the models’ stability and their capacity to retain accuracy during extended simulations or when conditions change, thus ensuring that the results do not significantly stray from the true response.

The method for generating blind loading paths is detailed in Section \ref{Sec.Blindpath}, where we outline the procedure for creating paths that are not part of the training set, but are used for model testing. According to the main components of the WG model outlined in Section \ref{WGModel}, we present the steps required to use trained models and infer them to predict the blind strain-controlled loading path, as follows:

\textbf{Stage 1:} 
    Initially, we set the state variables at the starting step, $n=0$, which includes \(e^{n}\), \(\boldsymbol{\sigma}^{n}\), \(\boldsymbol{\varepsilon}^{n}\), and \(\boldsymbol{\varepsilon}^{p, n}\), and determine the total number of loading steps (\(n_{\text{step}}\)). Here, we set \(n_{\text{step}} = 100\). The superscript \(n\) indicates that these variables correspond to the \(n\)-th step. 
    Given \(p^{\text{in}}\) and \(e^{\text{in}}\), we set \(\sigma_{ii}^n = p^{\text{in}}, e^{n} = e^{\text{in}}\), and \(\varepsilon_{ii}^{0} = \varepsilon_{ii}^{p, n} = 0\). This initialization establishes the starting conditions for our analysis and ensures that subsequent calculations accurately reflect the evolving state of the system throughout the loading process. By defining the state variables in this manner, we can systematically track changes and responses as we progress through each loading step.
 
\textbf{Stage 2:} 
     In the second stage, we define the strain increment, denoted as \(\Delta \boldsymbol{\varepsilon}^n\). To determine this increment, it is essential to know the final values of the strain components, specifically \(\boldsymbol{\varepsilon}^{\text{final}}\). With this information, we can calculate the strain increment using the formula \(\Delta \boldsymbol{\varepsilon}^n = (1 / n_{\text{step}}) \boldsymbol{\varepsilon}^{\text{final}}\). This calculation divides the total final strain by the number of loading steps, allowing us to distribute the strain incrementally across each step. This approach ensures that the strain evolution is handled systematically throughout the loading process.
     
\textbf{Stage 3:} 
     In the third stage, we utilize trained networks and infer them by providing \(e^{n}\), \(\boldsymbol{\sigma}^{n}\), \(\boldsymbol{\varepsilon}^{n}\), \(\boldsymbol{\varepsilon}^{p, n}\), and \(\Delta \boldsymbol{\varepsilon}^n\) as inputs, and compute the corresponding outputs, specifically \(\Delta e^{n}\), \(\Delta \boldsymbol{\sigma}^{n}\), and \(\Delta \boldsymbol{\varepsilon}^{p, n}\).
     
\textbf{Stage 4:} 
      Once the network outputs are obtained for step $n$, we then update the state variables for  step $n+1$ using,
      \begin{equation}
         \begin{array}{cc}
            e^{n+1} \gets e^{n} + \Delta e^{n},  & 
            \boldsymbol{\sigma}^{n+1} \gets \boldsymbol{\sigma}^{n} + \Delta \boldsymbol{\sigma}^{n},\\
             \boldsymbol{\varepsilon}^{n+1} \gets \boldsymbol{\varepsilon}^{n} + \Delta \boldsymbol{\varepsilon}^{n}, & 
             \boldsymbol{\varepsilon}^{p, n+1} \gets \boldsymbol{\varepsilon}^{p,n} + \Delta \boldsymbol{\varepsilon}^{p,n}.
         \end{array} 
      \end{equation}
      These updates are crucial for accurately reflecting the changes occurring within the system as we progress through the loading steps. By systematically adjusting these variables, we ensure that our model effectively captures the material's evolving response over time. Furthermore, as we compute these state variables, we can also derive other important quantities such as \(p\), \(q\), \(\varepsilon^{v, p}\), and \(\gamma^p\) using Eqs.~\eqref{Eq.QP}-\eqref{Eq.gammap} in Section \ref{WGModel}. This comprehensive updating process allows us to maintain a detailed understanding of the material's behavior under varying loading conditions, enabling more accurate predictions of its performance throughout the analysis.
      
\textbf{Stage 5:} 
Finally, we increment the loading step \(n\) by one unit and proceed to stage 3. This process continues until \(n\) reaches the total number of loading steps, denoted as \(n_{\text{step}}\). By systematically advancing \(n\) through each step, we ensure that all loading conditions are thoroughly evaluated, allowing us to capture the complete response of the system throughout the entire loading sequence. This iterative approach is essential for accurately modeling the behavior of the material under varying conditions.

In each test scenario, it is presumed that the strain-controlled path is linear, with the material incrementally loaded to an axial strain, i.e., deformation in the loading direction, of \(\varepsilon_{33} = 0.07\). These tests are carried out under controlled conditions with constant initial parameters throughout, specifically: the initial confining pressure \(p^{\text{in}} = 375 \, \text{kPa}\) and the initial void ratio \(e^{\text{in}} = 0.64\).
Moreover, besides the predictions made by trained models for these blind tests, the corresponding ground truth values, denoted by the exponent symbol (*), are also derived through direct numerical integration of the WG governing equations, refer to Section \ref{Sec.DataG}.

Figure \ref{Pic.Fig24_cKAN} illustrates a comparison between various cKAN architectures in predicting an axisymmetric triaxial test with \(\xi = -2.5\).
Panels (a) and (b) in Fig. \ref{Pic.Fig24_cKAN}, with their error plots in panels (d) and (e), reveal that for plastic strain, the Serial cKAN architecture performs best at certain strain levels, while both EPi-cKAN and Parallel cKAN display larger errors.
In panel (c) and its error plot in panel (f) in Fig. \ref{Pic.Fig24_cKAN}, all models exhibit similar behavior at lower strain levels, but as the strain increases, EPi-cKAN maintains the lowest error, further demonstrating its robustness.
Panels (g) and (h) in Fig. \ref{Pic.Fig24_cKAN}, along with their corresponding error plots in panels (j) and (k), show that the EPi-cKAN consistently achieves lower errors compared to the other cKAN methods, particularly at higher strain levels, highlighting its superior performance. 
Moreover, panel (i) in Fig. \ref{Pic.Fig24_cKAN} compares the stress-strain relationship between cKAN architectures and shows that EPi-cKAN is closer to the WG model, indicating higher accuracy in approximating the stress response under the conditions modeled.

Figure \ref{Pic.Fig24_EP} presents a comparison of various elasto-plastic architectures under an axisymmetric triaxial test with \(\xi = -2.5\). 
Panels (a) and (b) in Fig. \ref{Pic.Fig24_EP}, alongside their respective error plots in panels (d) and (e), suggest that EPNN achieves lower errors in predicting plastic strain at certain strain levels, whereas EPi-cKAN exhibits comparatively larger deviations.
In Fig. \ref{Pic.Fig24_EP}, panel (c) and its error in (f) indicate that EPi-cKAN provides a more precise approximation of the void ratio, maintaining minimal error throughout.
In panels (g) and (h) in Fig. \ref{Pic.Fig24_EP}, along with the corresponding error plots in (j) and (k), the EPi-cKAN model consistently shows reduced error when compared to EPNN, particularly at higher strain levels, demonstrating its enhanced accuracy.

In a similar manner to the previous analyses, we evaluate the performance of EPi-cKAN on a different path with \(\xi = -1.75\), comparing it against other cKAN architectures as well as EPNN. Figure \ref{Pic.Fig25_cKAN} illustrates these comparisons between the cKAN models. 
According to panels (a) and (b) in Fig. \ref{Pic.Fig25_cKAN} and their error plots in panels (d) and (e), the serial cKAN architecture demonstrates superior accuracy in predicting both volumetric and plastic strain at smaller strain levels. However, at higher axial strains, the difference in performance becomes smaller, with parallel cKAN catching up to or even doing better than serial cKAN, while EPi-cKAN remains less accurate than both.
In contrast, as shown in panels (c) and (f) in Fig. \ref{Pic.Fig25_cKAN}, EPi-cKAN outperforms the others in terms of void ratio approximation, especially at higher strain levels, where it maintains the smallest error.
Moreover, in Fig. \ref{Pic.Fig25_cKAN}, panels (g) trough (k) reveal that EPi-cKAN consistently delivers more accurate stress predictions, resulting in lower errors for both \(p\) and \(q\) relative to the other cKAN methods.

In Fig. \ref{Pic.Fig25_EP}, a comparison between the EPi-cKAN and EPNN models is presented. 
Panels (a) and (b) in Fig. \ref{Pic.Fig25_EP} and their error plots in panels (d) and (e), show that EPNN generally achieves lower absolute errors at certain axial strain levels compared to EPi-cKAN, particularly in predicting plastic strain and volumetric plastic strain. While EPi-cKAN exhibits slightly larger deviations in some cases, both models demonstrate similar trends as axial strain increases, with EPi-cKAN still performing reasonably well across the range.
Conversely,  according to panels (c) and and (f) in Fig. \ref{Pic.Fig25_EP}, EPi-cKAN provides a more precise approximation of the void ratio, especially when the axial strain increases.
 Similarly, From panels (g)-(k), it is evident that EPi-cKAN achieves superior accuracy in predicting stress, leading to consistently lower error values for both \(p\) and \(q\) when compared to the EPNN. This improved performance stems from EPi-cKAN’s ability to incorporate the underlying physical laws of elasto-plasticity more effectively, resulting in a more robust and insightful learning process.

Generally, based on Figs. \ref{Pic.Fig24_cKAN}-\ref{Pic.Fig25_EP}, panel (c), EPi-cKAN demonstrates strong performance in accurately predicting the void ratio sub-network. This highlights the model's capacity to capture the complex behavior of void ratio with high precision.
Moreover,according to Figs. \ref{Pic.Fig24_cKAN}-\ref{Pic.Fig25_EP}, panels (g) through (k), EPi-cKAN demonstrates a significant improvement in stress path predictions compared to other models. This enhanced performance can be attributed to its physics-informed architecture, which integrates domain knowledge into the learning process. Additionally, the advanced basis functions used in EPi-cKAN, particularly the combination of a ReLU activation function with Chebyshev polynomials, allow the network to capture complex, nonlinear stress-strain relationships more accurately. This synergy between the physics-driven design and the sophisticated mathematical framework ensures better generalization and precision in predicting the stress paths under various loading conditions.
However, in panels (a) and (b), which correspond to the plastic strain sub-network, EPi-cKAN does not perform as well, showing a noticeable gap in accuracy. This suggests that while the model excels in some aspects, such as void ratio and stress prediction, it may require further refinement or tuning to improve its ability to model plastic strain with the same level of accuracy.

In Fig. \ref{Pic.Size60_EP}, we compare the behavior of  EPi-cKAN and EPNN in approximation stress for different \(\xi\).
The test uses models trained on 187000 training data with the network structures reported in Table \ref{Tab.Settings}.
It can be seen that EPi-cKAN continues to outperform EPNN as the amount of training data increases. The accuracy of stress path predictions improves significantly in EPi-cKAN due to its physics-informed architecture and advanced basis functions, while EPNN struggles to achieve similar levels of precision. Table \ref{Tab.REQ22} presents the relative $\mathcal{L}^2$ error in predicting deviatoric stress components in triaxial axisymmetric tests for the EPi-cKAN and EPNN models compared to the WG method for varying values of \(\xi\). It can be seen that EPi-cKAN consistently surpasses EPNN for every \(\xi\). This demonstrates EPi-cKAN's superior ability to generalize and capture complex behaviors with a larger dataset, further strengthening its advantage over traditional neural network models.


\begin{table}[!h]
\centering
\caption{\label{Tab.REQ22}
Relative \(\mathcal{L}^2\) error in predicting deviatoric stress components in  triaxial axisymmetric tests for EPi-cKAN and EPNN models compared to the WG method across different values of \(\xi\).}
\renewcommand{\arraystretch}{1.2}
\setlength{\tabcolsep}{12pt}
\begin{tabular}{c|cccccc}
\toprule
\(\xi\) & \(-1.75\) & \(-2\) & \(-2.5\)& \(-4.0\) & \(-10.0\)& \(-1000.0\)\\
\midrule
EPi-cKAN & 4.73 \% & 2.58\% & 1.26\% & 0.17\% & 0.18\% & 0.25\% \\
EPNN & 60.48\% & 23.80\% & 7.11\% &6.67\% & 2.07\% & 1.93\% \\
\bottomrule
\end{tabular}
\end{table}


Figures \ref{Pic.Fig24_cKAN} to \ref{Pic.Fig25_EP} collectively illustrate the EPi-cKAN's strong performance in predicting the void ratio and stress paths, attributed to its physics-informed design and the spectral properties of Chebyshev polynomials. However, the results reveal a gap in accuracy for the plastic strain sub-network, suggesting room for improvement.
To enhance overall accuracy, one effective strategy is to design and experiment with more efficient activation functions tailored to cKANs. These functions could combine the strengths of both smooth nonlinear transformations and computational simplicity.

Also, despite the impressive accuracy of the EPi-cKAN architecture, cKAN architectures generally exhibit a higher computational overhead while training, as indicated in Table \ref{Tab.Settings}. This is primarily due to the intricate structure of cKANs, which introduces complexities in function compositions and Chebyshev polynomial evaluations. To enhance computational efficiency and reduce unnecessary calculations, several strategies can be employed. These include parallelizing computations across the architecture and utilizing sparse representations of Chebyshev polynomials. Such optimizations can significantly mitigate the computational burden associated with cKANs, making them more practical for large-scale applications. In addition, the EPi-cKAN method, according to Fig. \ref{Pic.Losses}, the loss function exhibits a slower reduction rate. Although this does not undermine the effectiveness of the EPi-cKAN approach, one potential strategy to accelerate convergence is the implementation of a step-decay schedule. This technique dynamically reduces the learning rate at predefined intervals, allowing the model to make larger updates initially and then refine the solution more gradually, ultimately improving the overall efficiency without compromising the model's accuracy.

This study primarily focused on evaluating the EPi-cKAN architecture by assessing its accuracy, computational time, and its ability to effectively generalize when approximating unseen paths.
Looking ahead, the interpretability of cKAN, an important feature, can be further investigated by examining how its structure, based on the Kolmogorov-Arnold representation, decomposes complex functions into interpretable univariate components. This exploration could include analyzing the relationship between individual Chebyshev polynomials and the underlying physical or mathematical processes they model. Moreover, employing techniques such as sensitivity analysis or visualizing the contributions of each polynomial to the overall solution may yield deeper insights into how cKANs capture and represent patterns, ultimately enhancing their practical interpretability.

\section{Conclusion}\label{Sec.Conclusion}






This work presented a new architecture known as the elasto-plasticity informed Chebyshev-based Kolmogorov-Arnold Network (EPi-cKAN), aimed at tackling the complexities of modeling non-linear elasto-plastic behavior in granular materials such as sand. By incorporating incremental elasticity and strain decomposition within the network's architecture and loss function, and employing augmented Chebyshev polynomials for parameterization, EPi-cKAN endeavored to significantly improve both accuracy and generalization. Comprehensive comparisons with MLP-based and c-KAN methods were performed to demonstrate EPi-cKAN's superior accuracy in modeling elasto-plastic behavior, its ability to generalize and predict strain-controlled paths beyond the training data distribution, and its efficiency in attaining higher accuracy with fewer parameters and limited data.   Although it carries a slightly higher computational cost stems from the sophisticated structure of cKANs, we demonstrated that EPi-cKAN substantially surpasses traditional MLPs, especially when predicting stress on unknown paths. For example, for a blind triaxial axisymmetric  test with \(\xi = -1.75\), the EPi-cKAN model predicted the components of the deviatoric stress with an error margin of 5\% relative to the ground truth, whereas the error was approximately 60\% for a physics-informed MLP-based conterpart, EPNN. This advantage is attributed to its physics-informed architecture and the integration of Chebyshev polynomials, which significantly enhance its generalization ability for blind strain-controlled paths. Our findings thus highlight EPi-cKAN's potential as a robust and reliable tool for replacing or supplanting constitutive models describing non-linear materials behavior various engineering and scientific domains.

\section{Acknowledgements}
S.A.F. acknowledges the supports by the U.S. Department of Energy’s Office of Environmental Management  (award no.: DE-EM0005314), and the American Chemical Society Petroleum Research Fund (award no. 66452-DNI9).

\section{Conflict of Interest}
The authors declare no conflict of interests.

\def\mybibdoicolor{\color{black}}
\newcommand*{\doi}[1]{\href{\detokenize{#1}} {\raggedright\mybibdoicolor{DOI: \detokenize{#1}}}}

\bibliographystyle{unsrtnat}
\bibliography{references.bib}

\end{document}